\documentclass[sensors,article,accept,moreauthors,pdftex,10pt,a4paper]{mdpi} 
\firstpage{1} 
\articlenumber{0000}
\doinum{10.3390/------}
\pubvolume{xx}
\pubyear{2016}
\copyrightyear{2016}
\externaleditor{Academic Editor: name}
\history{Received: date ; Accepted: date ; Published: date}

 \theoremstyle{mdpi}
 \newcounter{thm}
 \newcounter{ex}
 \newcounter{re}
 \theoremstyle{mdpidefinition}

\usepackage{supertabular}  
\usepackage{afterpage}
\usepackage{multirow}
\usepackage{longtable}
\usepackage{url}
\usepackage[table]{xcolor}
\usepackage{subcaption}
\usepackage{color}
\usepackage{diagbox}
\usepackage{gensymb}
\usepackage{float}
\usepackage{subcaption}
\usepackage{amssymb}

\usepackage{acronym}
\newacro{AoA}[AoA]{Angle of Arrival}
\newacro{API}[API]{Application Programming Interface}
\newacro{BI}[BI]{Business Intelligence}
\newacro{BLE}[BLE]{Bluetooth Low Energy}
\newacro{BT}[BT]{Bluetooth}
\newacro{COTS}[COTS]{Commercial Off-The-Shelf}
\newacro{CPS}[CPS]{Cyber-Physical Systems}
\newacro{FSM}[FSM]{Finite State Machine}
\newacro{HTTP}[HTTP]{Hyper Text Transfer Protocol}
\newacro{IoT}[IoT]{Internet of Things}
\newacro{LAN}[LAN]{Local Area Network}
\newacro{MES}[MES]{Manufacturing Execution System}
\newacro{MIMO}[MIMO]{Multiple-input Multiple-output}
\newacro{MRC}[MRC]{Maximum-Ratio Combiner}
\newacro{MSE}[MSE]{Minimum-Squared Error}
\newacro{RFID}[RFID]{Radio Frequency IDentification}
\newacro{REST}[REST]{REpresentational State Transfer}
\newacro{RSS}[RSS]{Received Signal Strength}
\newacro{RSSI}[RSSI]{Received Signal Strength Indicator}
\newacro{SBC}[SBC]{Single Board Computer}
\newacro{SC}[SC]{Selection Combiner}
\newacro{ScanC}[ScanC]{Scanner Combiner} 
\newacro{SSC}[SSC]{Switch-and-Stay Combiner}
\newacro{TDoA}[TDoA]{Time Different of Arrival}
\newacro{ToA}[ToA]{Time of Arrival}
\newacro{URI}[URI]{Uniform Resource Identifier}
\newacro{UWB}[UWB]{Ultra-Wide Band}
\newacro{Wi-Fi}[Wi-Fi]{Wireless Fidelity}
\newacro{WPAN}[WPAN]{Wireless Personal Area Network}
\newacro{WSN}[WSN]{Wireless Sensor Networks}

\extrafloats{100}

\Title{Smart Pipe System for a Shipyard 4.0}

\Author{Paula Fraga-Lamas, Diego Noceda-Davila, Tiago M. Fern\'andez-Caram\'es, Manuel A. D\'iaz-Bouza, Miguel Vilar-Montesinos}

\address{ \quad Unidad Mixta de Investigaci\'on Navantia-UDC, Universidade da Coru\~na, Edificio Talleres Tecnol\'ogicos, Mendiz\'abal s/n, 15403, Ferrol, Spain; 
E-mails: paula.fraga@udc.es, diego.noceda@udc.es, tiago.fernandez@udc.es, mdiaz@navantia.es, mvilar@navantia.es}

\corres{Correspondence: paula.fraga@udc.es; Tel.: +34 981167000 x 6051}

\abstract{As a result of the progressive implantation of the Industry 4.0 paradigm, many industries are experimenting a revolution that shipyards cannot ignore. Therefore, the application of the principles of Industry 4.0 to shipyards are leading to the creation of Shipyards 4.0. Due to this, Navantia, one of the ten largest shipbuilders in the world, is updating its whole inner workings to keep up with the near-future challenges that a Shipyard 4.0 will have to face. Such challenges can be divided into three groups: the vertical integration of production systems, the horizontal integration of a new generation of value creation networks, and the re-engineering of the entire production chain, making changes that affect the entire life cycle of each piece of a ship. Pipes, which exist in a huge number and varied typology on a ship, are one of the key pieces, and its monitoring constitutes a prospective cyber-physical system. Their improved identification, traceability, and indoor location, from production and through their life, can enhance shipyard productivity and safety. In order to perform such tasks, this article first conducts a thorough analysis of the shipyard environment. From this analysis, the essential hardware and software technical requirements are determined. Next, the concept of smart pipe is presented and defined as an object able to transmit signals periodically that allows for providing enhanced services in a shipyard. In order to build a smart pipe system, different technologies are selected and evaluated, concluding that passive and active RFID (Radio Frequency Identification) are currently the most appropriate technologies to create it. Furthermore, some promising indoor positioning results obtained in a pipe workshop are presented, showing that multi-antenna algorithms and Kalman filtering can help to stabilize Received Signal Strength (RSS) and improve the overall accuracy of the system.}


\begin{document}

\section{Introduction}
\label{sec:Intro}

After the triumph of the lean production systems in the 1970s, the outsourcing manufacturing phenomenon of the 1990s, and the automation that took off in the 2000s, the fourth major disruption in modern manufacturing is Industry 4.0. This industrial revolution can be defined as the next phase in the digitalization of the sector \cite{McKinsey2015}, driven by several emerging technologies: the ubiquitous use of sensors, the stunning rise in data volume, the increasing computational power, and connectivity; the emergence of analytics, cloud computing and business-intelligence capabilities; new forms of human-machine interaction such as augmented-reality systems; and advances in transferring digital instructions to the physical world, such as \ac{CPS}, \ac{IoT}, robotics, and 3-D printing. Most of these technologies are mature and they have been present for some time. Although some of them are not yet ready for a broader application, many are now at a position where their greater reliability and cost-effectiveness are starting to be appealing for industrial applications. 

In the short-term, Industry 4.0 is expected to have a major effect on global economies. PwC's 2016 Global Industry 4.0 Survey \cite{PwC} suggests that annual digital investments are expected to achieve US \$907 bn per year through 2020.
Survey respondents anticipate that those investments will lead to US \$493 bn in additional revenues annually. Furthermore, savings are estimated at US \$421 bn in costs and efficiency gains each year.

The foundations of the Industry 4.0 can be transferred straight to a Shipyard 4.0. The deployment of Cyber-Physical Systems in production systems gives birth to the "smart factory" and, analogously, to the "smart shipyard". Products, resources, and business and engineering processes are deeply integrated making production operate in a flexible, efficient, and green way with constant real-time quality control, and cost advantages in comparison with traditional production systems. Machinery and equipment will have the ability to improve processes through self-optimization and autonomous decision-making. Shipbuilders face the same challenges as industry \cite{Wang2016}, which can be classified into three main concerns: the vertical integration of production systems, the  horizontal integration of a new generation of networks that create added-value, and the acceleration of technologies that require the re-engineering of the entire production chain.

The vertical integration of production systems changes naval production chains. It entrusts the intelligent shipyards to ensure safe production. The smart ships, more environmental friendly, are capable of operating in network together with other ships and ground infrastructure. The horizontal integration of a new generation of value creation networks is critical as it provides an integrated way to satisfy the demands from the different stakeholders, allowing for the customization of ships in a short period of time.

The third challenge is the end-to-end digital integration of engineering across the entire value chain, ranging from design to after-sales service. This evolution implies introducing disrupting technologies that affect the entire life cycle of each piece of the ship, from acceleration technologies, such as artificial intelligence, robotics, virtual reality, driverless vehicles for the transport of parts, drones, remote sensing networks or 3D/4D printing, among others. The aim of these technologies is, primarily, to allow shipyards to collect more data and make better use of it, for example: 
\begin{itemize}
\item Naval Command, Control, Communications, Computers, Intelligence, Surveillance and Reconnaissance (C4ISR) capabilities will be impacted by the development of a number of technologies based on the information extracted from the emerging data.
\item Curved 3D organic light emitting diode (OLED) displays will be supported by form factors that take advantage
of capabilities such as voice, handwriting, touch, gesture, eye movement, or even brain control. Designers will be able to interact with their designs without a keyboard or mouse, Human-Computer Interfaces (HCI) will encourage innovation and efficient design workflow. Such interfaces will be able to support more natural modes of interaction and will be more intuitive and therefore easier to operate, reducing the need for training.
\item Data obtained from remote sensing and intelligent algorithms will accelerate the ship design process, and 2D design will be easily converted into 3D.
\item Complex construction and inspection tasks will be supported by augmented reality.
\item Graphene strips, with allocated sensors alongside the hull, will provide more accurate data about the hull's working conditions. These will monitor external (seawater temperature, impacts, and fouling) and internal factors (stresses, microbial induced corrosion, and bending). This information will enable a new approach called
Hull-Skin-Data centered decisions that would be adopted according to those working parameters.
\item An increasing number of embedded sensors will be fitted to pipes so that new laser technologies and robotics can speed up the cutting process. Adaptable hull forms will be developed to tackle better different speed profiles and changing load conditions. Robots will also control the curvature of materials more precisely, thus offering optimal hull form. Moreover, a ballast free design will be further developed to reduce the transfer of marine invasive species across different waters.
\item Instead of leaving the majority of outfitting tasks until the moment after launching, some outfitting, such as piping and heavy machinery, will be developed together with the hull structure speeding up the building process.
\item Progressive sensorization process will enable automated casting, forging, rolling, cutting, welding or cleaning \cite{cleaning}.
\item Time spent on the outfitting along the quay will be minimized. Robotics will capture 3D images throughout the vessel and will establish a reference dataset to support real-time ship operations and life maintenance.
\item Enhanced crane-lifting capabilities will speed up production time.
\end{itemize}

Furthermore, with the development of applications based on these emerging technologies, a Shipyard 4.0  can leverage smarter energy consumption,  greater inbound/outbound logistics and information storage (asset utilization, supply/demand match, inventories, time to market), workforce safety and control (automation of knowledge work, digital performance management, human-robot collaboration, remote monitoring) and real-time yield optimization.

Navantia \cite{Navantia} is a Spanish naval company that offers integral solutions to its clients and which has the capacity required to assume responsibility over any naval program in the world, delivering fully operational vessels, and support throughout the service life of the product. Its main working areas are the design and construction of hi-tech military and civil vessels, the design and manufacturing of control and combat systems, overhauls and alterations of military and civil vessels, diesel engine manufacturing, and turbine manufacturing. Although Navantia has developed naval programs all around the world, at a domestic scale, Navantia's main customer is the Spanish Navy (this collaboration dates back 250 years). The high level of the Spanish Navy, with a worldwide operating capacity and collaborations with the most modern navies, allows Navantia to offer value added products.
Specifically, this article reviews the advances in one of the research lines of the Joint Research Unit Navantia-UDC (University of A Coru\~na).  

Pipes are a key part of ships: a regular ship contains between 15.000 and 40.000 pipes, whose use goes from fuel transportation or coolant for engines, to carry drinking water or waste. With such a huge number and varied typology, it is important to maintain the traceability and status of the pipes, what speeds up their maintenance procedures, accelerates locating them, and allows for obtaining easily their characteristics when building and installing them. 

This need for controlling and monitoring pipes can be approached by Cyber-Physical Systems (CPSs). A smart pipe system is a novel example of the benefits of CPS, providing a reliable remote monitoring platform to leverage environment, safety, strategic and economic benefits. While the physical plane focuses on the designs for sensing, data-retrieving, event-handling, communication, and coverage problems, the cyber plane focuses on the development of cross-layered and cross-domain intelligence from multiple environments and the interactions between the virtual and the physical world. In this paper, the physical plane is based on the concept of smart pipe, a sort of pipe able to transmit signals periodically that allows for providing useful services in a shipyard.

Today, the pipe management process varies depending on the shipyard, but, in general, it is performed in three different scenarios: the pipe workshop where they are built, and the block outfitting and the ship, where assembly takes place. This paper is focused on the pipe workshop, which is handled in a similar way in most shipyards.

\begin{figure}[b]
    \centering
        \includegraphics[width=1\columnwidth] {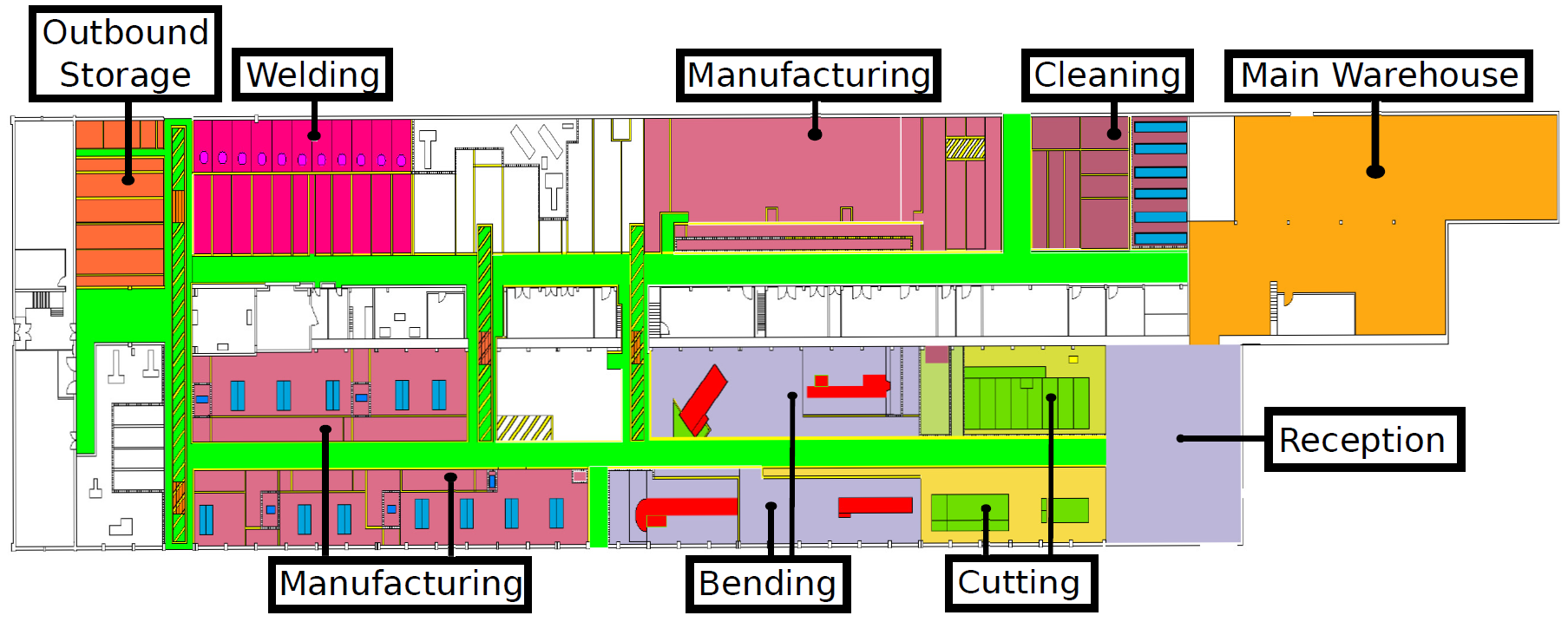}
		\caption{Floor map of the workshop.} 
        \label{figure:WorkshopFloorMap}
    \end{figure}

In Figure \ref{figure:WorkshopFloorMap} it is represented the floor map of the pipe workshop that Navantia owns in Ferrol (Galicia, Spain). The areas colored represent the main operative areas, while in white are offices and other secondary auxiliary areas. The following are the most relevant areas:

\begin{itemize}
\item Pipe reception. In this area raw pipes are stacked by the suppliers. It is divided into two different areas: small pipes are stored in a robotic storage, while large pipes are placed on the floor on diverse spots.

\item Cutting. This is where pipes are cut according to the engineering requirements.

\item Bending. Some pipes need to be bent to adapt them to the characteristics of the place where they will be installed on the ship.

\item Manufacturing. These are actually three areas of the workshop where operators add accessories and where pipes made of multiple sub-pipes are joined.

\item Provider's outbound storage. In times of excessive production load, some is derived to external providers. The outbound storage area is where providers collect the pipes and return them after their processing.

\item Welding. There are different booths where operators carry out welding tasks.

\item Cleaning. Before manufacturing, pipes have to be cleaned. This area contains bathtubs to expose pipes to hot water, acids, or pressurized water.

\item Main warehouse. This is where accessories and tool supplies are stored.

\end{itemize}

In this scenario, the way pipes are currently built (detailed next in Section \ref{SecDesc}) can be improved and optimized. In this article it is proposed a system of smart pipes that avoids paperwork and automates pipe identification, tracking, and traceability control. The system consists of a network of beacons that collect information about the location of the pipes continuously. Such information is provided by RFID tags that also contain information that allows operators to identify each pipe and determine how to process it at every stage.

The present paper is aimed at applying the latest research and the best technologies to build a smart pipe system for a shipyard, but it also includes the following novel contributions, which, as of writing, have not been found together in the literature:

\begin{itemize}
\item It presents the concept of Shipyard 4.0, which consists in the application of the principles of Industry 4.0 to a shipyard.

\item It describes in detail how a shipyard pipe workshop works and what are the requirements for building a smart pipe system.

\item The paper indicates how to build a positioning system from scratch in an environment as harsh in terms of communications as a shipyard. Furthermore, it was not found in the literature any practical analysis on the application of RFID technology in any similar application and scenario.

\item It defines the concept of smart pipe and shows an example of its implementation and the architecture that supports it. 

\item The article proposes the use of spatial diversity techniques to stabilize \ac{RSS} values, a kind of technique whose application in RFID systems has not been found in the literature. 

\end{itemize}

The remainder of this paper is organized as follows. Section \ref{WorkRelatedStateofTheArt} describes the process of pipe manufacturing in a modern shipyard and analyzes the technologies that can be used for identifying pipes. Section \ref{Design} details the design of the system, including the operational and hardware requirements, and the communications architecture.
Section \ref{Implementation} reviews the system modules and the RSS stabilization techniques proposed.
Section \ref{Experiments} describes the experimental setup and the tests performed with the technologies selected.  
 Finally, Section \ref{SecCon} is devoted to the conclusions.

\section{Related Work} \label{WorkRelatedStateofTheArt}

\subsection{Pipe manufacturing in a modern shipyard} \label{SecDesc}

The current procedure for managing the pipes in the workshop consists of the following steps:

\begin{enumerate}
\item Initially, pipes are placed in a storage area, where they will be collected by operators according to production needs. In the case of the shipyard that Navantia owns in Ferrol, two zones can be distinguished: one for small pipes and another for the large ones. The area for small pipes is an intelligent warehouse where an operator registers the pipes that arrive and then extracts them on demand according to the characteristics specified. Figure \ref{figure:stacking_large_pipes} shows the stacking area for large pipes, whose occupancy level is not determined automatically.

	\begin{figure}[b]
		\centering
		\includegraphics[width=0.6\columnwidth]{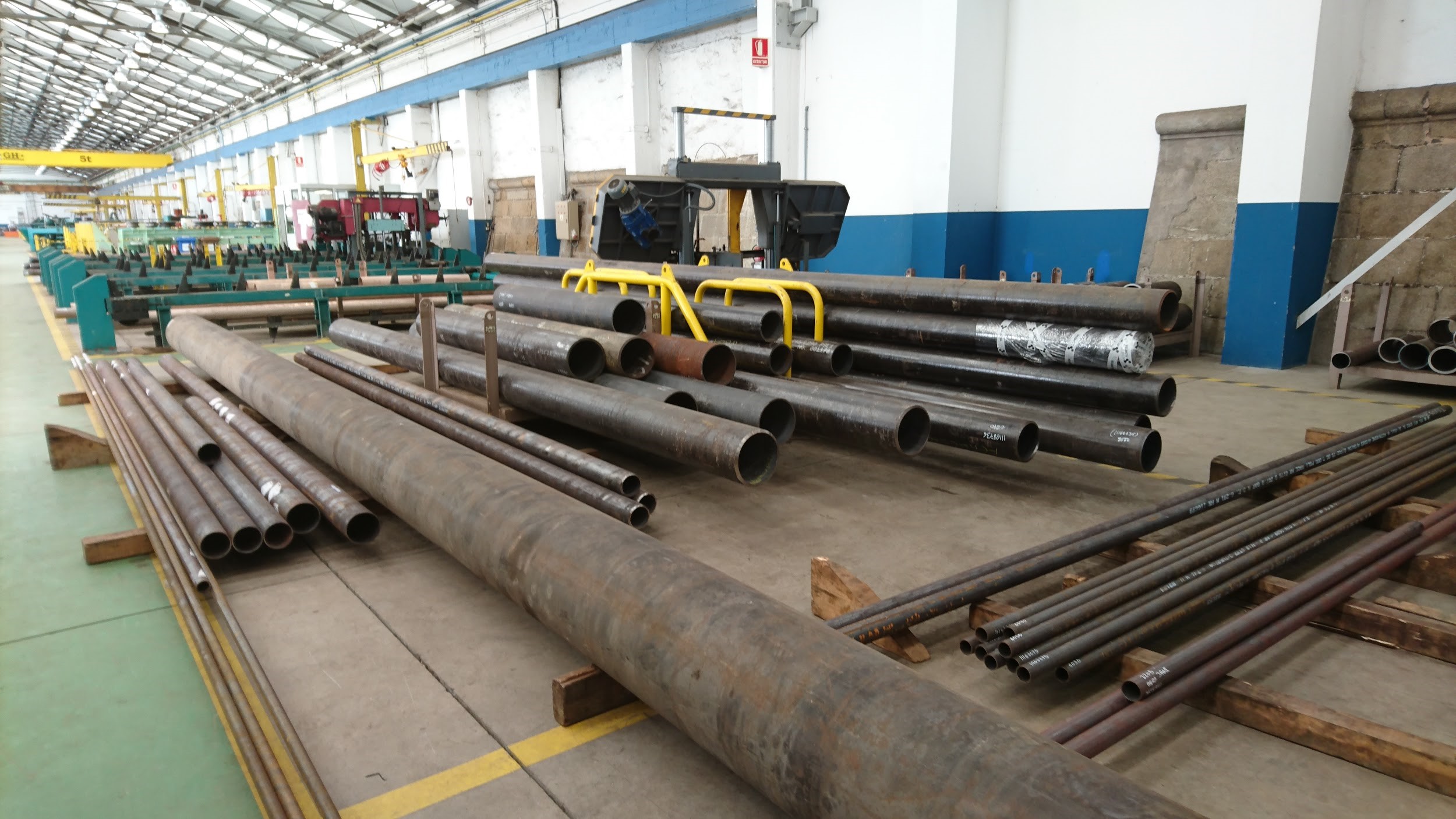}
		\caption{Stacking area for large pipes.} 
		\label{figure:stacking_large_pipes}
	\end{figure}

	\begin{figure}[t]
		\centering
		\includegraphics[width=0.6\columnwidth]{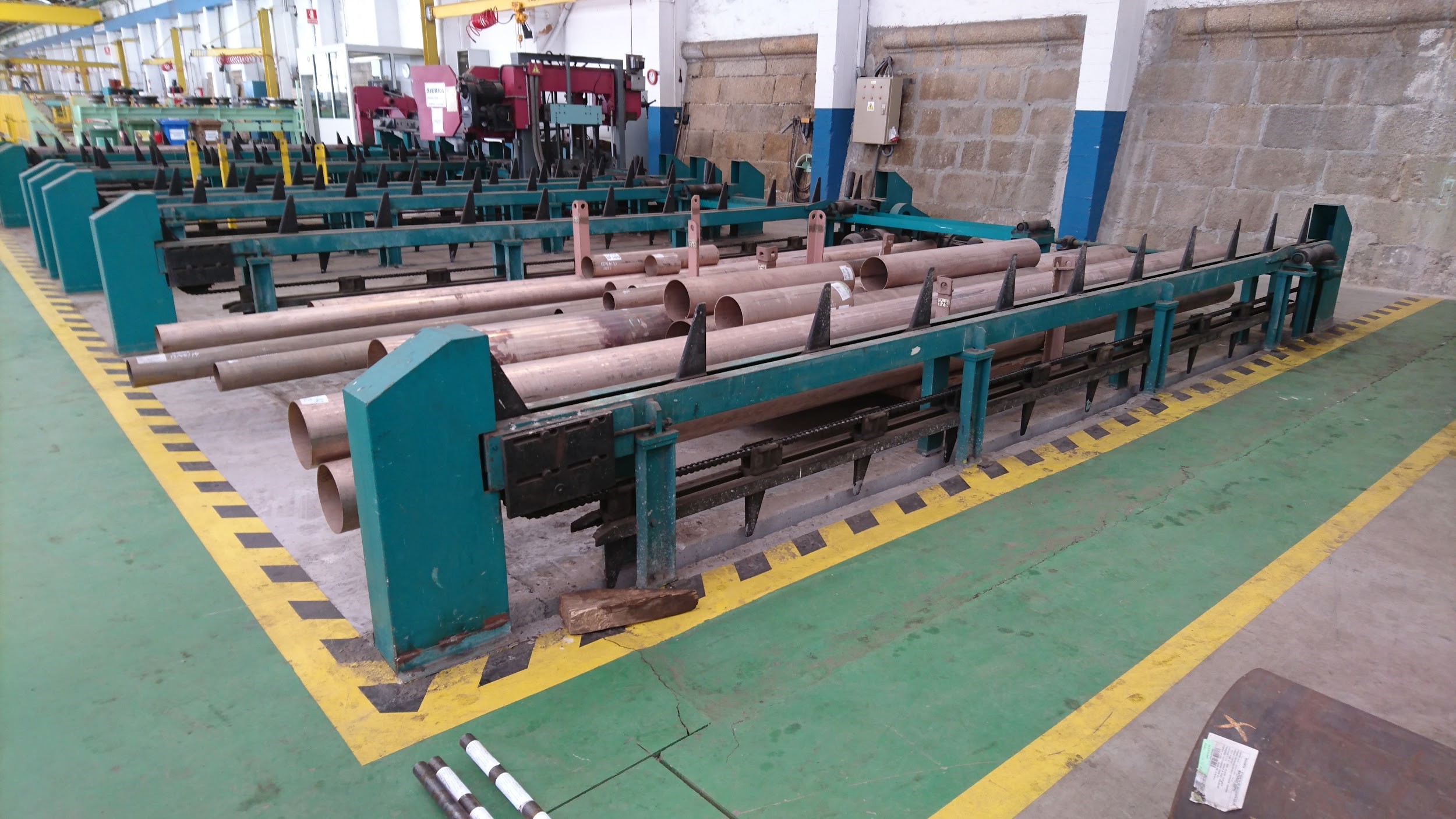}
		\caption{Cutting area of the workshop.} 
		\label{figure:cutting_area}
	\end{figure}

\item The first pipe processing point is the cutting area (in Figure \ref{figure:cutting_area}). In production, as soon as the first cut of a pipe is made, operators place a plastic label that is attached using electric cable (this kind of cable is used because it has to resist being exposed to acids and hot water). 
This label contains alphanumeric identification information and includes a barcode. Pipes are stacked on pallets, which allow for moving them easily between the different stages of the production chain. Regarding such pallets, it is important to note that:

\begin{itemize}
\item Operators distinguish visually each pallet through an identifier painted on it. 

\item Pallets are moved by cranes through the workshop. They are not usually moved until they are considered to be full. When a pallet is moved to a new section, pipes are checked by operators who, by reading the label barcode with a scanner, get information on the process that should be carried out on the pipe. At the same time, the barcode reading operation allows for registering its location, since every scanner is associated with a specific place.

\item Each pallet carries paper documentation related to the pipes contained.

\end{itemize}

\item The second stage of the pipes is bending (if it is required). There are three benders in the workshop, which can be controlled from a Windows-based PC that is also able to receive and load design files from the engineering department.

\item Before manufacturing, pipes might need to be cleaned. For such a purpose, there is an area for degreasing and rinsing pipes by using water or certain acids.

\item Next, pipes are moved to the manufacturing area, where accessories are added. These elements are transported in metal pallets from the workshop warehouse. There is not a quick communication between the warehouse and manufacturing to indicate when the accessories associated with a pipe are available (i.e. operators have to walk to the warehouse and check the availability of the accessories). 

\item After manufacturing, pipes are packed with others on pallets. This packaging is registered before the pallet leaves the manufacturing area.

\item Large pipes can be stored temporarily in a reserved area located at one end of the workshop. Although there are more stacking areas, both indoors and outdoors, there is no real-time control of the occupancy percentage of the areas (i.e. the number of pipes in them).

\item Next to the temporary storage, there is an area of welding stations with plastic separations and other auxiliary areas, mainly dedicated to store pipes.

\item Once the pallet leaves the production area, the traceability of the pipes is lost, and there are no records of their movements and/or location in the different storage areas. The largest storage area is outdoors, next to the workshop, in a nearby dock (as it is shown in Figure \ref{figure:external_storage}).

	\begin{figure}[b]
		\centering
		\includegraphics[width=1\columnwidth]{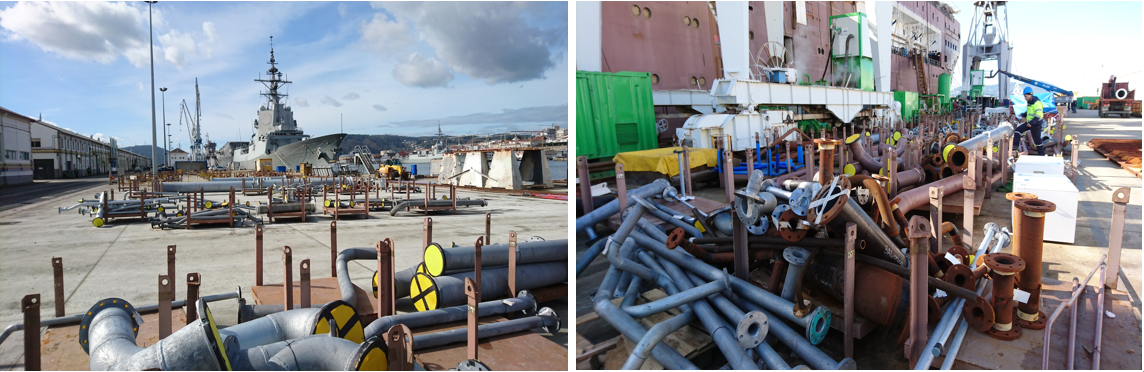}
		\caption{External storage areas in the dock.} 
		\label{figure:external_storage}
	\end{figure}

\end{enumerate}


\subsection{Identification, tracking and location systems for shipyards and smart manufacturing}

In recent years, several authors have studied and proposed various alternatives that address tasks in ship construction, including hull blasting \cite{Faina2009} and welding \cite{Kuss2016,Mun2015,Lee2011},
that can be improved through the application of technological solutions. For example, Kim et al. \cite{NeuralNavigationRobots} propose an automated welding machine for shipyards, in which mobile robots use neural networks to recognize the work environment. Similarly, the same authors propose the use of smart robots for welding in a shipyard \cite{VisualRecognitionEnvironment}, but in this case they design a display system for the recognition of the areas to be welded.

The problem of locating people in a shipyard has been studied by Kawakubo et al. in \cite{WirelessHumanPositioning}. In such a paper, the authors use Bluetooth technology for the location by means of fixed and mobile stations. Thus, the authors achieve a precision of 1.2\,m using a fixed network of readers in which each of the readers is placed at a distance of about 8 meters. 

Sensor networks have also been proposed for monitoring different construction tasks \cite{Perez2014}. For instance, a practical example of a real-time monitoring system for the concentration of CO is described in \cite{yang2015}. A more specific development for the construction of ships and maritime platforms in a shipyard is detailed in \cite{HyperEnvironmentTracking}. There, the authors describe a system of hyper-environments that use sensor networks, virtual reality, and RFID to improve the process of supply tacking.

In environments where the presence of metals is high, radio frequency (RF) communications are clearly affected. 
This impact is well illustrated in \cite{metalNoiseFactorEffect}, where a series of experiments with diverse tags showed that the signal strength decays when the tags are placed on a copper metal plate. In this regard, in \cite{improvingPerformanceRFID} several techniques are analyzed to improve the performance of RFID tags on metal, showing that the length of the antenna is a variable that can improve impedance adaptation. In an environment close to the shipyard, the authors in \cite{helicalPipesRFID} analyze the feasibility of adhering passive RFID tags on metal bent pipes.

In order to overcome harsh environments, multiple tags and components have been designed to enable RFID communications in metallic environments. Examples are \cite{RFIDtagMetal}, \cite{RFIDTransponderAutoID} or \cite{RFIDTagConcreteFloor}, where UHF RFID tags are specifically designed to be used on various metal surfaces and containers. If conditions such as high temperatures are added to the presence of metals, RF communication is even more complicated. Therefore, components need to be adapted to harsh communications scenarios. 
An example is studied in \cite{HighTempRFID}, where the authors analyze some of the complications faced by hardware in the complicated conditions mentioned, such as data memory retention for long periods of time.

\subsection{Technologies for Identifying Pipes} \label{Technologies}

This subsection analyzes different technologies that can perform pipe identification and monitoring in a workshop. Only the most relevant tag-based identification technologies are cited, but other approaches (e.g. dead reckoning or image-based technologies) are available. The technologies selected are described briefly to indicate their general characteristics, before being analyzed and compared in Section \ref{sec:analysisTech}. A summary of the basic characteristics of the technologies is shown in Table \ref{TabProtocol}.
 
\begin{table}[b]

\caption{\textbf{Main characteristics of the identification technologies selected.}}
	\small 
     \resizebox{\textwidth}{!}{  
    \begin{tabular}{c | c | c | c | c }
		\toprule
		\textbf{Technology}&\textbf{Frequency band}&\textbf{Range}&\textbf{Features}&\textbf{Popular Applications}\\  
		\midrule
        Barcode/QR & - &<\,4\,m &LOS, very low cost, visual decoding&Asset tracking and marketing \\
        LF RFID &30-300\,KHz (125\,KHz) &1-5\,cm (<\,10\,cm)&N-LOS, durability, low cost &Smart Industry and security access \\
        HF RFID &3-30\,MHz (13.56\,MHz) &30\,cm (<\,1\,m)&N-LOS, durability, low cost  &Smart Industry and asset tracking \\
        UHF RFID &30\,MHz-3\,GHz
        & 10\,m &N-LOS, durability, low cost&Smart Industry and toll roads \\
       
        NFC &13.56\,MHz & 4-10\,cm (<\,20\,cm)&Low cost, no power&Ticketing and payments\\
        BLE &2.4\,GHz &<\,50\,m& Low power &Wireless headsets \\
        Wi-Fi&2.4-5\,GHz & <\,100\,m&High-speed, ubiquity&LAN, internet access, broadband\\
        Infrared (IrDA) &800 to 1000\,$\mu$\,m &<\,1\,m&Security, high-speed&Remote control, data transfer \\
        UWB&3.1 to 10.6\,GHz&<\,10\,m&Low power, high-speed data&Radar, video streaming\\
        Ultrasound&>20\,kHz (2-10\,MHz)&<3\,m &Inspection of industrial materials&Medicine, positioning and location\\
        ZigBee&868\,MHz (EU), 2.4\,GHz&<\,10\,m&Mesh network&Smart Home and Industry\\
        DASH7&315-915\,MHz&<5\,Km&BLAST network technology &Smart industry and military \\
        ANT+&2.4\,GHz &<\,10\,m &Low power&Health, sport monitoring\\
        Z-Wave&868\,MHz (EU) &<\,30\,m&Simple protocol&Smart Home\\
        WirelessHART&2.4\,GHz&<\,10\,m&HART protocol&Smart Industry\\
        LoRa&2.4\,GHz&>\,15\,m&Long battery life and range&Smart city, M2M\\
        SigFox&868\,MHz&3-50\,Km&Global cellular&Internet of Things, M2M \\
        RuBee&131\,KHz&1-30 m (15\,m)&Harsh environments&Mission-critical scenarios\\
       \bottomrule
	\end{tabular}
     \label{TabProtocol}
	}
\end{table}

Navantia's current pipe monitoring system is based on barcodes, which represent a set of parallel lines of different thickness and spacing that, as a whole, contain certain information. Barcode readers are devices that translate optical impulses into electrical signals, so it is essential to place the code so that good visibility and readability are achieved. The usual reading distance is tens of centimeters, although there is specialized equipment that can reach several meters. 

There are two types of barcodes: linear and two-dimensional codes. Linear codes represent alphanumeric information (e.g. Code 39, Code 93) or numbers (EAN, European Article Number). Two-dimensional codes are able to encode more information per unit of area than linear codes. An example of two-dimensional codes are QR (Quick Response) codes, which were originally designed for the Japanese automotive industry. The code consists of black squares distributed through a grid with white background that can be read by an optical device. Large boxes at the corners allow to detect the code position, having a fourth one for the alignment and orientation. 

QR code reading distance depends on the size of the code: increased distances are obtained thanks to increasing QR code size in proportion. It is usually assumed that code size has to be one tenth of the reading distance (for instance, if a code has to be read at about 20 meters, it should have a size of at least 2 meters). Regarding the storage capacity of a QR code, it depends on the type of data encoded, the version, and the error correction level.

Besides the automotive industry, it is easy to find nowadays QR codes applied in other fields. For example, a QR code traceability system for mitigating food supply chain risks is described in \cite{QR2014}. 

Another identification technology that has experienced a huge growth in the last years is RFID (Radio Frequency Identification). Such a technology consists of readers and electronic tag (also called transponders). These tags are very low power components that react to waves emitted by radio readers by providing the stored information.
 
RFID systems are usually classified according to two characteristics: their frequency of operation and the way they are powered. Depending on the frequency, RFID systems may be classified in radio bands. Each band differences from the others in its propagation behavior and spectrum regulations.  There are three main RFID bands:

\begin{itemize}
\item LF (Low Frequency) RFID. According to the ITU (International Telecommunications Union), the LF band goes between 30\,KHz and 300\,KHz. Frequency and power in this band are not regulated globally in the same way: most systems operate at 125\,KHz, but there are some at 134\,KHz. The reading range provided is short (between 1 and 5\,cm and, generally, up to 10\,cm), so, in practice, LF devices are not usually sensitive to radio interference. The most popular LF RFID applications are access control and animal identification (mainly for pets and livestock). 
They can be used for communications in harsh environments (presence of metals and liquids) due to their long wavelengths.  However, they present low reading speeds and they are not recommended for environments where many tags are together in a small space.

\item HF (High Frequency) RFID. Although the HF band goes from 3\,MHz to 30\,MHz, most systems operate at 13.56\,MHz. HF systems can reach a reading distance of up to 1 meter (they usually reach roughly 30\,cm), what can lead to interference and to the implementation of MAC (Medium-Access Control) mechanisms. This sort of RFID systems is massively used in transportation, payment, ticketing, and access control. HF systems present more interference than LF in unfavorable environments, but they allow for a larger reader range, more speed, and more space for information.

\item UHF (Ultra-High Frequency) RFID. The UHF band actually covers from 300\,MHz to 3\,GHz, but most systems operate in the Industrial, Scientific and Medical (ISM) bands around 860-960\,MHz and 2.45\,GHz. UHF tags can be easily read at 10 meters, so they are ideal for inventory management and item tracking. It has better performance in both transmission speed and
distance with respect to LF and HF systems. However, UHF tags consume more power, they are more sensitive to harsh environments, and, generally, they store less
information that HF tags.

\end{itemize}

All these RFID systems can be also classified according to the way the tags are powered:

\begin{itemize}
\item Passive systems. They do not need internal batteries to operate, since they rectify the energy sent through the reader's antenna. There are passive LF, HF, and UHF systems, which nowadays can be easily read at a 10-meter distance.

\item Active systems. They include batteries, what allow them to reach further distances (usually up to 100 meters). Due to power regulations, almost all active systems operate in the UHF band.

\item Semi-active, semi-passive or BAP (Battery-Assisted Passive) systems. They decrease power consumption by using batteries just for powering the tags for certain functionality. Commonly, batteries are used to power up the basic electronics, while the energy obtained from the reader is used for powering the communications interface. Tags have a battery to power part of the circuitry, thus, they are more reliable than passive, but require more maintenance.

\end{itemize}

As it was mentioned, RFID has been used previously for identifying and tracking items in different industries. For instance, an example of a RFID-enabled real-time manufacturing execution system is presented in \cite{RFID2013}. There the authors describe devices that are deployed systematically to track manufacturing objects and that collect real-time production data. The paper also details a case study with a company that manufactures large-scale and heavy-duty machinery, whose efficiency (planning and scheduling decisions) is evaluated with real-life industrial data. Other applications of RFID include tracking protective equipment \cite{smartEpis}, appliances \cite{smartsocket}, or maritime freight containers \cite{freightcontainers}.

NFC (Near-Field Communication) is a technology that evolves from RFID. NFC devices can be passive (tags) or active (for example, smartphones). The technology operates at 13.56 MHz, with a power that allows for the communication between elements at a distance of less than 20\,cm. Different authors have studied the use of NFC for identification and positioning. A good example is described in \cite{NFC2015}, where it is proposed a navigation system for identifying and tracking tags indoors, where there are no GPS signals available.

Bluetooth Low Energy (BLE), also known as Bluetooth Smart, is a \ac{WPAN} technology oriented to short-range applications (around 10\,m) and small devices that is optimized for energy efficiency. It works at 2.4 \,GHz, sharing the frequency band with other technologies like Wi-Fi. Bluetooth is not designed for a particular application: it defines a series of profiles representing a default solution for a particular use and establishes the requirements for interoperation between devices. Each Bluetooth device can support one or more of these profiles, being the most common the ones that establish links between devices and send data between them.
Such devices include beacons (devices that broadcast certain information periodically), which serve as a reference in an indoor location scenario. An example of the use of beacons for tracking is detailed in \cite{BLE2016}. The authors describe a real-time simulator using workers' position data to support manager's decision making in a manufacturing system. In such a system, BLE beacons are used to collect data easily in an experimental manufacturing line. 

Similarly to Bluetooth, Wi-Fi (IEEE 802.11 a/b/g/n/ac) is also a widespread and popular technology. It may work at 2.4\,GHz or/and 5\,GHz. Due to its popularity, many researchers have studied its use for providing location and tracking services. There are several techniques for indoor location over Wi-Fi, which are based, in general, in determining the position of the clients respect to the access points using the angle or the time of arrival of the signals \cite{TimeOfArrival2016}, the RSSI (Received Signal Strength Indicator) \cite{yang2015} or fingerprinting \cite{fingerprinting2016}.

Another well-known technology is infrared communications. Infrared radiation is a form of electromagnetic and thermal radiation of a frequency lower than the visible light perceived by the human eye. Infrared sensors are opto-electronic devices capable of measuring the radiation of the bodies found in their field of view (it requires direct LOS (Line-of-Sight) between the reader and the object). An example of a system that uses infrared for tracking is described in \cite{Infrared2016}, where it is presented a sensing device that can simultaneously monitor urban flash floods and traffic congestion.

Ultrasounds have also been used extensively for positioning and tracking. As its name implies, ultrasounds are sound (mechanical) waves whose frequency is above the threshold of human hearing. Ultrasonic sensors are devices capable of converting sound signals to electrical signals. These devices operate like radar or sonar, evaluating the echo produced by the waves to estimate the distance between the reader and objects. A good overview of indoor ultrasonic positioning systems can be found in \cite{Ultrasonic2013}.
 
One of the latest technologies that can be applied to identification is \ac{UWB}. UWB is a short range radio technology that allows for the transmission of large amounts of
information over a wide spectrum of frequencies, achieving a very low power density and very short duration pulses. A detailed review of UWB indoor positioning systems and algorithms is provided in \cite{UWB2016}.

ZigBee can be also used in positioning and tracking applications. ZigBee is a technology for creating low-power and low-cost wireless sensor networks. It is able create mesh networks of intermediate devices that allow for achieving large coverage distances. All ZigBee devices are designed for low consumption and high security (encryption). ZigBee transmission rate depends on the operating frequency, which may differ among regions, varying between 20\,kbits/s and 250\,kbits/s. An example of the use of ZigBee in smart manufacturing is provided in \cite{ZigBee2011}. There the authors described a system that uses RFID devices as data collectors and a ZigBee wireless network to serve as the communication system to transmit the data to the different levels of the enterprise management.

DASH7 is a standard evolved from RFID that provides long range wireless communications and which is designed for low power applications that require low bandwidth (up to 200\,kbits/s). It works at frequencies between 315\,MHz and 915\,MHz, and allows for the connection with objects on the move. In the literature there are not many well-documented DASH7-based developments. An exception is \cite{Dash72015}, where it is analyzed the DASH7 Alliance Protocol v1.0 specification and two practical applications are detailed: bird tracking and a greenhouse monitoring application.
 
ANT+ is a subsystem of the ANT base protocol (a technology designed for wireless sensor networks, similar to BLE, but oriented to the use with sensors) that defines a protocol stack that allows for the operation in the 2.4\,GHz band. It is designed for interoperability and data transfer over a network. Most ANT+ developments are related to fitness and healthcare. An example is detailed in \cite{Ant+2013}, where it is studied mobile health monitoring systems in elderly patients. Such a paper proposes a proof of concept solution that allows patients to measure their weight and blood pressure with ANT+ sensors connected to their Android smartphones.

Z-Wave is a specification for wireless communications oriented to home automation. Its aim is to minimize the consumption of the devices to make it suitable to use batteries, reaching transmission speeds of up 100\,kbits/s. 
It works in the frequency rage around 900\,MHz and it has a theoretical range (in open space) of up to 100\,m. Z-Wave creates a mesh infrastructure with at least one controller and an end device. No academic sources that propose a Z-Wave based system for identification or tracking were found.

WirelessHART is a wireless technology based on HART (Highway Addressable Remote Transducer Protocol). HART is the implementation of a protocol of industrial automation. This wireless technology works in the 2.4 GHz band, creating networks through a mesh architecture capable of self-organization. An example of the use of WirelessHART in a CPS system can be found in \cite{WirelessHART2015}, where it is presented a system simulator to evaluate the performance of wireless real-time mesh networks.

LoRa (Long-Range Wide Area Network) is a low consumption wireless network protocol designed for secure and low cost communications in the field of Internet of Things (IoT). 
It uses a frequency range just below 1\,GHz, being designed to communicate sensors in unfavorable environments.
It uses a star topology, where a device (gateway) forwards messages between end devices. The connection with these devices in terms of frequency is negotiated according to the distance and the message length, so as not to interfere with each other. Thus, transmission speeds between 0.3\,kbits/s and 50\,kbits/s are achieved. A comprehensive analysis of the LoRa performance is analyzed in \cite{LoRa2016}.

Another long-range technology is SigFox, which is actually a telecommunications network that follows the style of cellular networks. It is designed to provide low cost and low speed transmissions. Being a network operated by a company, a subscription must be paid for its service that allows up to 140 messages per device per day, 12 Bytes per message and a transmission rate of 100 bits/s at 868\,MHz. In \cite{LoRaandSigFox2015} it can be found a good reference for understanding the insights on the application of SigFox in industrial applications.

Finally, RuBee (IEEE 1902.1) is a point-to-point wireless communication standard based on magnetic waves. It works in low frequencies (131\,KHz), what implies that it has a long wavelength (around 2000\,m) and a low transmission speed (1200\,baud). Its main feature is that it does not use radio waves, but it emits magnetic waves, allowing for the communication in unfavorable environments (with the constant presence of liquid, metal, and NLOS (Non-Line-of-Sight) communications). By using such a low frequency, RuBee consumes little power (a tag may last between 5 and 15 years) with a range of up to 15\,m. An example of an end-to-end asset visibility model for military logistics using RuBee is described in \cite{RuBee2014}.


\section{Design of the System} \label{Design}

\subsection{Operational requirements of smart shipyard pipes} \label{OperativeReq}

The military \cite{ICMCIS,NATOReport,TacticalWimax} and mission-critical infrastructures \cite{Nets4cars, AranduconSecurity, AranduconSurvey} have unique operational requirements. Security \cite{1-Li2012, Fernandez2016}, safety, robustness, interoperability challenges, as well as bureaucratic and cultural barriers, stand in the way of the broad adoption of new CPS and IoT applications \cite{fraga2016}.
Based on the study of the real shipyard environment described in the previous section, different research lines have been detected to improve the efficiency in the pipe processing chain and have a significant impact on the shipyard productivity.
In this section, a set of operational requirements grouped by functionality are assessed in order to cover the scenario previously described.

\subsubsection{Automating the identification of the pipes in the workshop}
Nowadays, the identification of pipes is performed manually, which means that operators have to spend part of their working time reading barcodes. This process requires direct Line Of Sight (LOS) between the reader, and the tag and it is susceptible to reading errors. Likewise, this approach is burdensome and poses risks due to human errors (it is susceptible of not being performed or being performed at incorrect time instants). Nevertheless, operators require information on how to process pipes, thus they have to perform their identification and read the information associated with each of them.
The system proposed should allow for carrying out the identification of the pipes with the smallest possible error in order to avoid manual tasks from the operators involved in the process. Additionally, the system should offer the operators dynamic information about the work to be done on the pipe.

\subsubsection{Location of the pipes}
The present system can determine the location of a pipe in the workshop visually or at the instants in which the quality control processes are performed by middle management. However, the remaining time the pipe is in an unknown position. This unawareness of the exact or approximate position of the pipes causes loss of time due to the seeking for pipes. The system proposed should locate in real or near-real time the pipes circulating in the workshop, not only the ones being processed, but also those that are stored. Thus, the system should trace and monitor in real-time the pipes within the workshop and the ones that left the manufacturing area. 

Note also that the location awareness of the pipes would help to automate different tasks, like notifying an operator on the arrival of a pipe to certain workshop stage.

\subsubsection{Pipe tracking}
The aim of pipe tracking is to find a system that identifies a pipe during the production process. 

Today's operation faces many demands, the pipes can be on the workshop for many years and they suffer from very aggressive processes (e.g. treatment with acids) during its manufacturing. Pipes are made of metallic materials and they are grouped on pallets in significant amounts (tens of units), which makes difficult their visual and/or electronic identification.

\subsubsection{Optimization of the manufacturing time}

In the shipyard it has been observed that the pipes manufactured in the workshop have different storage times: the oldest pipe may rust at the time of assembly, while others (most of them) show no signs of external corrosion. Knowing the real needs of demand for construction, or the available pipes and the workshop capabilities, enhances the storage times to avoid problems of excess of stock (i.e. space problems) and corrosion (e.g. rust). Thus, the system should minimize part stock time, and consequently, it should decrease the likelihood of exposure to external elements.

\subsubsection{Route optimization}
 
Once created a visualization system for locating the pipes, it will be possible to improve the system's capacity for providing additional recommendations thanks to the identification information and the location data collected.
A good example is the optimization of routes for the transfer of pipes. For instance, given a pipe placed in the cutting area, it would be interesting to know what is the fastest route to move it to its storage spot in the dock.
The ultimate goal is to optimize manufacturing and assembly times by obtaining the best routes for the transportation and final installation of the pipes.


\subsection{Technical requirements of smart shipyard pipes} \label{TechReq}

\subsubsection {Hardware requirements}  

This subsection reviews the main hardware needs focusing on the tagging system and on the concerns regarding the deployment.

\subsubsection*{\textbf{Tagging system}}

Electronic tags require a number of features to optimize their performance in aggressive environments in terms of electro-magnetic propagation and exposure to external interference (i.e. shocks,  hits, pressure, acids, high temperature liquids, among others). The following are the main constraints faced by an electronic tag-based system when operating in the shipyard scenarios previously described:

\begin{itemize}
\item Deployment. Tags are deployed on a workshop, where there are different areas (a detailed description of the workshop is given previously performed in Section \ref{sec:Intro}). In addition, it must be emphasized that tags should be as small as possible so as not to cause problems during the treatment and handling of the pipes.

\item Presence of metals. In the workshop there are many metallic elements that originate signal reflections and interfere in RF communications in the HF and higher bands. Therefore, only technologies prepared to tolerate the presence of a significant level of metallic elements should be considered.

\item Presence of water. Navantia's pipe workshop is not particularly cold, but it is next to the sea, so humidity levels are relatively high (between 40\% and 95\%).

\item Exposure to liquids, acids, salinity, fuel or other corrosive substances. The tags selected should support the degreasing, metal pickling, and rinsing processes that are carried out in the workshop during the manufacturing of the pipes. Specifically, Table \ref{tab:chemical} indicates examples of the exposure to different chemical solutions, temperatures, and time durations that should be supported.

\begin{table}[!hbt]
\centering
\caption{Procedures for pipe cleaning.}
\label{tab:chemical}
\resizebox{\textwidth}{!}{  
\begin{tabular}{|c|c|c|c|c|c|}  
\hline
\textbf{Pipe type} & \textbf{Process} & \textbf{Solution} & \textbf{Temperature} & \textbf{Duration} & \textbf{Observation} \\ 
\hline
\multirow{2}{*}{Bent pipes}
&Degreasing&Water and T-149-E (6.25\% $\pm$ 1)&70\degree C$\pm$10\degree C&15$\pm$5 min&     \\
&Pickling&Hot water&60\degree C$\pm$10\degree C& &PH: from 5 to 8  \\
\hline
\multirow{3}{*}{Carbon steel pipes}
&Degreasing&Water and T-149-E (6.25\% $\pm$ 1)&70\degree C$\pm$10\degree C&45$\pm$5 min&  \\

&Pickling&Water and DECAPINOX C (10\% $\pm$ 2)&25\degree C$\pm$10\degree C&15$\pm$5 min&   \\

&Rinsing&Hot water&60\degree C$\pm$10\degree C& &PH: from 5 to 8 \\
\hline
\multirow{3}{*}{Cooper alloys}
&Degreasing&Water and T-149-E (6.25\% $\pm$ 1)&70\degree C$\pm$10\degree C&15$\pm$5 min&  \\

&Passive Pickling&Water and SCALE-GO (4,37\% $\pm$ 1)&50\degree C$\pm$10\degree C&15$\pm$5 min&  \\

&Rinsing&Hot water&60\degree C±10\degree C& &PH: from 5 to 8 \\
\hline
\multirow{3}{*}{Stainless steel} 
&Degreasing&Water and T-149-E (6.25\% $\pm$ 1)&70\degree C$\pm$10\degree C&15$\pm$5 min&  \\

&Passive Pickling&Water and DECAPINOX C (10\% $\pm$ 2).&25\degree C$\pm$10\degree C&15$\pm$5 min&  \\

&Rinsing&Hot water&60\degree C$\pm$10\degree C& &PH: from 5 to 8 \\
\hline

\end{tabular}
}
\end{table}
 
 Additionally, the following situations should be considered:
\begin{itemize}
\item During cleaning, pipes might be exposed to pressurized water through a pressure washing machine.
\item During testing, pipes could be also exposed to hot air, water and oil (in hydraulic systems).
\item If the tags do not support the aggressive processes described in Table \ref{tab:chemical}, it has to be considered the addition of an external protection.
\item The encapsulation of tags/readers must be able to resists acids, salinity, fuel and other substances that may corrode them.
 \end{itemize}
 
\item Potential communications interference. The technology selected must be able to transmit in the presence of the most common sources of electromagnetic interference (e.g. Wi-Fi, Bluetooth, the use of mechanical saws) and other unusual sources (e.g. radar tests, whose frequency ranges from tens of MHz to GHz, and their power can reach several KW).

\item Reading distances. The monitoring system must be able to provide access to location data from a remote computer. Such identification/location information must be as accurate as possible, regardless of the distance required to read the pipes/pallets to be monitored. It is important to note that the workshop is 205\,m long, so a network of readers would probably need to be created to cover the whole building.

\item Tolerance to high temperatures. During manufacturing, pipes can be exposed to high temperatures in two processes: while washing them in water/acids, or during welding.

\item Pressure. Both during the storage and the transfer of the pipes it is possible that they (and their tagging system) will be exposed to pressure due to the accumulation of weight and collisions. Pressure varies depending on the weight and strength supported by the base material. Note that, in general, between 30 and 35 pipes are moved into each pallet, and that such pallets support up to 2 tons.
 
The following are the most common situations where external pressures are produced:
\begin{itemize}
\item When moving a pallet with other pipes on top. There is no standard criteria for stacking pipes, but, usually, the heaviest are placed at the bottom of the pallet.

\item When lifting a block to its mounting position.

\item During manufacturing, it may be necessary to round heads/ends, thus strokes can be applied, what involve deformations.
\end{itemize}

\item Battery duration. Since pipes arrive to the workshop, up to three years may go by until they depart from the storage areas to be installed on a ship. Therefore, battery should last at least such a period of time.

\item Mobility. The technological solution selected must provide portable readers for dynamic and in-situ operation on the various identification, location, and traceability systems.\\
\end{itemize}

\subsubsection*{\textbf{Readers/scanners location}}
Readers should be located in places where there is access to both a data network and electricity. Similarly, those locations should be in places where they interfere the operator work as little as possible.

\begin{itemize}
\item Electricity. The typical workshop usually has numerous electrical outlets in different locations, thus it should not be a problem to power the hardware of the system.
\item Data. Ethernet and Wi-Fi networks should be available to allow for receiving and sending data to the readers deployed. In Navantia's workshop, while these networks are available in most of the workshop, the number of Ethernet sockets, except in certain locations, is scarce, and they are almost always associated to control equipment (e.g. the pipe storage robot or the bending machines).
For such a reason, it is almost essential to place switches or hubs that allow for adding new Ethernet devices easily.

\end{itemize}

\subsubsection {Software requirements}

The system should include the following basic functionality regarding user features:

\begin{itemize}
\item The system should display the location of the pipes in the workshop in real or near-real time. Ideally, the visualization should be implemented in a multi-platform system, which should allow for monitoring the whole workshop from a remote computer, a tablet, and even a smartphone.

\item Easy interaction with the basic pipe information. In addition to viewing pipes in a map, it is desirable that users can access certain information about them.

\item Filtering the pipes displayed. Once the system is operating, numerous pipes would be displayed while moving through the workshop. Therefore, it is convenient that a user can filter them in order to only show a specific pipe or a subset that meet certain criteria.

\item The system should be able to issue different notifications about relevant events that happen in the workshop. For instance, a \ac{BI} module should issue a notification when a pipe goes from one workshop stage to another (e.g. from the cutting area to bending).

\end{itemize}

\subsection{Selection of the identification technology}

\subsubsection{Analysis of the technologies}
\label{sec:analysisTech}
 
After exposing the requirements, the technologies best adapted to the application environment can be determined. A comparison that considers all the factors mentioned such as deployment, presence of metals, presence of water, exposure to liquids, acids, salinity, fuel or other corrosive substances, potential communications interference, reading distances, tolerance to high temperatures, pressure, battery duration, mobility, and cost, is shown in Table \ref{tab:analysis}.

\definecolor{yellow}{rgb}{1,0.88,0.21}

\begin{table}[b]
\centering
\caption{Comparison of the different identification technologies.}
\label{tab:analysis}
\resizebox{\textwidth}{!}{  
\begin{tabular}{|*{12}{c|}}  
\hline
\scriptsize{\textbf{Factor}}&\scriptsize{\textbf{Deployment}}&\scriptsize{\textbf{Metal}}&\scriptsize{\textbf{Water}}&\scriptsize{\textbf{Corrosion}}&\scriptsize{\textbf{Interference}}&\scriptsize{\textbf{Reading dist.}}&\scriptsize{\textbf{Temperature}}&\scriptsize{\textbf{Pressure}}&\scriptsize{\textbf{Battery}}&\scriptsize{\textbf{Mobility}}&\scriptsize{\textbf{Cost}}\\ 
\hline
LF RFID&\cellcolor{green}&\cellcolor{green}& \cellcolor{green}&\cellcolor{yellow}{*}&\cellcolor{green}& \cellcolor{red}&\cellcolor{yellow}{*}&\cellcolor{yellow}{*}&\cellcolor{green}&\cellcolor{green}&\cellcolor{green}\\ 
\hline
HF RFID&\cellcolor{green}&\cellcolor{green}& \cellcolor{green}&\cellcolor{yellow}{*}&\cellcolor{green}& \cellcolor{red}&\cellcolor{yellow}{*}&\cellcolor{yellow}{*}&\cellcolor{green}&\cellcolor{green}&\cellcolor{green}\\ 
\hline
UHF RFID&\cellcolor{green}&\cellcolor{yellow}& \cellcolor{yellow}&\cellcolor{yellow}{*}&\cellcolor{yellow}& \cellcolor{yellow}&\cellcolor{yellow}{*}&\cellcolor{yellow}{*}&\cellcolor{yellow}&\cellcolor{green}&\cellcolor{yellow}\\ 
\hline
BLE&\cellcolor{red}&\cellcolor{red}& \cellcolor{red}&\cellcolor{yellow}{*}&\cellcolor{red}& \cellcolor{yellow}&\cellcolor{yellow}{*}&\cellcolor{yellow}{*}& \cellcolor{yellow}&\cellcolor{green}&\cellcolor{yellow}\\ 
\hline
Wi-Fi&\cellcolor{red}&\cellcolor{red}& \cellcolor{red}&\cellcolor{red}&\cellcolor{red}& \cellcolor{green}&\cellcolor{red}&\cellcolor{red}& \cellcolor{yellow}&\cellcolor{green}&\cellcolor{yellow}\\ 
\hline
UWB&\cellcolor{yellow}&\cellcolor{red}& \cellcolor{red}&\cellcolor{red}&\cellcolor{red}& \cellcolor{red}&\cellcolor{red}&\cellcolor{red}& \cellcolor{yellow}&\cellcolor{yellow}&\cellcolor{red}\\ 
\hline
Ultrasounds&\cellcolor{green}&\cellcolor{green}& \cellcolor{yellow}&\cellcolor{red}&\cellcolor{green}& \cellcolor{yellow}&\cellcolor{red}&\cellcolor{red}& \cellcolor{green}&\cellcolor{green}&\cellcolor{yellow}\\
\hline
ZigBee&\cellcolor{red}&\cellcolor{red}& \cellcolor{red}&\cellcolor{red}&\cellcolor{red}& \cellcolor{green}&\cellcolor{red}&\cellcolor{red}& \cellcolor{yellow}&\cellcolor{yellow}&\cellcolor{yellow}\\
\hline
DASH7&\cellcolor{green}&\cellcolor{yellow}& \cellcolor{yellow}&\cellcolor{yellow}{*}&\cellcolor{yellow}& \cellcolor{green}&\cellcolor{yellow}{*}&\cellcolor{yellow}{*}&\cellcolor{yellow}&\cellcolor{green}&\cellcolor{yellow}\\
\hline
Z-Wave&\cellcolor{red}&\cellcolor{green}& \cellcolor{green}&\cellcolor{red}{*}&\cellcolor{green}& \cellcolor{green}&\cellcolor{red}&\cellcolor{red}& \cellcolor{yellow}&\cellcolor{yellow}&\cellcolor{yellow}\\
\hline
WirelessHART&\cellcolor{red}&\cellcolor{red}& \cellcolor{red}&\cellcolor{red}&\cellcolor{red}& \cellcolor{green}&\cellcolor{red}&\cellcolor{red}& \cellcolor{yellow}&\cellcolor{yellow}&\cellcolor{yellow}\\
\hline
RuBee&\cellcolor{green}&\cellcolor{green}&\cellcolor{green}&\cellcolor{green}&\cellcolor{green}&\cellcolor{yellow}&\cellcolor{green}&\cellcolor{green}& \cellcolor{green}&\cellcolor{green}&\cellcolor{yellow}\\
\hline
\end{tabular}
}
\end{table}

At the view of the table, it can be observed that some of the technologies explained in Section \ref{Technologies} can be directly discarded, since they do not fulfill any of the significant requirements of the system.
The technologies that are fully compliant with the operational and technical requirements described in Sections \ref{OperativeReq} and \ref{TechReq} are in green color, the ones that partially fulfill the requirements are shown in yellow, and the non-compliant ones are colored in red.

Note that, in Table \ref{tab:analysis} an asterisk means that custom tags are required and that they are currently on the market. For instance, high temperatures are only supported by customized tags. The practical experience of Navantia has shown that the upper peak temperature can be set at 105\degree C. This temperature can be easily supported: regular electronic tags usually tolerate peaks of 125\degree C, but specialized tags can reach up to 200\degree C or 300\degree C. The same happens with corrosive chemicals (including water) or high pressure: the tags required have to be designed to support them.

\subsubsection{Discussion} \label {Feasibility}

This subsection is focused on describing and choosing the technology more appropriate for automating the identification and location of the pipes in the shipyard workshop. First, a number of technologies were discarded following Table \ref{tab:analysis}, since they do not properly address some of the most relevant requirements of the system:

\begin{itemize}
\item LF and HF RFID: the reading distance they reach is not enough for building a ubiquitous real-time CPS system.
\item 2.4\,GHz RFID: due to its operating frequency, its performance decreases in the presence of metals, liquids, or the interference from other systems that work on the same frequency band.

\item BLE and Wi-Fi: these technologies share operating frequency with the 2.4\,GHz RFID, so they suffer from the same  problems in terms of interference. 
 
\item ZigBee, Z-Wave, and WirelessHART. These technologies are aimed at creating sensor networks: their application in location is possible, but they are not suitable for the shipyard environment described in this paper. In addition, ZigBee and WirelessHART work in the 2.4\,GHz band, which must be avoided.
\end{itemize}

Next, it is possible to discard technologies that, by their nature, they cannot be rejected at a first instance, but which pose risks:

\begin{itemize}
\item Ultrasounds. Its biggest disadvantage is the need for direct LOS between the reader and the tags. Although the frequencies used do not interfere directly with the test environment, they can interfere with sensor communications and/or armament of the ships, and even with marine animals.

\item UWB. They obtain an excellent precision in location applications, but it is difficult to adapt to the shipyard environment described due to its short range and its problems with the presence of metal objects.
\end{itemize}

Finally, three technologies were chosen for their theoretical characteristics, and because they are suitable for the shipyard environment. These are:

\begin{itemize}
\item RuBee. It does not suffer from electromagnetic interference. Moreover, RuBee tags are designed to withstand adverse conditions and their battery lasts up to 15 years. It has been also tested in
weapon control environments, where it was shown that its use is safe \cite{Pantex}.
 
\item DASH7/active UHF RFID. Both UHF RFID and DASH7 work in a frequency range to some extent sensitive to the interference present in the shipyard environment, but that it is slightly less aggressive than in the 2.4\,GHz band. They have a theoretical reading distance of up to 100\,m.

\item UHF RFID. As it has been already mentioned, the use of frequencies below 1\,GHz decreases the influence of the environment conditions. UHF technology has the advantage of having been tested thoroughly in location and tracking applications. In addition, tags are usually inexpensive.

\end{itemize}

\subsection {Communications Architecture} \label {PilotScenarios}
 
\begin{figure}[t]
		\includegraphics[width=1\columnwidth]{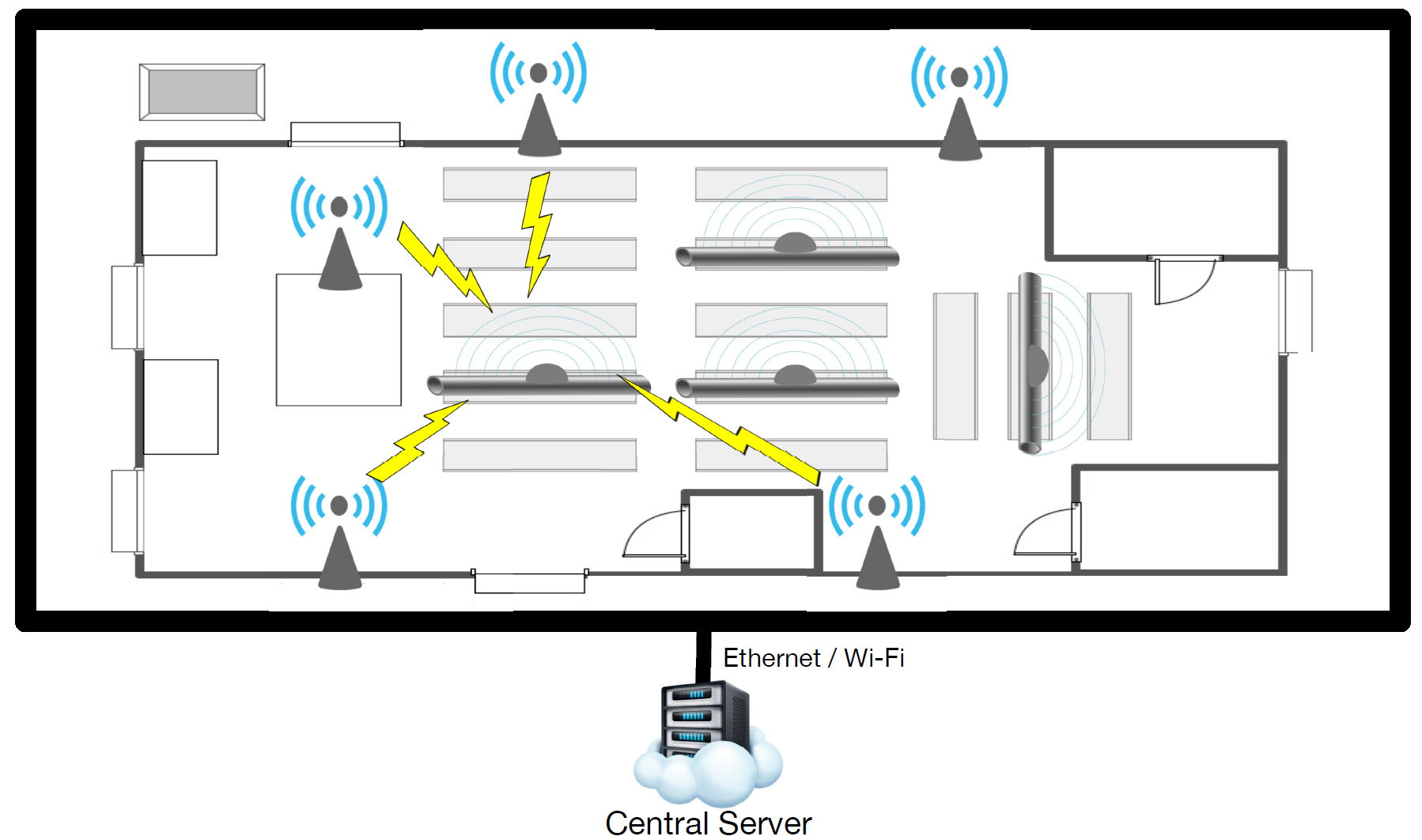}
		\caption{Communications architecture of the smart pipe system.} 
		\label{figure:smart_areas}
	\end{figure}

The communications architecture proposed relies on an infrastructure of beacons that identify pipes throughout different areas. Such an architecture is illustrated in Figure \ref{figure:smart_areas}.

Each of the beacons reads the identifiers of the pipes/pallets and estimates its position. Contrary to the current Navantia system, it is possible to put aside the concept of pallet, allowing the system to focus only on pipes. The removal of the concept of the pallet simplifies the deployment. Tags are only on the pipes, concentrating the logic of identification, location and tracking in the readers that are deployed.
Therefore, a careful planning of the location infrastructure is required. It is possible to increase the accuracy of the system by scaling it, increasing its granularity (i.e. the number of readers deployed), in exchange for an increase in the economic cost.

\section{Implementation} \label{Implementation}

In the previous Section it was concluded that there were three technologies with high potential to carry out the identification and location of the pipes: RuBee (IEEE 1902.1), DASH7 (active UHF RFID) and passive UHF RFID. After obtaining these conclusions, various suppliers of each of the technologies were contacted. In the case of RuBee an unexpected restriction arose: the only worldwide supplier refused to sell the hardware (as a distributor would do), because recently they had changed their business model (as of writing, they only sell completely closed projects, from design to implementation).  Thus, the range of technologies was reduced to active and passive RFID.

	\begin{figure}[t]
    \centering
		\includegraphics[width=0.8\columnwidth]{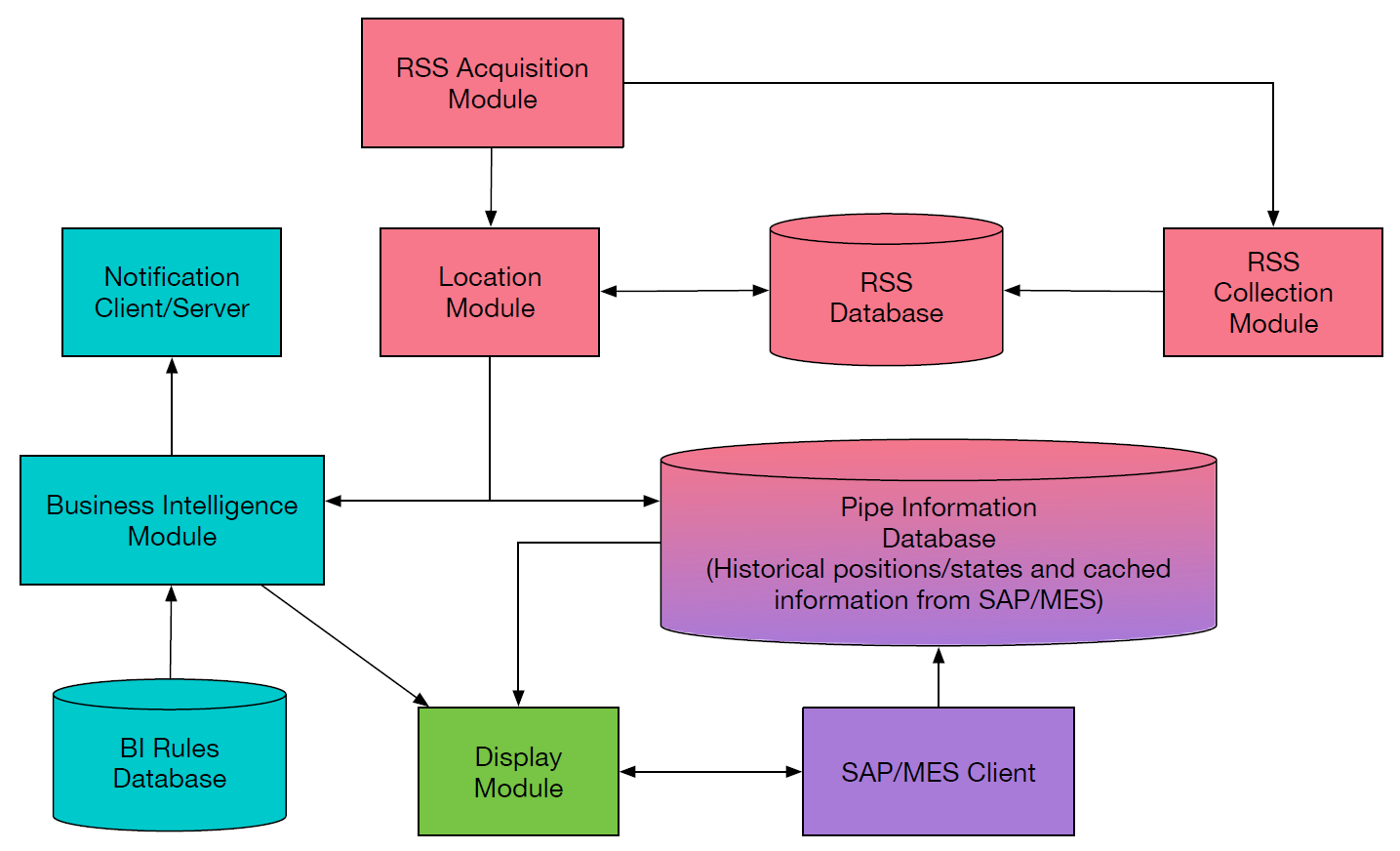}
		\caption{Modules of the smart pipe system proposed.} 
		\label{figure:system_modules}
	\end{figure}

\subsection{System modules}

The system was designed assuming an RFID-based implementation and it consists of the software modules shown in Figure \ref{figure:system_modules}. Such modules perform the following tasks:
\begin{itemize}

\item Location module: it is the core of the system. It obtains the coordinates of each tag after processing the signal strength.

\item RSS acquisition module: it is the interface with the RFID readers. It allows for obtaining the signal strength from each tag.

\item RSS collection module: it is in charge of storing the signal strength values in the RSS database.

\item Business Intelligence module: it decides which notifications should be launched depending on the current position of a pipe, its historical position (where it has been in the past), and the states through which it has passed.

\item Display module: it displays the positions of the pipes and the relevant notifications on a user screen. It allows operators to filter the pipes based on various parameters for easy viewing.

\item SAP/\ac{MES} client: it allows for obtaining the data about the characteristics of the pipes, which are stored in different remote repositories.

\end{itemize}

The processing and collection modules were implemented in the programming language Python. Django was used as a web framework and Nginx as a web server. The display module was implemented using websockets, Javascript, and jQuery.
 
There are also three databases that collect the information needed by the system (in the implementation presented in this paper, they all are SQLite databases):
\begin{itemize}

\item \ac{RSS} database: it stores the signal strength values collected by the RSS Collection Module. At the same time, it can be used by the real-time Location Module to determine the location of the tag.

\item Pipe information database: it stores the information received about the pipes from third-party systems such as SAP or MES.

\item Business intelligence rules database: it contains the necessary rules for the BI system to determine when to launch notifications.
\end{itemize}

\subsection{RSS-based Location Techniques}
\label{RSSStabilizationTechniques}

This section describes the models and techniques used in the implementation of the Location module. Its experimental behavior will be analyzed in the tests performed in Section \ref{Experiments}.
The algorithms include modeling \ac{RSS} respect to the distance, a Kalman filter to reduce the noise, or spatial diversity techniques to increase \ac{RSS} stability. Note that, in this paper, the term RSS is used instead of \ac{RSSI}: RSS is more generic than RSSI, which is commonly used for estimating signal strength in Bluetooth and Wi-Fi devices.

\subsubsection{RSS Mathematical Model}\label{Model}

RSS can be used to determine the location of each tag by relating it directly to the signal propagation. However, this method depends on the variability of RSS, which is influenced by obstacles and by the presence of metallic elements. In fact, indoors, the propagation of radio waves is mainly affected by two types of losses: the path loss and the losses due to small and large scale fading. Fading is usually associated with reflections, diffraction or absorption, commonly present in real environments. The small scale fading arises due to the multipath propagation effect, while large scale fading is related to the shadowing effect. 
From the RSS values it is possible to derive the mathematical model that allows us to relate them with the distance. The simplest approach is to average the RSS values for each distance. Another approach is based on the construction of a mathematical formula that model the system behavior.
The proposed formula which closely reflects the path loss in the indoor environment is the log distance path loss model. This model can be seen as a generalization of both the free space propagation model and the two-ray ground propagation model. Thus, the behavior of the signal can be simplified through the following model \cite{Rappaport2002}:

\begin{equation}
\centering
PL(d)[dB] = PL(d_0) + 10\,n\,log( d ) + X_\sigma  
\end{equation}

where:

\begin{itemize}

\item $PL(d)[dB]$  is the attenuation in decibels suffered at distance $d$.

\item $PL(d_0)[dB]$  is the attenuation in decibels at a reference distance $d_0$. Typically obtained through measurements.

\item \textit{n} is a calculated value that minimizes the \ac{MSE} difference between the model and the empirical results.

\item $X_\sigma$  is a Gaussian variable with mean zero and standard deviation $\sigma$.

\end{itemize}

Considering the model and using the RSS as inputs, the distance $d$ can be easily obtained with the expression:

\begin{equation}
\centering
d =  10^{\frac{RSS - RSS_{d_0}}{10*n}}
\end{equation}

\subsubsection{Kalman filtering}
\label{sec:KalmanFiltering}
The fundamental problem that faces the RSS indoors is the noise, which causes inaccuracies when determining the location. There are different methods to filter noise, among which it was chosen one that is popularly used for guidance, navigation and control of vehicles (mainly spaceships): the Kalman filter. A detailed explanation of how Kalman filtering works is out of the scope of this paper, but the interested reader can check \cite{Grewal2008} for a detailed description of the theory behind it. For the experiments shown in Section \ref{Experiments}, Kalman filtering was adapted to reduce noise on RSS, based on the work described in \cite{Bulten2016}. 

First, it should be clarified that a regular Kalman filter assumes that its model is linear (there is an extended version for non-linear models), so the transition from the current state to the next state should be performed through a linear transformation. Hence, it can be formulated first the general Kalman filter expression for a transition:

\begin{equation}
\centering
x_t = A_t x_{t-1} + B_t u_t + \epsilon _t
\end{equation}

where $x_t$ and ($x_{t-1}$) are the current and previous states, respectively, $A$ is a transformation matrix, $u_t$ is a control input, $B$ is the control input model, and $\epsilon$ is the noise. 

This general expression can be simplified by assuming that pipes remain in specific places during the measurements performed. Thus, RSS is expected to be constant, having a varying contribution from the noise. Then, the control input $u_t$ can be ignored and it can be assumed $A_t$ to be the identity matrix, what results in the following expression:

\begin{equation}
\centering
 x_t = x_{t-1} + \epsilon _t
\end{equation}

Regarding the observation model, in general, it is expressed as:

\begin{equation}
\centering
z_t = C_t x_t + \delta _t
\end{equation}

where C is a transformation matrix and $\delta$ the noise related to faulty measurements. Since RSS observations and states are equal (RSS is estimated by using old RSS values), the transformation matrix C becomes the identity matrix, so the expression ends up being as follows:

\begin{equation}
\centering
z_t = x_t + \delta _t
\end{equation}

Once transitions have been defined, the prediction step can be formulated. Such a step indicates what is expected for the next state without taking measurements into account. Since RSS is expected to be constant, the expression is simple:

\begin{equation}
\centering
\overline{\mu_t} = \mu_{t-1}
\end{equation}

\begin{equation}
\centering
\overline{\Sigma _t} = \Sigma _{t-1} + R_t
\end{equation}

In contrast to $x$, which is the true value, $u$ is the value predicted. $\Sigma$ is the certainty of the prediction. Such a certainty depends on the previous certainty and on the noise $R$.
In addition, the bar over $\overline{u_t}$ and $\overline{\Sigma _t}$ means that it is still needed to incorporate the information added by the measurement. 

With the prediction estimate $\Sigma$ it is possible to calculate the Kalman gain, which is defined as:

\begin{equation}
\centering
K_t = \overline{\Sigma_t}(\Sigma_t Q_t)^{-1}
\end{equation}

where $Q$ is the measurement noise.
Then, the update step can be obtained, where it is calculated the prediction $\mu$ and the certainty $\Sigma$:

\begin{equation}
\centering
\mu_t = \overline{\mu_t} + K_t(z_t - \mu_t)
\end{equation}

\begin{equation}
\centering
\Sigma_t = \overline{\Sigma_t} - K_t \Sigma_t
\end{equation}

From this equation it can be observed that, the larger the Kalman gain, the bigger the influence of the measurement on the estimation. In the same way, if the Kalman gain is low, it is trusted more the prediction than the measurement.

\subsubsection{Spatial diversity techniques}
\label{sec:spatial}

\ac{MIMO} technology offers substantial performance gains in wireless links \cite{MIMO2013}. Thanks to the use of multiple antennas for transmission and/or reception, the serious effects produced by fading on the RSS can be decreased. Likewise, it is possible to use spatial diversity to improve the stability of the mathematical model proposed thanks to the fact that there are several receiving antennas. To reduce the impairments caused in the signal by these environments, spatial diversity takes advantage of the fact that there exists a low probability of getting simultaneously a deep fading on all the signal paths. Then, assuming uncorrelated channels and using spatial diversity techniques, it is possible to combine the RSSs in such a way that the effects of fading and multipath can be reduced. 

These techniques are based on the classic algorithms of combination and selection, typically used in systems that try to increase the capacity or to stabilize the signals received. Although in this article the same names as the ones used in the traditional algorithms are used, this is not totally precise, because the implementations made are adapted to the \ac{RSS} parameter \cite{Fernandez2007}, modifying the principles of the classic schemes, which are focused on data processing.


\subsubsection*{\textbf{Combination Methods}}
 \begin{itemize}
\item Equal Gain Combiner (EGC). This method weights equally all the antennas, performing the average of all the RSSs, following the equation:

\begin{equation}
\centering
EGC=\frac{1}{N} \sum_{i=1}^{N} RSS_i
\end{equation}

where $N$ is the number of antennas and $RSS_i$ is the level of received signal for the $i$-th antenna. For the sake of clarity, in the curves shown in the experiments section, the term "Mean" was used instead of "EGC", but its computation is identical.

\item Maximum-Ratio Combiner (MRC). Unlike EGC, this method weights each RSS depending on its signal quality. That is, it gives more weight to the most positive RSSs than to the lower ones. To carry out this computation the following expression was chosen:

\begin{equation}
\centering
MRC=\sum_{i=1}^{N}  \left\lbrace \frac{RSS_i - RSS_{min}}{\sum_{i=1}^{N} \left\lbrace RSS_j - RSS_{min}\right\rbrace}\right\rbrace RSS_i
\end{equation}

where $RSS_{min}$ is the smallest RSS that can be obtained. 
\end{itemize}

\subsubsection*{\textbf{Selection methods}}
\begin{itemize}
\item Selection Combiner (SC). This technique consists in sorting the available measurements from higher to lower, choosing only one value. The aim of this method is to achieve the best RSS level for each instant, without considering the RSS fluctuations. In practical applications, the highest value is usually chosen, which is the criterion used in the implementation evaluated in the Experiments Section.
  
\item Switch-and-Stay Combiner (SSC). This algorithm first chooses one antenna, through which the RSS is received until it falls under a threshold, which has been previously set. In that moment, the algorithm switches to the next antenna without verifying its RSS. There exists the possibility of changing the threshold dynamically, but in the present paper it is only considered the case when the threshold remains constant. It is important to emphasize that the switching between antennas is performed in a blind way, so it is not guaranteed to get a better RSS with the switching. 
Without a doubt, the most important issue for this method is the calibration of the optimum switching threshold. To carry out this task, the optimum threshold was decided according to an estimator of the signal dispersion. Specifically, in the experiments, the threshold was set depending on the signal variance.

\item Scanner Combining (ScanC). This method is similar to SCC, but instead of making a blind switching when the threshold is exceeded, it checks each antenna until it finds one over the threshold. If there are not any antennas above the threshold, the method keeps on using the same antenna. Like with SCC, the value of the switching threshold is very important, because it decides the system behavior. Besides, as in the SCC algorithm, during the experiments, it was considered a switching threshold that minimizes the variance.
 \end{itemize}

\section {Experiments} \label{Experiments}

This Section presents the results of several tests conducted to validate the technologies selected and the CPS software developed. Regarding the tests, they were performed to determine which of the two technologies selected, passive and active RFID, adapted better to the peculiarities of the environment and the characteristics of pipe workshop.

\subsection{Selected Hardware} 
Based on the requirements enumerated in Sections \ref{OperativeReq} and \ref{TechReq}, the most promising readers and tags for both technologies were selected. They are described in the next subsections.

\subsubsection{Passive RFID hardware}
The passive UHF reader selected was a Speedway Revolution R420 from Impinj \cite{Speedway}. The reader has connections for up to 4 antennas (four panel high-gain antennas were used during the tests) and the ability to exchange data via Ethernet, USB, RS-232, or a GPIO port. Reader data is accessed through its native \ac{REST} \ac{API}. 
Similarly, a mobile reader (A6-UHF Long Range) based on Windows CE was chosen to provide mobile identification to operators \cite{A6-UHFLongRange}.

A wide range of tags allows for carrying
out a reliable validation of the technology thanks to its diverse nature/objectives. Specifically, the following models from Omni-ID \cite{Omni-ID} were selected: 
Fit 400 UHF Tag on-metal,
Exo UHF Tag on-metal family (Exo 600, Exo 750, and Exo 800),
Dura UHF Tag family (Dura 600, Dura 1500, and Dura 3000),
and Adept 360\degree-ID UHF Tag on-metal. Its main physical specifications can be seen in Table \ref{tab:tags}. 

\begin{table}[b]
\centering
\caption{Specifications of the passive RFID tags selected.}
\label{tab:tags}
\resizebox{\textwidth}{!}{  
\begin{tabular}{|c|c|c|c|c|}  
\hline
\textbf{Family} & \textbf{Model} & \textbf{Max. Reading Range} & \textbf{Dimensions} & \textbf{Weight}   \\ 
\hline
\multirow{1}{*}{Fit}
&400 &4\,m &13.1$\times$7.1$\times$3.1\,mm&380\,g    \\
\hline
\multirow{3}{*}{Exo}
&600 &Fixed reader: 6\,m, handheld: 3\,m &With holes: 80$\times$15$\times$12\,mm, without: 60$\times$15$\times$ 12\,mm&12\,g  \\
&750 &Fixed reader: 7\,m, handheld: 3.5\,m &51$\times$48$\times$12.5\,mm  &25.6\,g    \\
&800 &Fixed reader: 8\,m, handheld: 4\,m  &110$\times$25$\times$13\,mm &26\,g   \\
\hline
\multirow{3}{*}{Dura}
&1500 &Fixed reader: 15\,m, handheld: 7.5\,m  &140$\times$66$\times$14\,mm &75\,g   \\
&3000 &Fixed reader: 35\,m, handheld: 20\,m  &210$\times$110$\times$21\,mm  &265\,g   \\
&600 &Fixed reader: 5\,m, handheld: 2.5\,m  &49$\times$38$\times$9.5\,mm &12\,g  \\
\hline
\multirow{1}{*}{Adept}
&360 &10\,m &136.5$\times$48$\times$5.5\,mm &126\,g    \\

\hline
\end{tabular}
}
\end{table}

The Fit 400 is a small form factor, high performance RFID tag, it is the perfect solution when space is limited, but performance is demanded. 
With respect to the Exo family, Omni-ID Exo 600, with a small footprint, is well suited for being attached to metal bars.  Omni-ID Exo 750 is designed with a broad read angle and with a global RF response. Omni-ID Exo 800 is a long range passive UHF RFID tag capable of being read on, off, and near metal surfaces. Designed in a surprisingly small form factor, it features a rugged design for long term use outdoors and in industrial environments.

In the Omni-ID Dura family, Dura 600 is a small form factor RFID tag, with extreme impact resistance, and superior on-metal performance. Omni-ID Dura 1500 is a durable and long range tag. Designed with heavy industry in mind, it features extreme impact resistance and high temperature ratings. Omni-ID Dura 3000 is designed for heavy industry and outdoor applications. Its features include high impact resistance, water proof, and a durable case (it is optimized for tracking large assets in open storage environments, without worrying about batteries). 

The Omni-ID Adept 360\degree is an UHF RFID tag for the harshest environmental applications. The tag is encased in an industrial steel frame with a tether attachment designed to meet the needs of heavy industry applications.

\subsubsection{Active RFID hardware}
The active reader chosen was NPR ActiveTrack-2 \cite{NPR}, that, according to the manufacturer, has a coverage radius of 45 meters with standard antennas. High gain antennas were acquired to extend its coverage to about 90 meters.  

Different tags can be used with the chosen reader (i.e. different sizes, different features). Among all the models, the Active RuggedTag-175S \cite{Active RuggedTag-175S} was chosen, since it is designed to stand aggressive environments and is sonically welded (what helps to resist the effects of maritime environments). According to the manufacturer, its lithium CR2032 battery lasts more than 4 years. Its dimensions are 63.75$\times$37.72$\times$25.4\,mm and a weight of 51\,g.

\subsection{Test methodology}

The tests were conducted with the readers inside Navantia's pipe workshop. These tests were focused primarily on assessing the most favorable cases for determining how far RFID tags can be read in the best case scenario: if the results for the best case are not as good as expected, then, obviously, the system will perform worse in more complex situations.

Two different kinds of experiments were performed. First, it was obtained the maximum reading distance, taking diverse factors into account (e.g. the type of tag, the number of reading antennas, type of antennas, or the shape of the antenna array). The second kind of experiments were associated with obtaining a mathematical function that relates signal strength with distance in order to locate accurately the tags.

\begin{figure*}[!t]
     \caption{Measurements with passive UHF reader with two antennas.}
\centering
    \begin{subfigure}[t]{0.5\textwidth}
        \centering
        \includegraphics[width=1\columnwidth]{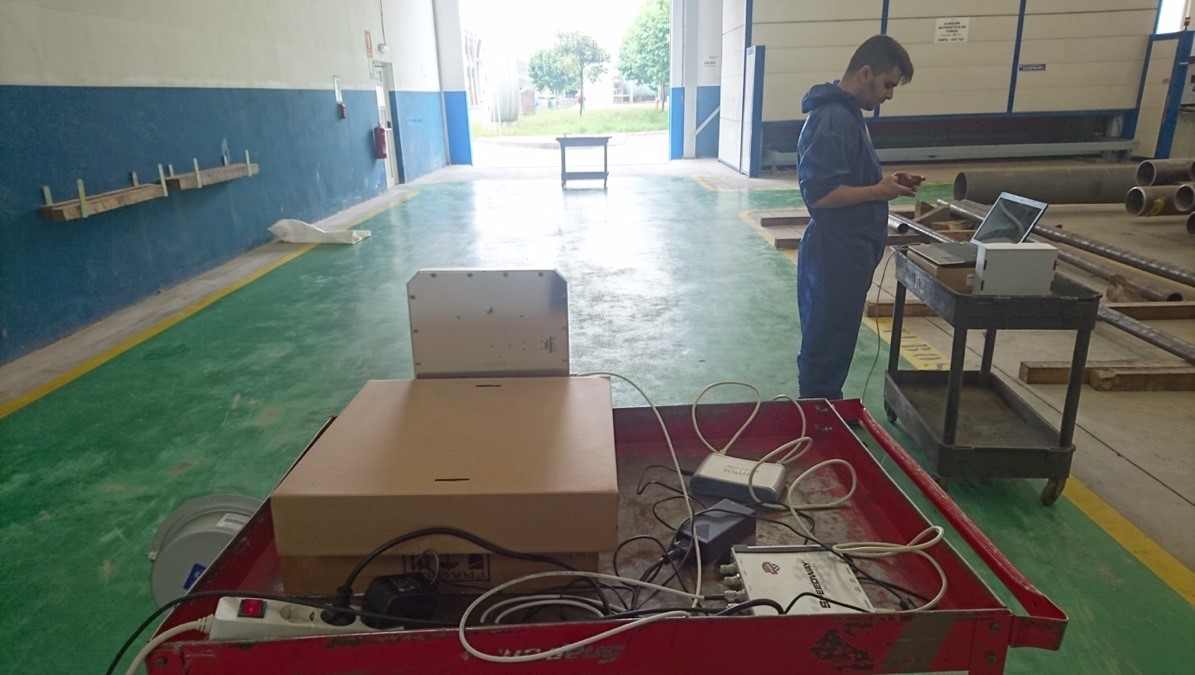}
 
    \end{subfigure}%
    ~ 
    \begin{subfigure}[t]{0.5\textwidth}
        \centering
        \includegraphics[width=1\columnwidth]{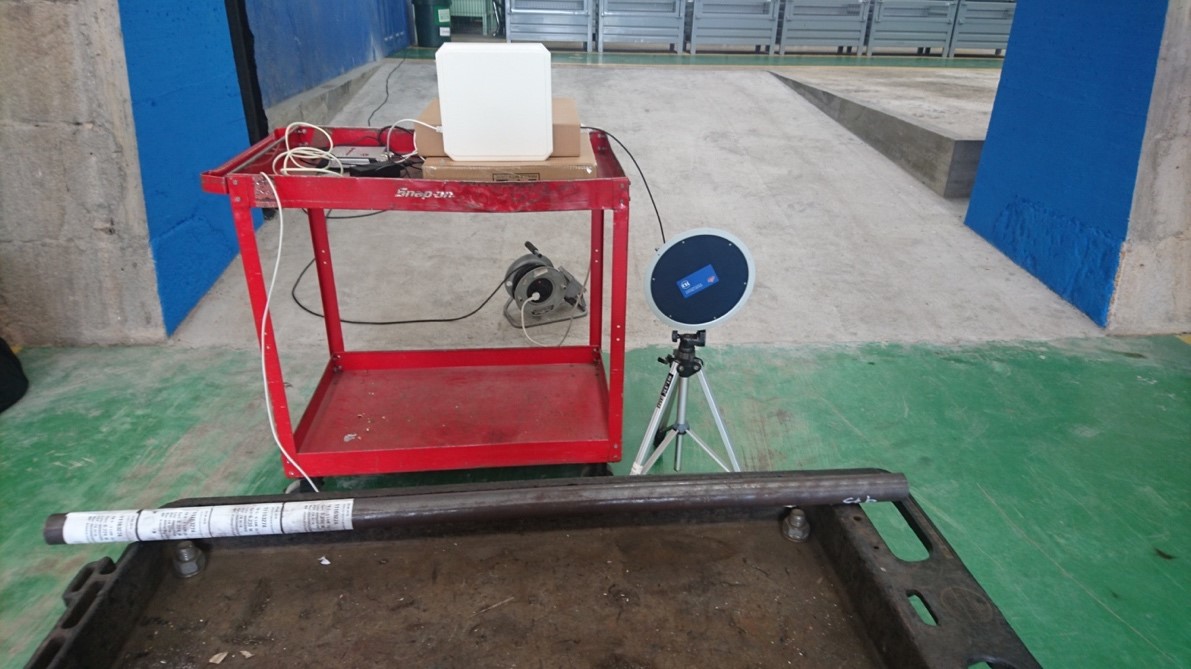}
    \end{subfigure}
 \label{fig:UHF2antennas}
\end{figure*}

\addtocounter{figure}{-1}
\begin{figure*}[!t]
     \caption{An example of tags used for measurements.}
\centering
    \begin{subfigure}[t]{0.49\textwidth}
        \centering
        \includegraphics[width=1\columnwidth]{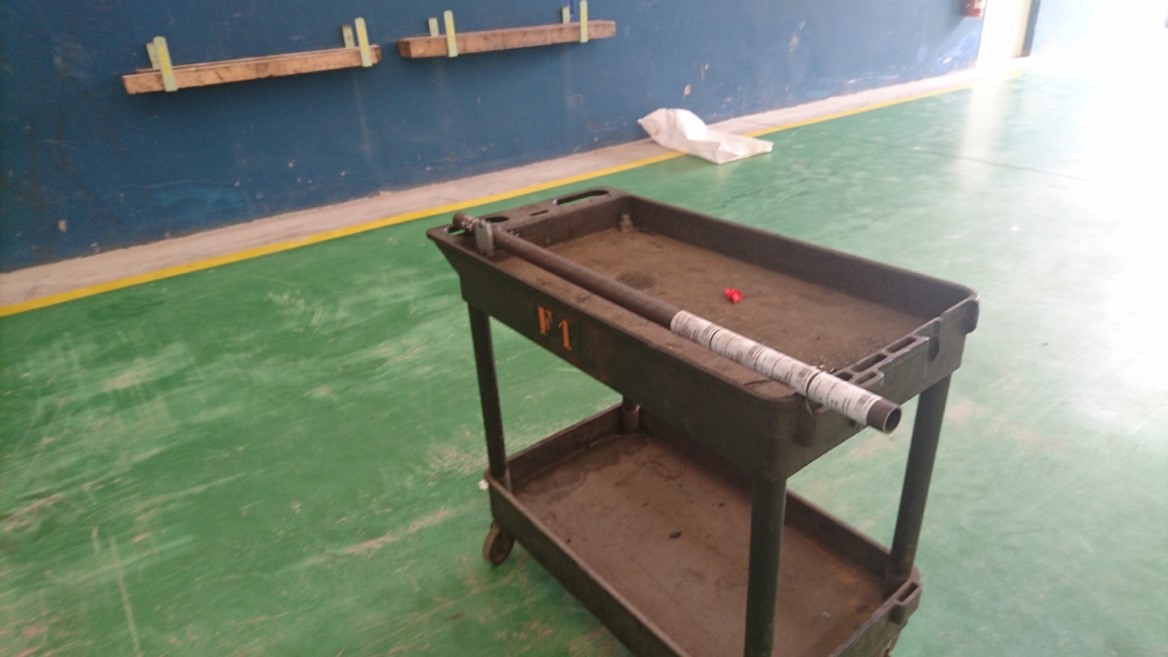}
 
         \caption{Exo 750 UHF Tag.}
    \end{subfigure}%
    ~ 
    \begin{subfigure}[t]{0.49\textwidth}
        \centering
        \includegraphics[width=1\columnwidth]{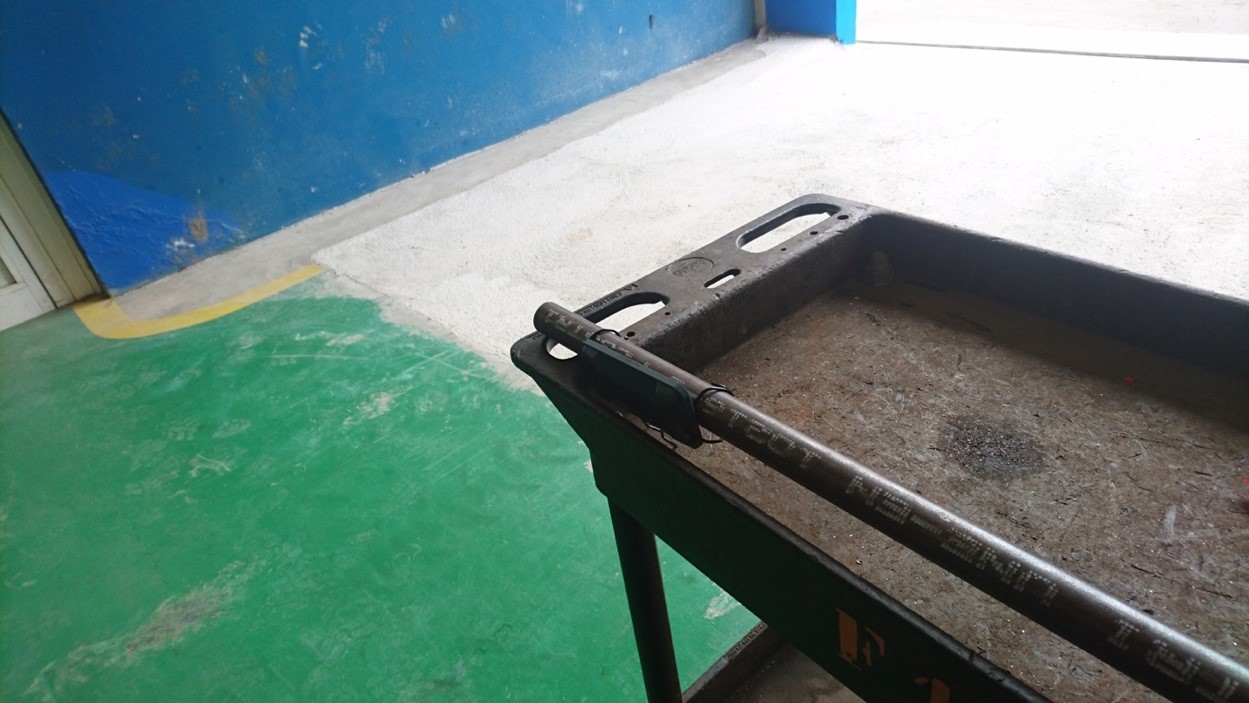}
 	    \caption{Dura 1500 UHF Tag.}
    \end{subfigure}
      ~ 
    \begin{subfigure}[t]{0.49\textwidth}
        \centering
        \includegraphics[width=1\columnwidth]{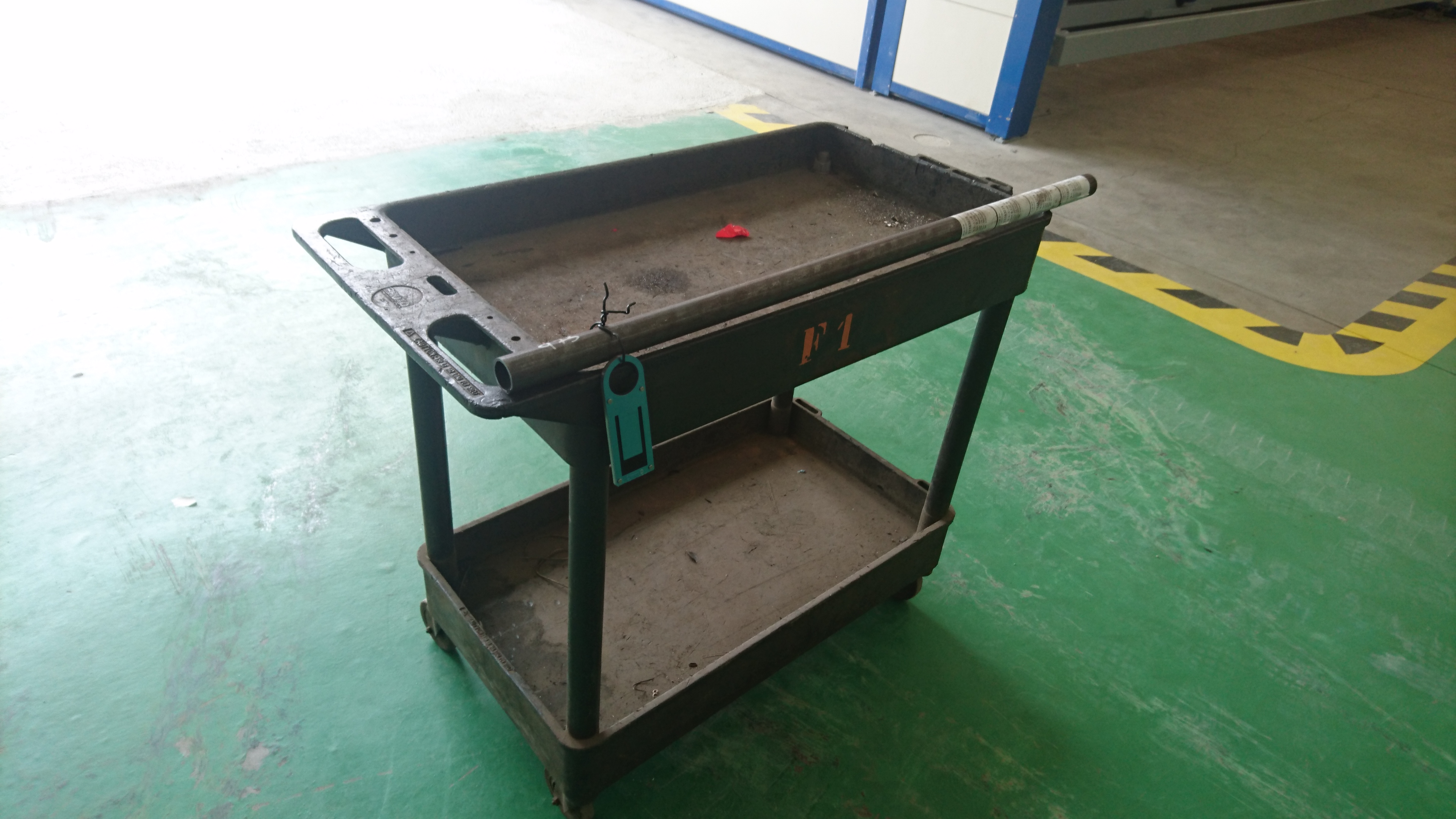}
  	    \caption{Adept 360 UHF Tag.}
    \end{subfigure}
 \label{fig:UHFtags}
\end{figure*}

\subsection{Passive RFID tests}

\subsubsection{Maximum reading distance}

These tests were conducted to determine, for each of the tags acquired, the maximum distance at which the tags can be read when they are oriented in the most favorable position (parallel to the reader antennas). 
Specifically, these tests were carried out in a side of the pipe workshop, taking advantage of its width (about 17 meters). 
At one end of the workshop, the passive UHF reader was placed with their antennas and a pipe of 31\,mm diameter with the different tags adhered was placed in a wheel cart.

The layout of the different elements used in the measurements can be seen in Figures \ref{fig:UHF2antennas} and \ref{fig:UHFtags}.
In these first measurements, which were focused on the determination of the maximum reading distance, only two antennas were used. When tests were performed to include spatial diversity, four antennas were used (with more antennas distributed over the reading area, it is more likely to capture the signal or some reflections). Thus, on the left of Figure \ref{figure:arrayAntennas} it is shown the system while capturing with an array of four antennas.  

\addtocounter{figure}{-1}
\begin{figure}[!hbt]
		\includegraphics[width=1\columnwidth]{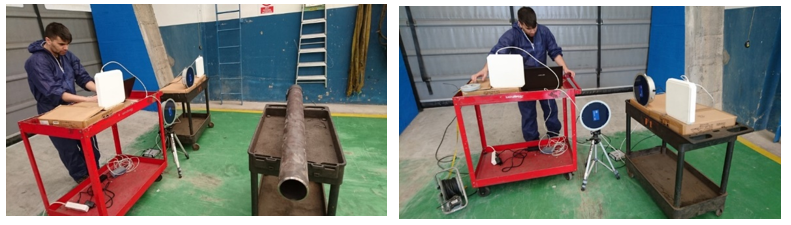}
		\caption{Measurements with passive UHF reader with four antennas.} 
		\label{figure:arrayAntennas}
\end{figure}

Table \ref{table:ReadingDistances} shows a summary of the reading distances achieved when reading the passive tags selected at different distances through the two panel antennas. The distances at which a good percentage of the readings is achieved (i.e. when readings are obtained more than 95\% of the time) are colored in green.
The distances where no readings were obtained, or where a reading was achieved sporadically, are in red. Table \ref{table:ReadingDistances} allows us to conclude that Exo 800, Dura 1500, Dura 3000 and Adept 360\degree are the tags that can be read further. However, Dura 3000, and Adept 360\degree have to be excluded from the final selection for this specific application: the former is so big (210\,mm x 110\,mm x 21\,mm) that it would hinder the operator work on the pipe, while the latter has an inadequate form factor (it is designed to be hanged).

In the rest of the paper, for the sake of brevity, most of the curves represented will refer only to Exo 800 tags, since the conclusions draw from the experiments are identical to the ones obtained with the Dura 1500.

\begin{table}[!htb]
\centering
\caption{Reading distances achieved with the different tags.}
\label{table:ReadingDistances}
\resizebox{\textwidth}{!}{  
\begin{tabular}{|*{18}{c|}}  
\hline
\scriptsize{\backslashbox{\textbf{Tag}}{\textbf{Reading Distance} }}&\scriptsize{\textbf{1\,m}}&\scriptsize{\textbf{2\,m}}&\scriptsize{\textbf{3\,m}}&\scriptsize{\textbf{4\,m}}&\scriptsize{\textbf{5\,m}}&\scriptsize{\textbf{6\,m}}&\scriptsize{\textbf{7\,m}}&\scriptsize{\textbf{8\,m}}&\scriptsize{\textbf{9\,m}}&\scriptsize{\textbf{10\,m}}&\scriptsize{\textbf{11\,m}}&\scriptsize{\textbf{12\,m}}&\scriptsize{\textbf{13\,m}}&\scriptsize{\textbf{14\,m}}&\scriptsize{\textbf{15\,m}}&\scriptsize{\textbf{16\,m}}&\scriptsize{\textbf{17\,m}}\\ 
\hline
 Fit 400&\cellcolor{green}&\cellcolor{green}&\cellcolor{red}&\cellcolor{red}&\cellcolor{red}&\cellcolor{red}&\cellcolor{red}&\cellcolor{red}&\cellcolor{red}&\cellcolor{red}&\cellcolor{red}&\cellcolor{red}&\cellcolor{red}&\cellcolor{red}&\cellcolor{red}&\cellcolor{red}&\cellcolor{red}\\
\hline
Exo 600&\cellcolor{green}&\cellcolor{green}&\cellcolor{green}&\cellcolor{green}&\cellcolor{red}&\cellcolor{red}&\cellcolor{red}&\cellcolor{red}&\cellcolor{red}&\cellcolor{red}&\cellcolor{red}&\cellcolor{red}&\cellcolor{red}&\cellcolor{red}&\cellcolor{red}&\cellcolor{red}&\cellcolor{red}\\
 \hline
Exo 750&\cellcolor{green}&\cellcolor{green}&\cellcolor{green}&\cellcolor{green}&\cellcolor{green}&\cellcolor{green}&\cellcolor{red}&\cellcolor{red}&\cellcolor{red}&\cellcolor{red}&\cellcolor{red}&\cellcolor{red}&\cellcolor{red}&\cellcolor{red}&\cellcolor{red}&\cellcolor{red}&\cellcolor{red}\\
 \hline
Exo 800&\cellcolor{green}&\cellcolor{green}&\cellcolor{green}&\cellcolor{green}&\cellcolor{green}&\cellcolor{green}&\cellcolor{green}&\cellcolor{green}&\cellcolor{green}&\cellcolor{green}&\cellcolor{green}&\cellcolor{green}&\cellcolor{red}&\cellcolor{red}&\cellcolor{red}&\cellcolor{red}&\cellcolor{red}\\
 \hline
Dura 1500&\cellcolor{green}&\cellcolor{green}&\cellcolor{green}&\cellcolor{green}&\cellcolor{green}&\cellcolor{green}&\cellcolor{green}&\cellcolor{green}&\cellcolor{green}&\cellcolor{green}&\cellcolor{green}&\cellcolor{green}&\cellcolor{green}&\cellcolor{green}&\cellcolor{green}&\cellcolor{red}&\cellcolor{red}\\
 \hline
Dura 3000&\cellcolor{green}&\cellcolor{green}&\cellcolor{green}&\cellcolor{green}&\cellcolor{green}&\cellcolor{green}&\cellcolor{green}&\cellcolor{green}&\cellcolor{green}&\cellcolor{green}&\cellcolor{green}&\cellcolor{green}&\cellcolor{green}&\cellcolor{green}&\cellcolor{green}&\cellcolor{red}&\cellcolor{red}\\
 \hline
Dura 600&\cellcolor{green}&\cellcolor{green}&\cellcolor{red}&\cellcolor{red}&\cellcolor{red}&\cellcolor{red}&\cellcolor{red}&\cellcolor{red}&\cellcolor{red}&\cellcolor{red}&\cellcolor{red}&\cellcolor{red}&\cellcolor{red}&\cellcolor{red}&\cellcolor{red}&\cellcolor{red}&\cellcolor{red}\\
 \hline
Adept 360\degree&\cellcolor{green}&\cellcolor{green}&\cellcolor{green}&\cellcolor{green}&\cellcolor{green}&\cellcolor{green}&\cellcolor{green}&\cellcolor{green}&\cellcolor{green}&\cellcolor{green}&\cellcolor{green}&\cellcolor{red}&\cellcolor{red}&\cellcolor{red}&\cellcolor{red}&\cellcolor{red}&\cellcolor{red}\\
 \hline
\end{tabular}
}
\end{table}

\subsubsection{Improving the antenna's reading angle}

The beam of the panel antennas of the UHF system is relatively narrow on purpose, in order to increase reading distance. However, in exchange, the maximum reading angle in which a tag can be read is reduced.
To quantify such an angle, two antenna configurations were tested: a linear array (on the left in Figure \ref{figure:arrayAntennas}) and an 'L-shaped' array (in Figure \ref{figure:arrayAntennas}, on the right). The study of the system coverage was performed for the two tags that offered a better balance between range and physical size (i.e. Exo 800 and Dura 1500). 

Table \ref{table:DistanceAngleDura1500Linear} shows the reading range for the Dura 1500 when using the linear array. Table \ref{table:DistanceAngleExo800Linear} shows the results of the experiment for the Exo 800 tag. In the horizontal axis of the table it is indicated the tag movement respect to the array of antennas, being 90\degree the existing angle when facing the four antennas. The conclusions drawn are identical for both: it can be clearly seen that, when moving the tag to obtuse angles (between 90\degree and 180\degree), reading distance decreases. Some improvement in reading may be achieved at certain distances, but this is due to punctual reflections that occur in the specific location where the measurements were performed. Therefore, the table indicates that a configuration of antennas aligned obtains a good maximum distance as long as tags are located preferably frontwards, but, as the tag moves toward more obtuse angles (and similarly to more acute angles), the system loses much of its reading range. This means that the system would also work in places to control the movement of the tags through certain areas (for example, when moving from one room to another), but it would not work for a constant monitoring of all the workshop tags.

The results for the 'L-shape' array are shown in Table \ref{table:DistanceAngleExo800L}: they show that, in exchange for losing reading distance (because two frontal antennas are used instead of four), the probability of the reading one side of the array is improved significantly.

\begin{table}[H]
\centering
\caption{Distance/angle respect to the different antennas configuration.}
\label{table:DistanceAngle}
\begin{subtable}{.45\linewidth} 
     \begin{tabular}{|c|c|c|c|c|c|c|}
\hline
\tiny{\textbf{Distance/angle}}&\scriptsize{\textbf{180\degree}}&\scriptsize{\textbf{160\degree}}&\scriptsize{\textbf{135\degree}}&\scriptsize{\textbf{130\degree}}&\scriptsize{\textbf{110\degree}}&\scriptsize{\textbf{90\degree}}\\ 
\hline
0.5\,m&\cellcolor{green}&\cellcolor{green}&\cellcolor{green}&\cellcolor{green}&\cellcolor{green}&\cellcolor{green}\\
\hline
1\,m&\cellcolor{green}&\cellcolor{green}&\cellcolor{green}&\cellcolor{green}&\cellcolor{green}&\cellcolor{green}\\
\hline
2\,m&\cellcolor{red}&\cellcolor{red}&\cellcolor{green}&\cellcolor{green}&\cellcolor{green}&\cellcolor{green}\\
\hline
3\,m&\cellcolor{red}&\cellcolor{red}&\cellcolor{green}&\cellcolor{green}&\cellcolor{green}&\cellcolor{green}\\ 
\hline
4\,m&\cellcolor{red}&\cellcolor{red}&\cellcolor{green}&\cellcolor{green}&\cellcolor{green}&\cellcolor{green}\\ 
\hline
5\,m&\cellcolor{red}&\cellcolor{red}&\cellcolor{red}&\cellcolor{green}&\cellcolor{green}&\cellcolor{green}\\ 
\hline
6\,m&\cellcolor{red}&\cellcolor{red}&\cellcolor{red}&\cellcolor{green}&\cellcolor{green}&\cellcolor{green}\\ 
\hline
7\,m&\cellcolor{red}&\cellcolor{red}&\cellcolor{red}&\cellcolor{green}&\cellcolor{green}&\cellcolor{green}\\ 
\hline
8\,m&\cellcolor{red}&\cellcolor{red}&\cellcolor{red}&\cellcolor{green}&\cellcolor{green}&\cellcolor{green}\\ 
\hline
9\,m&\cellcolor{red}&\cellcolor{red}&\cellcolor{red}&\cellcolor{green}&\cellcolor{green}&\cellcolor{green}\\ 
\hline
10\,m&\cellcolor{red}&\cellcolor{red}&\cellcolor{red}&\cellcolor{green}&\cellcolor{green}&\cellcolor{green}\\ 
\hline
11\,m&\cellcolor{red}&\cellcolor{red}&\cellcolor{red}&\cellcolor{green}&\cellcolor{green}&\cellcolor{green}\\ 
\hline
12\,m&\cellcolor{red}&\cellcolor{red}&\cellcolor{red}&\cellcolor{green}&\cellcolor{red}&\cellcolor{green}\\ 
\hline
13\,m&\cellcolor{red}&\cellcolor{red}&\cellcolor{red}&\cellcolor{green}&\cellcolor{red}&\cellcolor{green}\\ 
\hline
14\,m&\cellcolor{red}&\cellcolor{red}&\cellcolor{red}&\cellcolor{red}&\cellcolor{red}&\cellcolor{green}\\ 
\hline
15\,m&\cellcolor{red}&\cellcolor{red}&\cellcolor{red}&\cellcolor{red}&\cellcolor{red}&\cellcolor{green}\\
\hline
\end{tabular}
\caption{Dura 1500 tag with linear array.}
\label{table:DistanceAngleDura1500Linear}
    \end{subtable}
\hfill{}
\begin{subtable}{0.45\linewidth} 
      \begin{tabular}{|c|c|c|c|c|c|c|} 
\hline
\tiny{\textbf{Distance/angle}}&\scriptsize{\textbf{180\degree}}&\scriptsize{\textbf{160\degree}}&\scriptsize{\textbf{135\degree}}&\scriptsize{\textbf{130\degree}}&\scriptsize{\textbf{110\degree}}&\scriptsize{\textbf{90\degree}}\\ 
\hline
0.5\,m&\cellcolor{green}&\cellcolor{green}&\cellcolor{green}&\cellcolor{green}&\cellcolor{green}&\cellcolor{green}\\
\hline
1\,m&\cellcolor{red}&\cellcolor{green}&\cellcolor{green}&\cellcolor{green}&\cellcolor{green}&\cellcolor{green}\\
\hline
2\,m&\cellcolor{red}&\cellcolor{red}&\cellcolor{green}&\cellcolor{green}&\cellcolor{green}&\cellcolor{green}\\
\hline
3\,m&\cellcolor{red}&\cellcolor{red}&\cellcolor{green}&\cellcolor{green}&\cellcolor{green}&\cellcolor{green}\\ 
\hline
4\,m&\cellcolor{red}&\cellcolor{red}&\cellcolor{green}&\cellcolor{green}&\cellcolor{green}&\cellcolor{green}\\ 
\hline
5\,m&\cellcolor{red}&\cellcolor{red}&\cellcolor{red}&\cellcolor{red}&\cellcolor{green}&\cellcolor{green}\\ 
\hline
6\,m&\cellcolor{red}&\cellcolor{red}&\cellcolor{red}&\cellcolor{red}&\cellcolor{green}&\cellcolor{green}\\ 
\hline
7\,m&\cellcolor{red}&\cellcolor{red}&\cellcolor{red}&\cellcolor{red}&\cellcolor{green}&\cellcolor{green}\\ 
\hline
8\,m&\cellcolor{red}&\cellcolor{red}&\cellcolor{red}&\cellcolor{red}&\cellcolor{green}&\cellcolor{green}\\ 
\hline
9\,m&\cellcolor{red}&\cellcolor{red}&\cellcolor{red}&\cellcolor{red}&\cellcolor{green}&\cellcolor{green}\\ 
\hline
10\,m&\cellcolor{red}&\cellcolor{red}&\cellcolor{red}&\cellcolor{red}&\cellcolor{green}&\cellcolor{green}\\ 
\hline
11\,m&\cellcolor{red}&\cellcolor{red}&\cellcolor{red}&\cellcolor{red}&\cellcolor{green}&\cellcolor{green}\\ 
\hline
12\,m&\cellcolor{red}&\cellcolor{red}&\cellcolor{red}&\cellcolor{red}&\cellcolor{green}&\cellcolor{green}\\ 
\hline
13\,m&\cellcolor{red}&\cellcolor{red}&\cellcolor{red}&\cellcolor{red}&\cellcolor{green}&\cellcolor{red}\\ 
\hline
14\,m&\cellcolor{red}&\cellcolor{red}&\cellcolor{red}&\cellcolor{red}&\cellcolor{red}&\cellcolor{red}\\ 
\hline
15\,m&\cellcolor{red}&\cellcolor{red}&\cellcolor{red}&\cellcolor{red}&\cellcolor{red}&\cellcolor{red}\\ 
\hline
 \end{tabular}
\caption{Exo 800 tag with linear array.}
\label{table:DistanceAngleExo800Linear}
	\end{subtable}
\hfill{}

\begin{subtable}{0.45\linewidth} 
      \begin{tabular}{|c|c|c|c|c|c|c|} 
\hline

\tiny{\textbf{Distance/angle}}&\scriptsize{\textbf{180\degree}}&\scriptsize{\textbf{160\degree}}&\scriptsize{\textbf{135\degree}}&\scriptsize{\textbf{130\degree}}&\scriptsize{\textbf{110\degree}}&\scriptsize{\textbf{90\degree}}\\ 
\hline
0.5\,m&\cellcolor{green}&\cellcolor{green}&\cellcolor{green}&\cellcolor{green}&\cellcolor{green}&\cellcolor{green}\\
\hline
1\,m&\cellcolor{green}&\cellcolor{green}&\cellcolor{green}&\cellcolor{green}&\cellcolor{green}&\cellcolor{green}\\
\hline
2\,m&\cellcolor{green}&\cellcolor{green}&\cellcolor{green}&\cellcolor{green}&\cellcolor{green}&\cellcolor{green}\\
\hline
3\,m&\cellcolor{green}&\cellcolor{green}&\cellcolor{green}&\cellcolor{green}&\cellcolor{green}&\cellcolor{green}\\ 
\hline
4\,m&\cellcolor{green}&\cellcolor{green}&\cellcolor{green}&\cellcolor{green}&\cellcolor{green}&\cellcolor{green}\\
\hline
5\,m&\cellcolor{green}&\cellcolor{green}&\cellcolor{green}&\cellcolor{green}&\cellcolor{green}&\cellcolor{green}\\
\hline
6\,m&\cellcolor{green}&\cellcolor{green}&\cellcolor{green}&\cellcolor{green}&\cellcolor{green}&\cellcolor{green}\\
\hline
7\,m&\cellcolor{green}&\cellcolor{red}&\cellcolor{red}&\cellcolor{green}&\cellcolor{red}&\cellcolor{green}\\ 
\hline
8\,m&\cellcolor{red}&\cellcolor{red}&\cellcolor{red}&\cellcolor{green}&\cellcolor{red}&\cellcolor{red}\\ 
\hline
9\,m&\cellcolor{red}&\cellcolor{red}&\cellcolor{red}&\cellcolor{red}&\cellcolor{red}&\cellcolor{red}\\ 
\hline
10\,m&\cellcolor{red}&\cellcolor{red}&\cellcolor{red}&\cellcolor{red}&\cellcolor{red}&\cellcolor{red}\\ 
\hline
11\,m&\cellcolor{red}&\cellcolor{red}&\cellcolor{red}&\cellcolor{red}&\cellcolor{red}&\cellcolor{red}\\ 
\hline
12\,m&\cellcolor{red}&\cellcolor{red}&\cellcolor{red}&\cellcolor{red}&\cellcolor{red}&\cellcolor{red}\\  
\hline
13\,m&\cellcolor{red}&\cellcolor{red}&\cellcolor{red}&\cellcolor{red}&\cellcolor{red}&\cellcolor{red}\\ 
\hline
14\,m&\cellcolor{red}&\cellcolor{red}&\cellcolor{red}&\cellcolor{red}&\cellcolor{red}&\cellcolor{red}\\ 
\hline
15\,m&\cellcolor{red}&\cellcolor{red}&\cellcolor{red}&\cellcolor{red}&\cellcolor{red}&\cellcolor{red}\\ 
\hline
 \end{tabular}
\caption{Exo 800 tag with 'L-shaped' array.}
\label{table:DistanceAngleExo800L}
	\end{subtable}    
  \end{table}

\subsubsection{Modeling RSS vs distance}

\begin{figure}[htb]
\centering
\includegraphics[width=1\columnwidth]{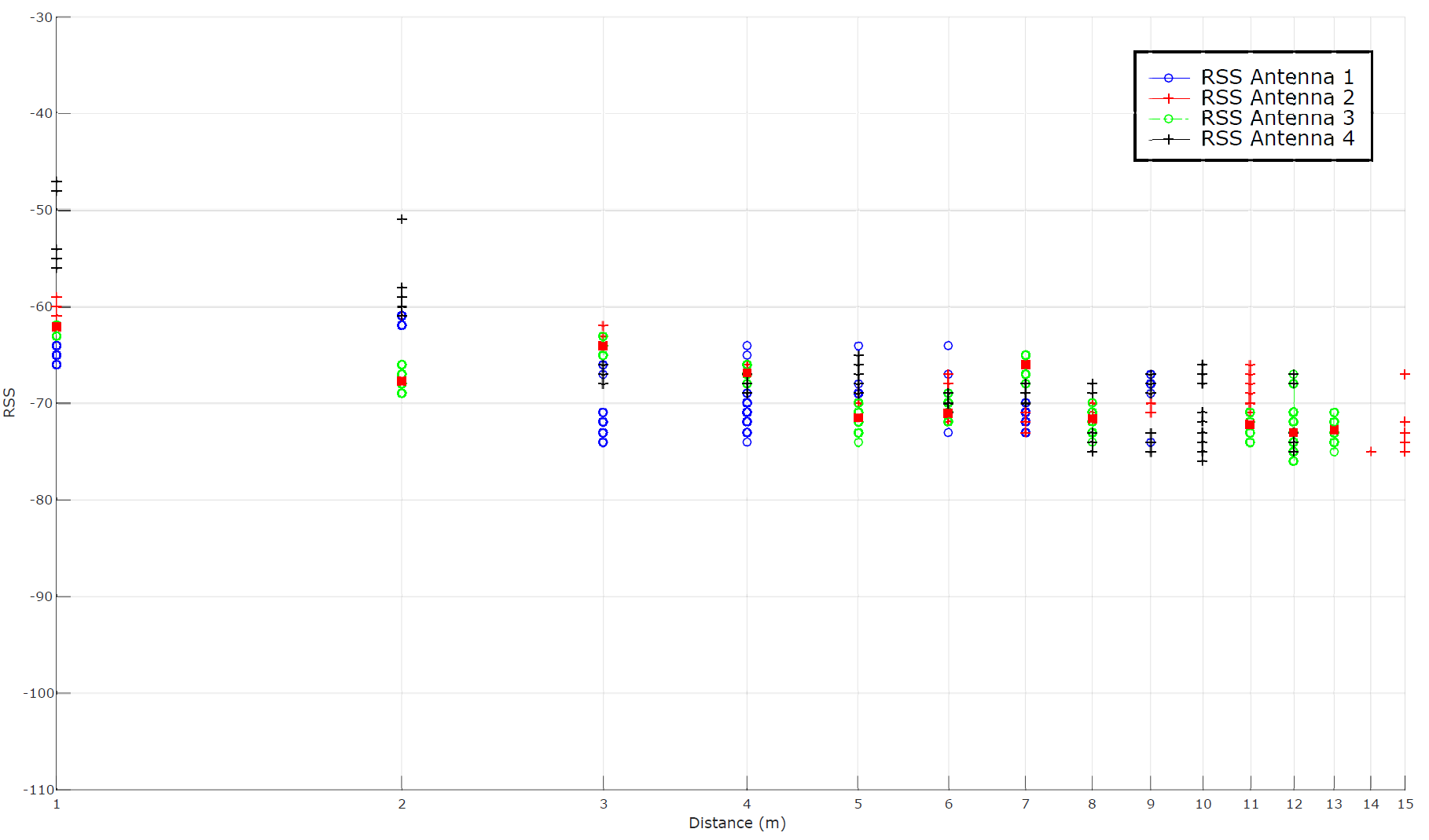}
		\caption{Exo 800: RSS for each antenna.} 
		\label{figure:Exo800RSS_SubFig1}
\end{figure}

As it was concluded previously, the two most promising tags are the Dura 1500 and the Exo 800 tags. It was evaluated their performance in a LOS scenario. Figure \ref{figure:Exo800RSS_SubFig1} shows the distribution of RSS (measured for this Figure and for the others presented in this paper in dBm) in relation to the distance when evaluating the Exo 800: it can be seen how the signal levels decline as the distance with the reader increases. From these RSS values it is possible to look for the mathematical model that allows us to relate them with the distance between the reader and the tag. Therefore, in order to determine the location of a pipe in the workshop, it has to be found a mathematical function that takes as an input the received signal level of a tag and that returns, as an output, the estimated distance to the reader.

\begin{figure}[htb]
\centering
\includegraphics[width=1\columnwidth]{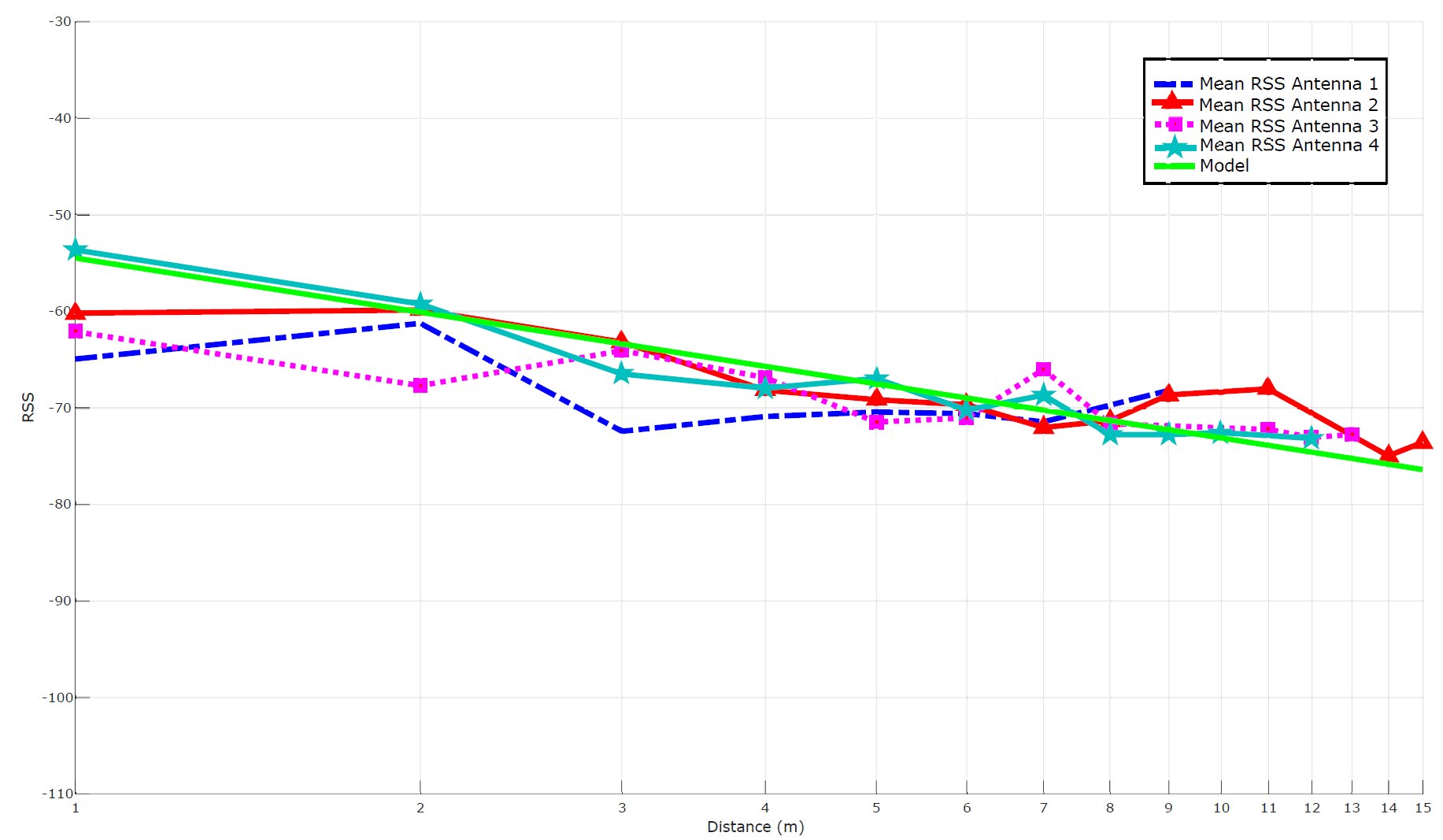}
		\caption{Exo 800: mean curves for each antenna and model obtained with the mean of the four antennas.} 
		\label{figure:Exo800RSS_SubFig2}
\end{figure}

As it was explained in Section \ref{RSSStabilizationTechniques}, there are different ways to obtain this mathematical function. The most direct way is to average the RSS for each distance, what is represented in Figure \ref{figure:Exo800RSS_SubFig2}. In the same Figure, the theoretical model is also adapted to the RSS received and plotted for $PL(d_0)$ and $n$ values equal to -54.5\,dBm and 1.8638, respectively. Ideally, the RSS-based curves should follow a model to obtain a mathematical function that relates RSS to distance, but it can be observed that, in practice, RSS oscillates through distance.

Additionally, it is relevant to highlight that there are antennas at the sides of the linear array that receive no signal at different distances (for instance, in this LOS scenario, antenna 2 is the only one that receives RSSs at a distance of 14\,m and 15\,m).

\subsubsection{Reducing noise: Kalman filtering}

Theoretically, RSS values only should depend on the distance between a tag and the reader. However, in reality, signal strength is influenced by the environment (for instance, the reflections created by metal objects or the absorption related to the presence of water in the air) and, consequently, signals include high levels of noise. In order to filter it, a Kalman filter was applied as described in Section \ref{sec:KalmanFiltering}.

\begin{figure}[!hbt]
		\includegraphics[width=1\columnwidth]{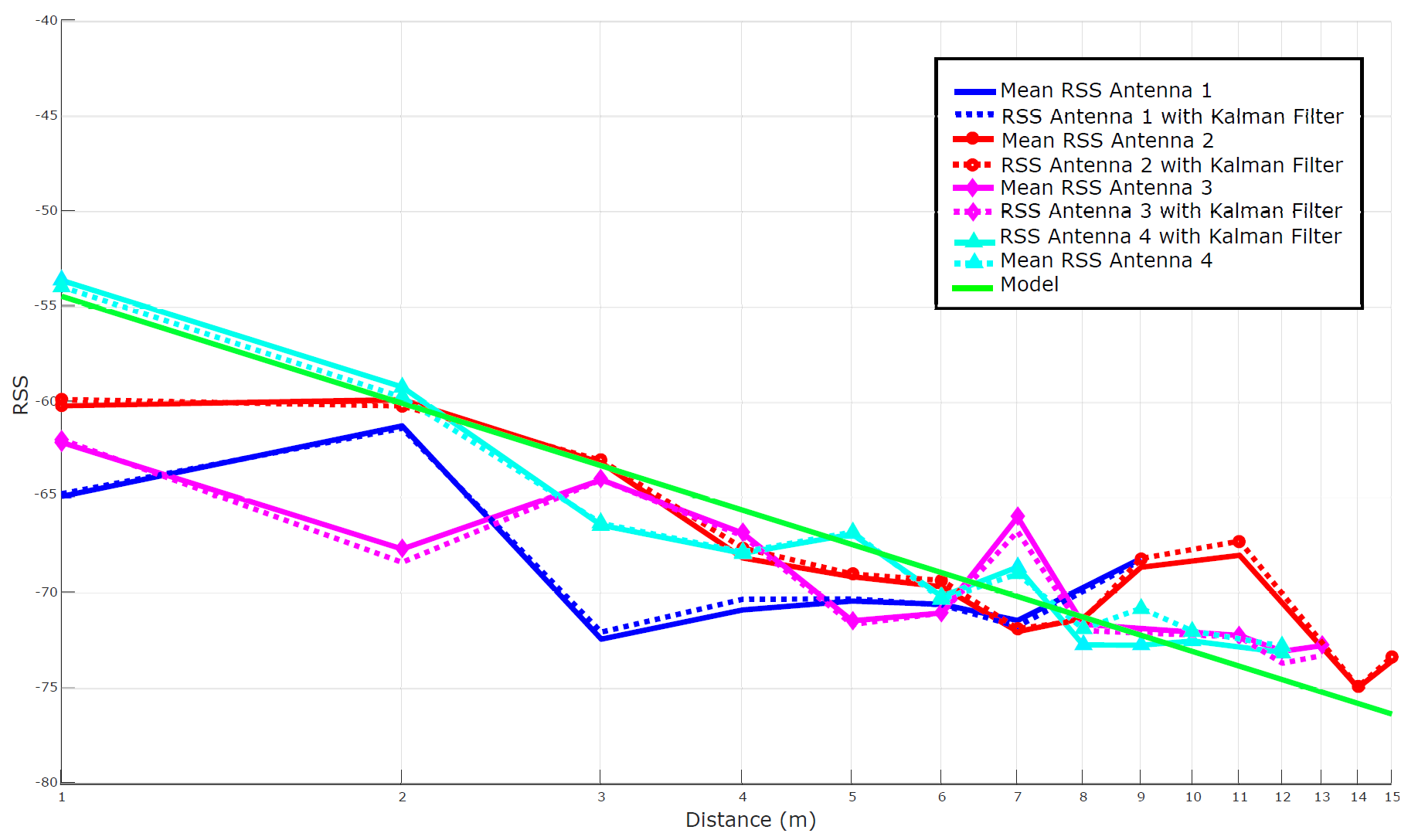}
		\caption{Exo 800: Comparison of the RSS curves with and without Kalman filtering.} 
		\label{figure:Exo800Kalman}
\end{figure}

Figure \ref{figure:Exo800Kalman} compares the results shown in Figure \ref{figure:Exo800RSS_SubFig2} with the ones obtained after applying Kalman filtering to each antenna. It can be seen how the filtered versions of the curves are, in general, slightly close to the theoretical model and seem to be smoother (i.e. more stable). However, at the sight of Figure \ref{figure:Exo800Kalman}, it is not obvious the improvement added by the filter (it can be perceived better in the next subsection, where multi-antenna techniques are also applied).

\subsubsection{Stabilizing RSS with spatial diversity techniques}

\begin{figure}[!hbt]
\centering
		\includegraphics[width=1\columnwidth]{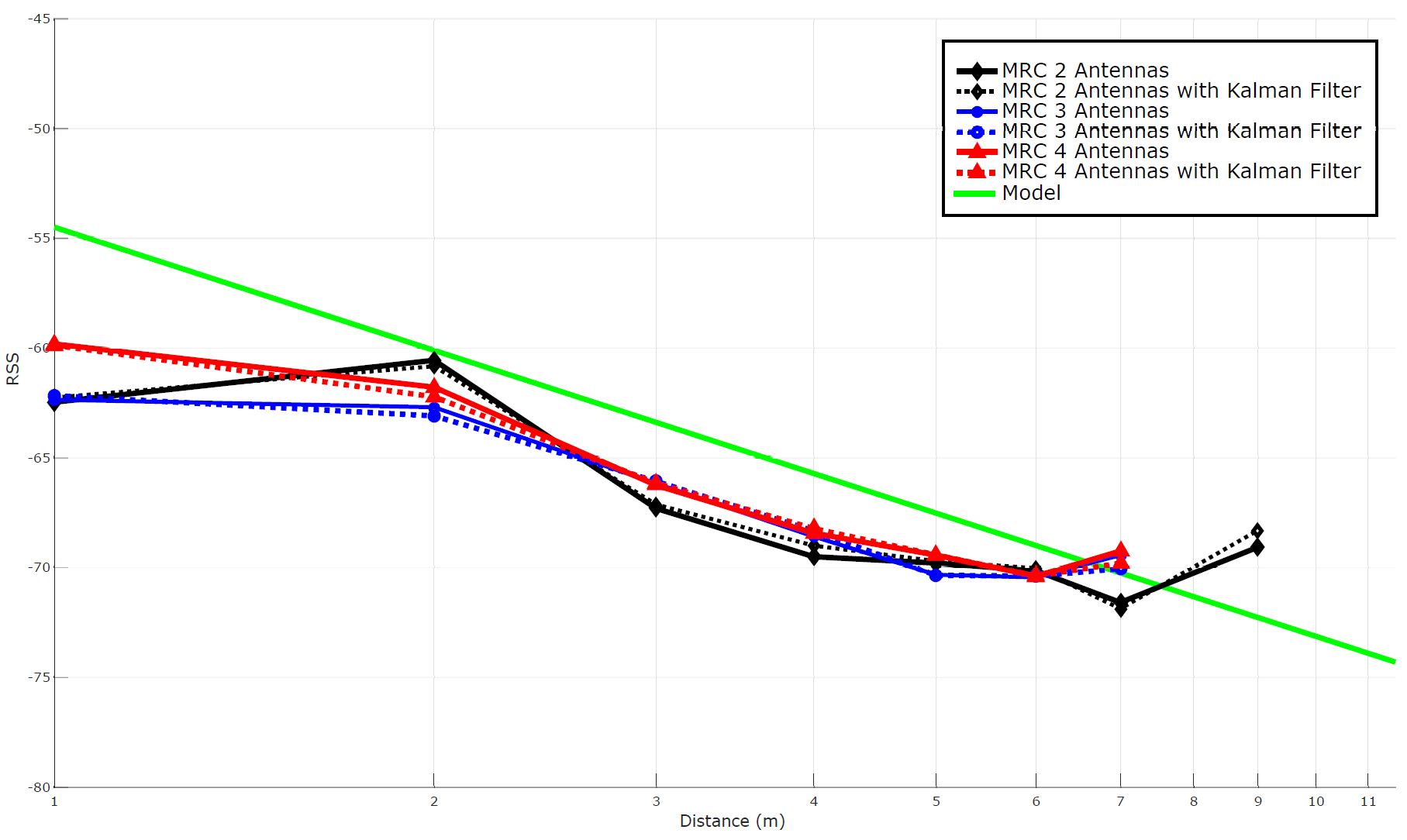}
		\caption{Stabilizing Exo 800 RSS with the MRC technique.} 
		\label{figure:Exo800MRC}
\end{figure}

\begin{figure}[!hbt]
\centering
		\includegraphics[width=1\columnwidth]{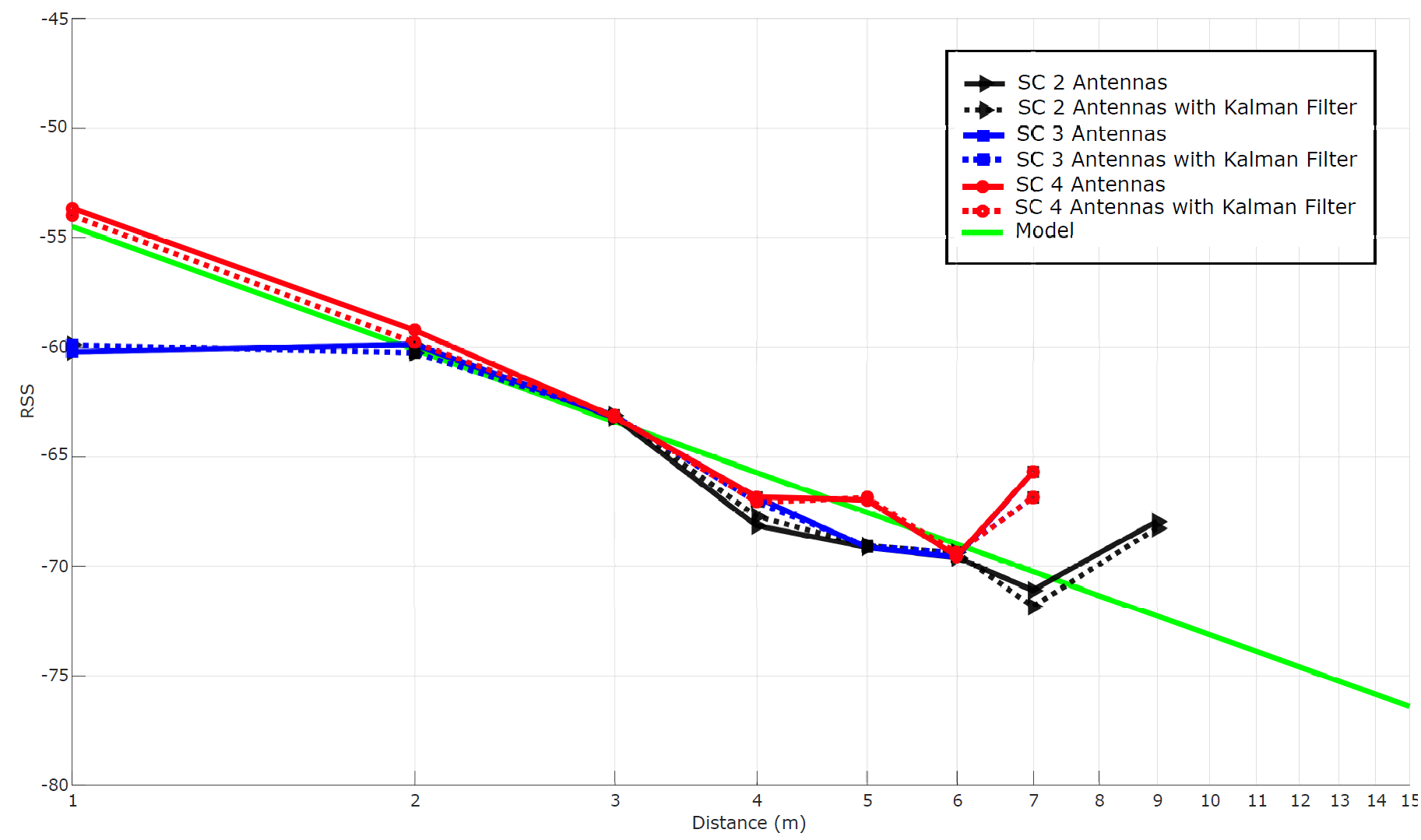}
		\caption{Stabilizing Exo 800 tag RSS with the SC technique.} 
        		\label{figure:Exo800SC}
\end{figure}

\begin{figure}[!hbt]
\centering
		\includegraphics[width=1\columnwidth]{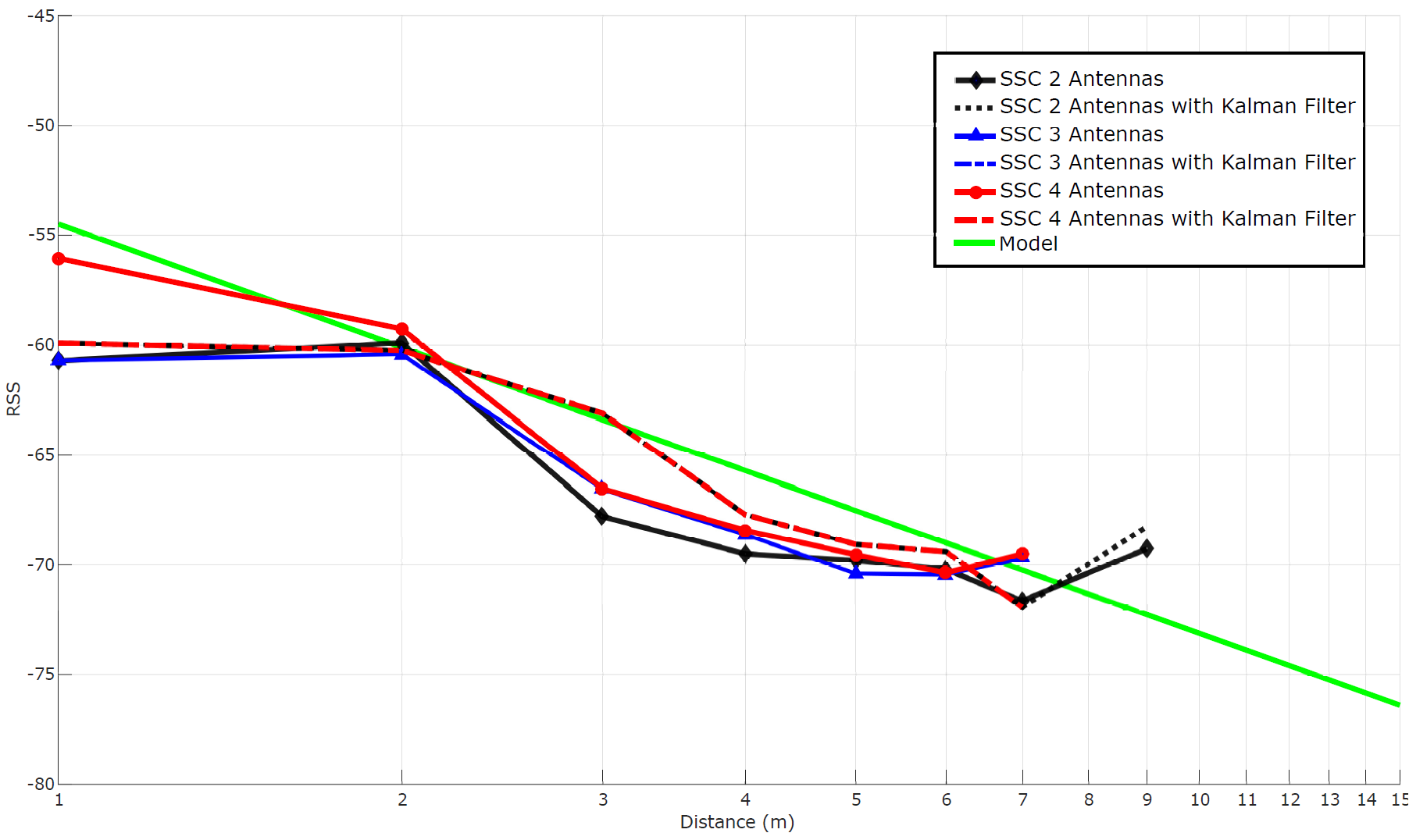}
		\caption{Stabilizing Exo 800 tag RSS with the SSC technique.} 
        		\label{figure:Exo80SSC}
\end{figure}

\begin{figure}[!hbt]
\centering
		\includegraphics[width=1\columnwidth]{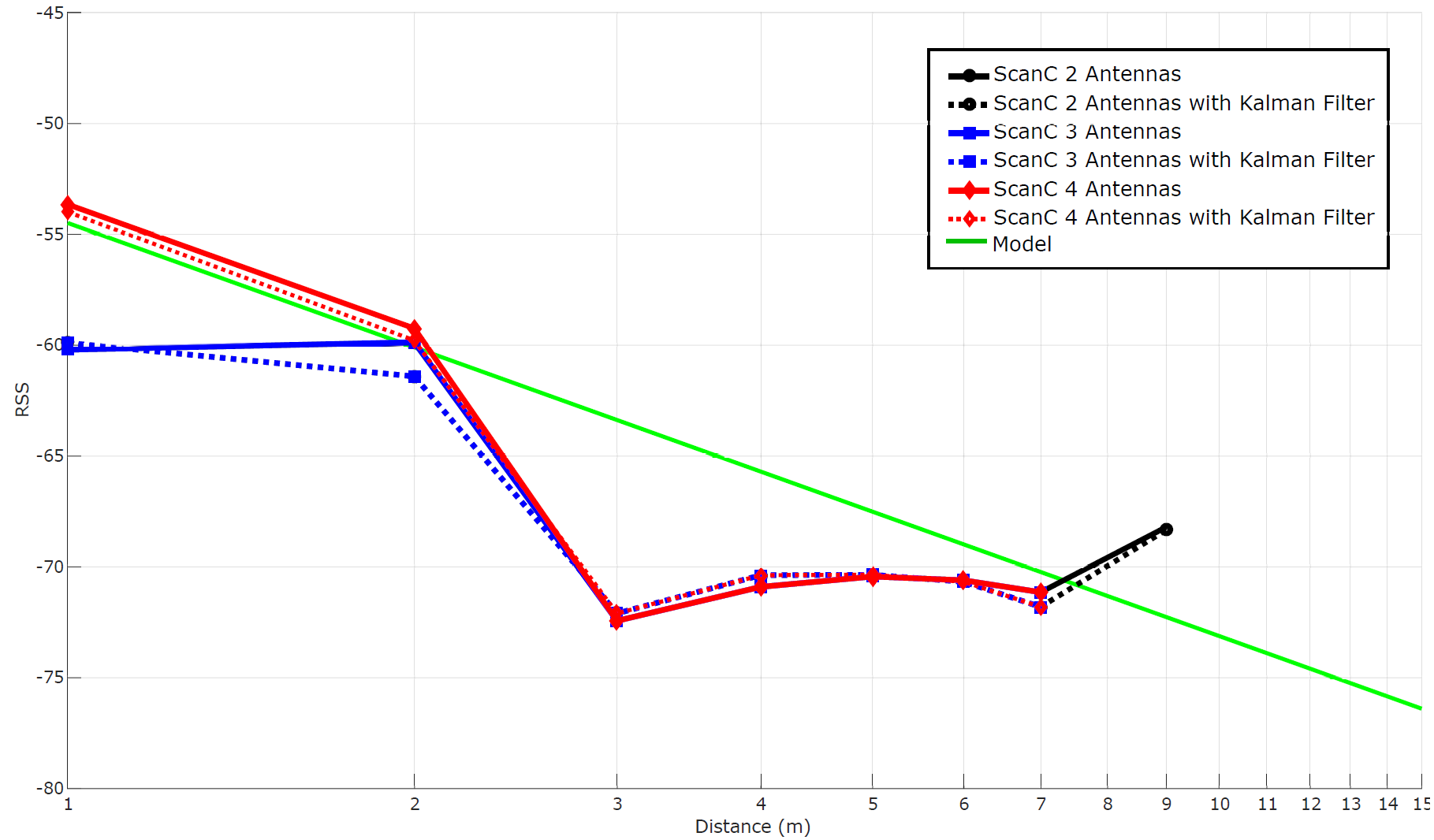}
		\caption{Stabilizing Exo 800 tag RSS with the ScanC technique.} 
        \label{figure:Exo800ScanC}
\end{figure}

The techniques detailed in Section \ref{sec:spatial} can be applied easily to the RSS signals collected by the four antennas of the passive UHF RFID system. Figures \ref{figure:Exo800MRC} to \ref{figure:Exo800ScanC} show the results obtained after applying MRC, SC, SSC and ScanC techniques. Note that, since there are no RSSs for some antennas at certain distances, the curves end before reaching 15\,m. The Figures show that, although the multi-antenna techniques yield curves relatively smooth for low distances, they oscillate for the highest distance values. At a plain sight, it can be observed that the curves for ScanC are the ones that oscillate the most, while the application of MRC, SC and SSC yields curves with an almost constant slope. 

\begin{figure}[!hbt]
\centering
		\includegraphics[width=0.9\columnwidth]{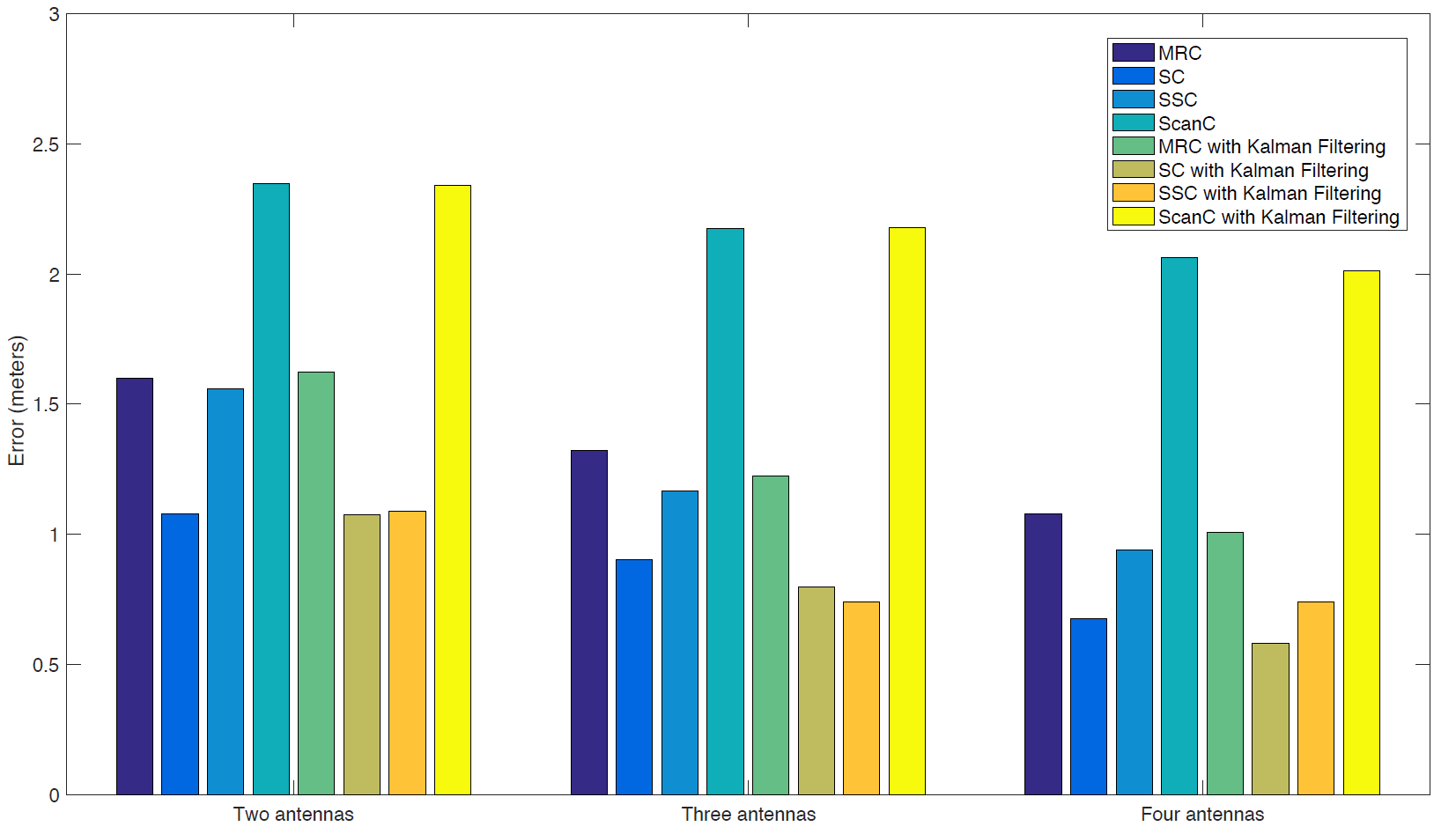}
		\caption{Comparison of RSS stabilization techniques applied to the Exo 800.} 
		\label{figure:ComparisonMethodsAndFiltered}
\end{figure}

Regarding the results obtained for the techniques after applying the Kalman filter designed, it can be stated that the resulting RSS curves are really similar. To quantify the stability of the RSS obtained and the level of improvement achieved thanks to filtering, the Euclidean distance of the curves obtained respect to the curve of the theoretical model was measured. The results are shown in Figure \ref{figure:ComparisonMethodsAndFiltered}. It can be concluded that the addition of antennas stabilizes the RSS and creates curves more similar to the model. Moreover, the use of Kalman filtering, as of implemented, enhances the stability of the multi-antenna techniques applied, but in some cases, the gaining is small. 

Table \ref{tab:passiveerror} shows the mean error of every technique when estimating the distance based on the RSS received for the Exo 800 tags. The table quantifies the improvement achieved by using Kalman filtering: except for the two-antenna MRC, the rest of techniques clearly improve their precision. Furthermore, it can be observed that the SC technique is the most accurate, while ScanC is the one that obtains the worst results, with a precision of roughly 2 meters.

\begin{table}[htb]
\centering
\caption{Mean error (in meters) of the different multi-antenna techniques.}

\begin{tabular}{|*{9}{c|}}
\hline
\scriptsize{\backslashbox{\#Antennas}{Technique}}&\scriptsize{\textbf{MRC}}&\scriptsize{\textbf{Filt. MRC}}&\scriptsize{\textbf{SC}}&\scriptsize{\textbf{Filt. SC}}&\scriptsize{\textbf{SSC}}&\scriptsize{\textbf{Filt. SSC}}&\scriptsize{\textbf{ScanC}}&\scriptsize{\textbf{Filt. ScanC}}\\ 
\hline
2 & 1.6006 & 1.6235 & 1.0791 & 1.0738 & 1.5583 & 1.0882 & 2.3467 & 2.3399\\
\hline
3 & 1.3210 & 1.2236 & 0.9034 &0.7967 & 1.1668 & 0.7398 & 2.1751 & 2.1790\\
\hline
4 & 1.0769 & 1.0061 & 0.6759 & 0.5802 & 0.9381 & 0.7398 & 2.0629 & 2.0119\\
\hline
\end{tabular}
\label{tab:passiveerror}
\end{table}

\subsubsection{Key findings}

After analyzing the results obtained indoors with LOS, it seems that the passive UHF RFID system is suitable only for situations where transitions between areas want to be controlled: due to its limited reading range, it is not useful for a ubiquitous real-time control. Moreover, such a limited range also increases hardware costs, since the larger the area to cover, the more readers and antennas would be needed for the deployment.

An 'L-shaped' antenna array helps to mitigate the reduced reading angle of the system, but it decreases to almost a half the reading distance.
Regarding the stabilization of the RSS, the results obtained show that it is possible to reduce the noise (and, therefore, increase the accuracy) of the system by exploiting spatial diversity and applying Kalman filtering.

\subsection{Active RFID}

Tests were conducted with the active reader following the same methodology as in the passive case: first it was measured the propagation with LOS, and then the tags were tested at different angles and locations throughout the workshop.

\subsubsection{Maximum reading distance}

Before performing the tests in the same way as with the passive system, the total reading range of the tag was measured: it was found that the active tag could be read 95\% of the time at a distance of 100\,m when using high-gain antennas. This figure is greater than the one indicated by the manufacturer in a LOS scenario (90\,m), but it must be taken into account that in the pipe workshop there are many metal objects (for instance, the ceiling is metallic), what generates reflections that, in this case, favor the propagation of wireless signals.

\begin{figure}[t]
		\centering
		\includegraphics[width=0.6\columnwidth]{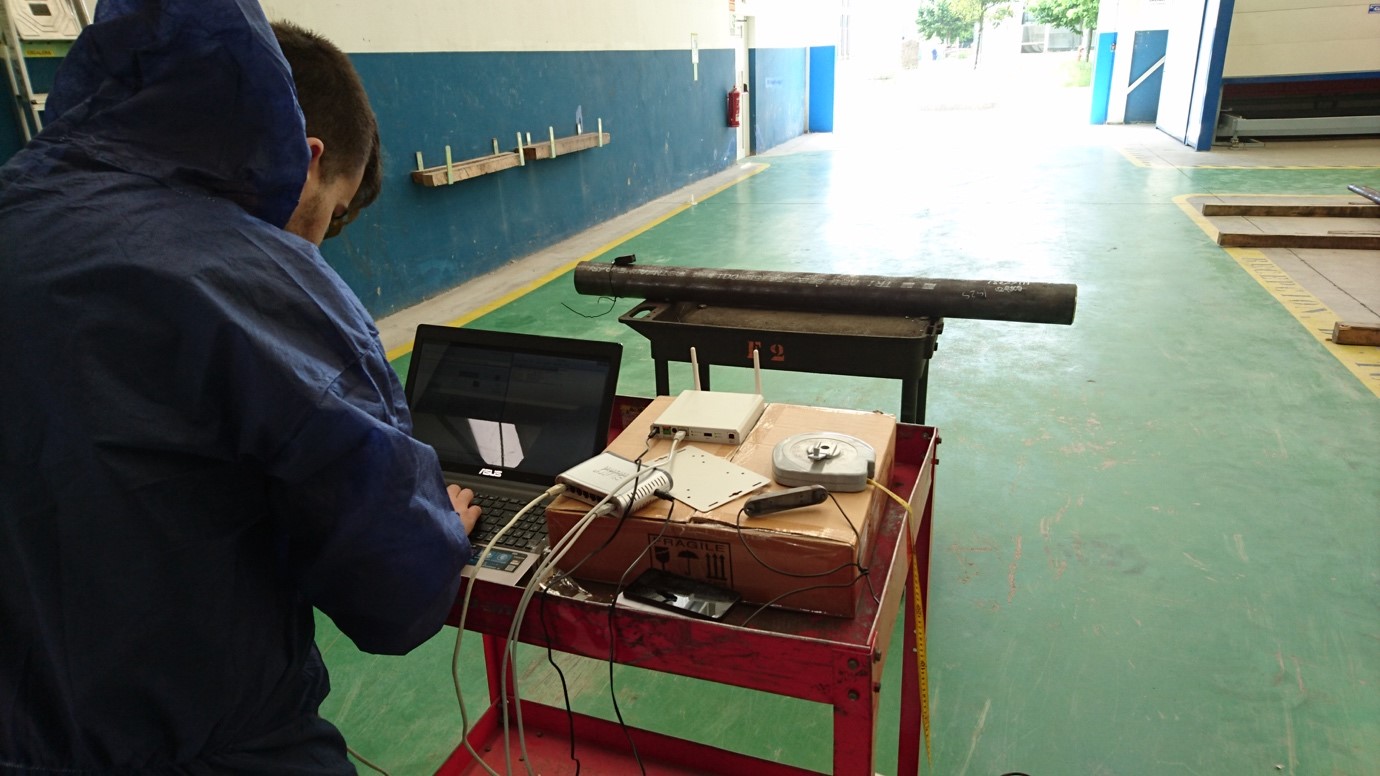}
		\caption{Measurements with the active UHF reader.} 
		\label{figure:ActiveRFID}
\end{figure}

Next, the reading distance was measured like it was previously done for the passive system in order to carry out a fair comparison between both RFID systems (one of the moments during the measurement campaign is shown in Figure \ref{figure:ActiveRFID}). Thus, in this process, the reader remained static at a point, while a pipe placed in a wheeled cart had an Active RuggedTag-175S attached.

Note that the reader can use omnidirectional and high-gain antennas, but since the latter provide more reading range, the results obtained when using them will be the only ones shown in this Section.

\subsubsection{Modeling RSS vs distance}

\begin{figure}[!hbt]
		\includegraphics[width=1\columnwidth]{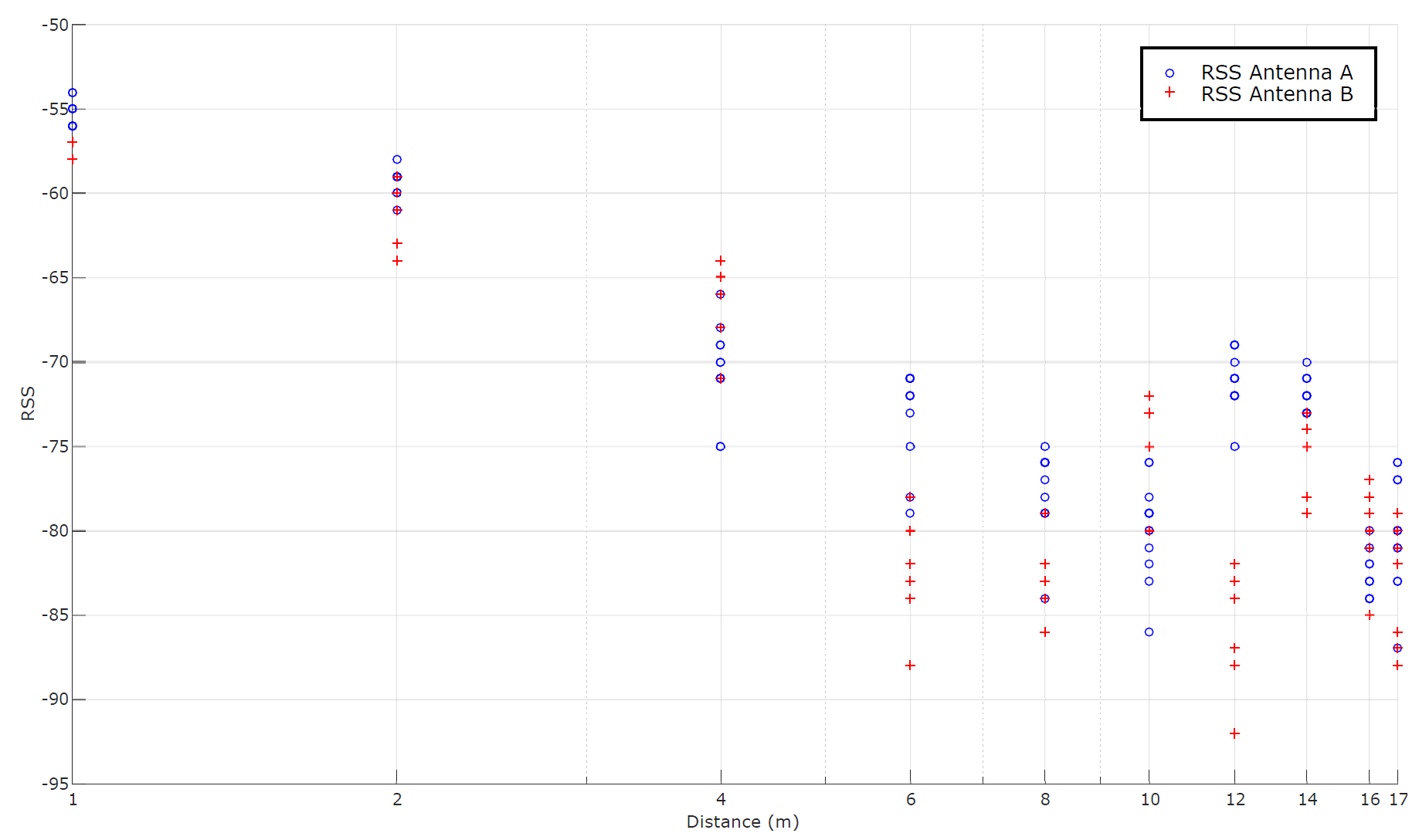}
\caption{RSS values when using high-gain antennas.} 	
\label{figure:RSSActive}
\end{figure}

Figure \ref{figure:RSSActive} shows the RSS values obtained for the active RFID system. Like in the passive system, in such a Figure it can be appreciated that signal levels decrease as the distance between the reader and the tag increases. It is relevant to note that between 10 and 14 meters a light stabilization occurs, which is associated almost certainly to the existence of reflections from metallic elements (at that point it is where the robotic storage of small pipes is placed).

\begin{figure}[!hbt]
		\includegraphics[width=1\columnwidth]{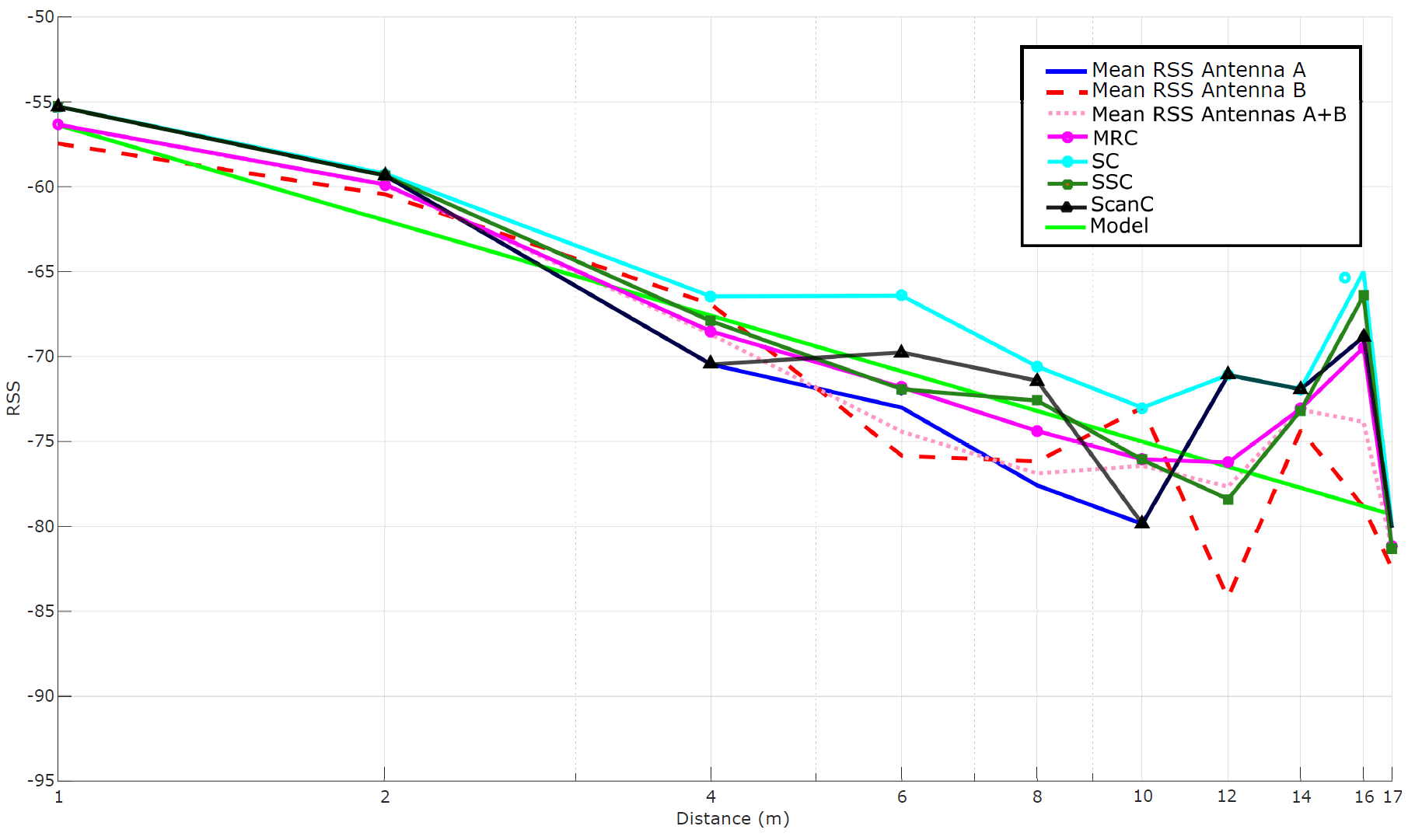}
\caption{RSS means and multi-antenna techniques when using high-gain antennas.} 		\label{figure:RSSActiveReaderMeanHighGain}
\end{figure}

The model explained in \ref{Model} can be applied (with $PL(d_0)$ equal to -56.5\,dBm and $n$ equal to 1.8261) to the RSS collected from the active reader. The curve for the model is shown in Figure \ref{figure:RSSActiveReaderMeanHighGain}, where it can be observed that, for low distance values, the mean RSS from both antennas follows the model closely, but as the distance increases, RSSs oscillate remarkably.

\subsubsection{Kalman filtering}

\begin{figure}[!hbt]
		\includegraphics[width=1\columnwidth]{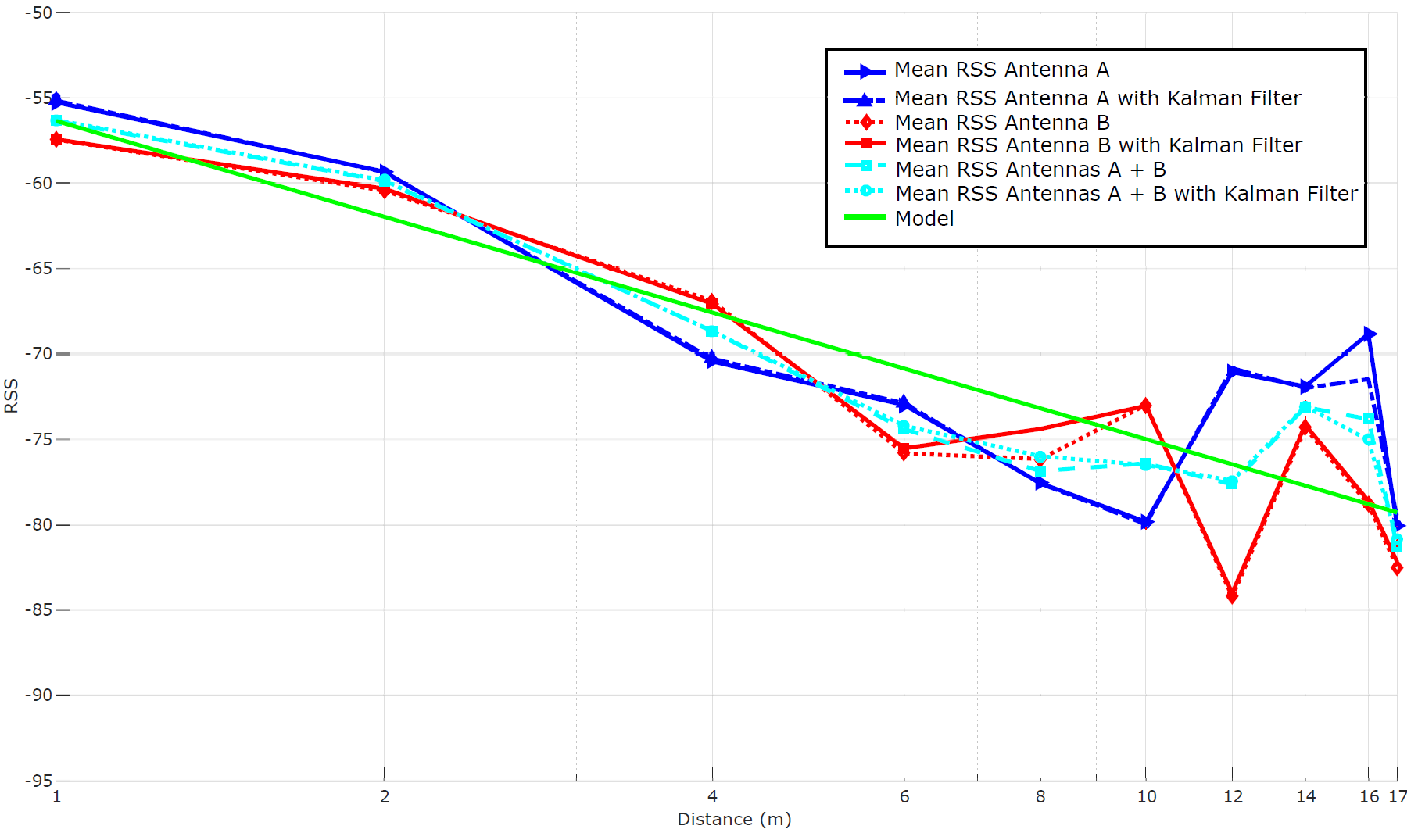}
\caption{Comparison of the RSS curves when using Kalman filtering in the active system.} 
\label{figure:activekalman}
\end{figure}

A Kalman filter can be apply to the mean RSS of each individual antenna and for the mean of both. The resulting curves are shown in Figure \ref{figure:activekalman}. Like in the passive case, it seems that the filtered RSSs generate slightly more stable curves. Nonetheless, the improvement added by the filter is observed better through the results obtained for the multi-antenna experiments described in the next subsection.

\subsubsection{Increasing RSS stability through spatial diversity techniques}

\begin{figure}[htb]
		\includegraphics[width=1\columnwidth]{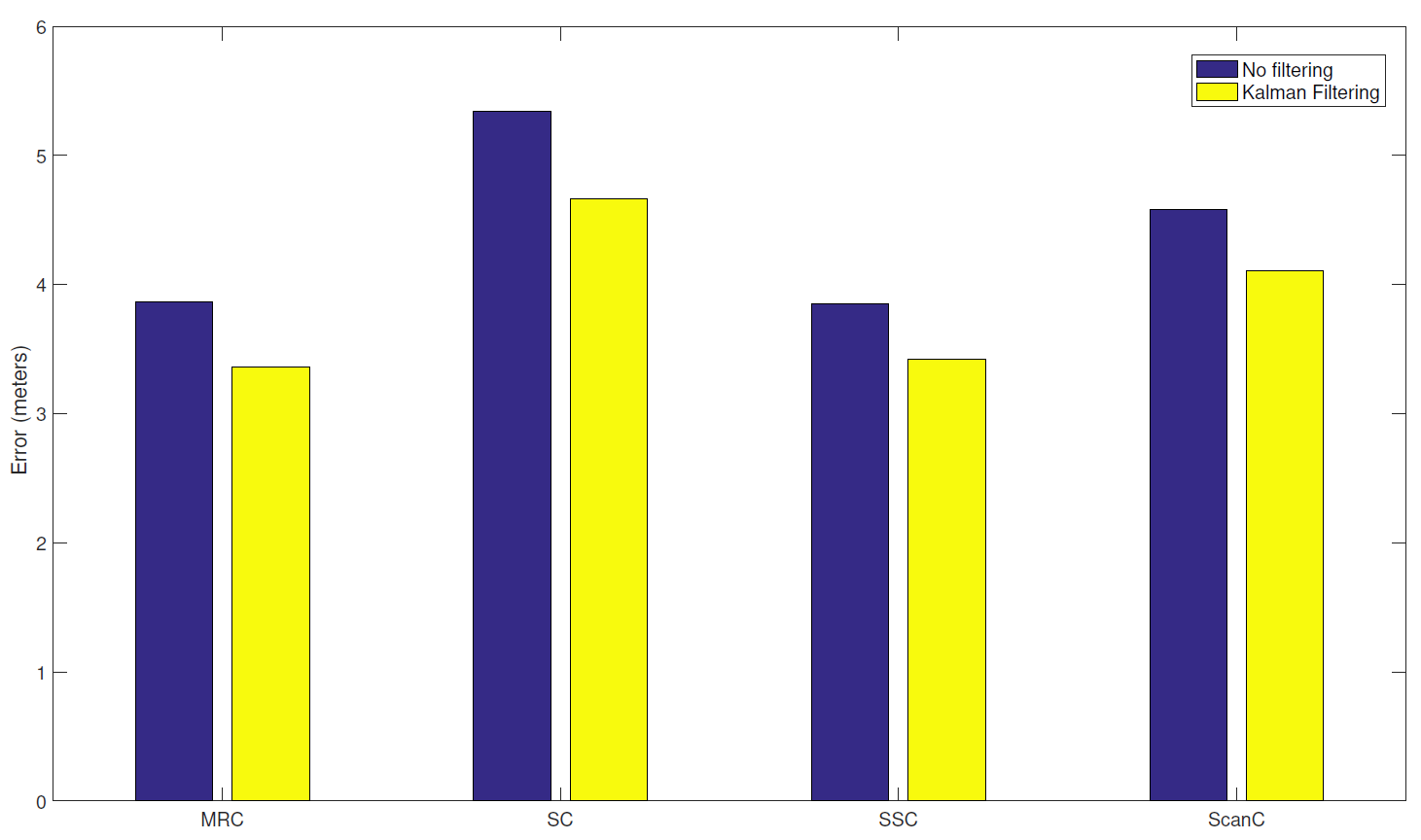}
\caption{Multi-antenna technique stability with and without Kalman filtering.} 
\label{figure:activePrecision}
\end{figure}

It is possible to take advantage of the spatial diversity offered by the two reader antennas to try to improve the stability of the RSS curves. The results obtained for the MRC, SC, SSC, and {\color{red}{ScanC}} techniques are also shown in Figure \ref{figure:RSSActiveReaderMeanHighGain}. Likewise, these results can be compared among them respect to the theoretical model, and when applying Kalman filtering: as it can be concluded from Figure \ref{figure:activePrecision}, MRC and SSC obtain similar results, and, in all cases Kalman filtering improves the stability of the RSS.

Table \ref{tab:activeerror} shows the mean error for every multi-antenna technique when estimating the distance based on the RSS received. In the table it is computed the mean error for different reading distances in order to verify what can be observed after analyzing the RSS curves visually: for distances of up 10\,m the RSS follows closely the theoretical model, but after that distance, RSS oscillations alter the estimations, deriving into a clear loss of accuracy. Thus, up to 10\,m the precision of system is around 1\,m, but it is increased to 3-4\,m for higher distances. The table also shows the worsening of the accuracy when increasing the reading distance from 16\,m to 17\,m: although there is only 1\,m between them, the system loses approximately 50\,cm of precision. Finally, Table \ref{tab:activeerror} allows us to conclude that Kalman filtering clearly improves the precision of the system and, among the different spatial diversity techniques, in this specific scenario, the filtered version of MRC is the most accurate technique.

\begin{table}[htb]
\centering
\caption{Mean error (in meters) of the different multi-antenna techniques for the active system.}

\begin{tabular}{|*{9}{c|}}
\hline
\scriptsize{\backslashbox{Max. Distance}{Technique}}&\scriptsize{\textbf{MRC}}&\scriptsize{\textbf{Filt. MRC}}&\scriptsize{\textbf{SC}}&\scriptsize{\textbf{Filt. SC}}&\scriptsize{\textbf{SSC}}&\scriptsize{\textbf{Filt. SSC}}&\scriptsize{\textbf{ScanC}}&\scriptsize{\textbf{Filt. ScanC}}\\ 
\hline
10\,m & 0.9057 & 0.6212 & 1.6425 & 1.1268 & 1.0485 & 1.1008 & 1.0060 & 0.5650\\
\hline
16\,m & 3.5396 & 2.9744 & 4.8282 & 4.0904 & 3.5497 & 3.2720 & 4.0793 & 3.5286\\
\hline
17\,m & 3.8662 & 3.3645 & 5.3449 & 4.6672 & 3.8546 & 3.422 & 4.5838 & 4.1054\\
\hline
\end{tabular}
\label{tab:activeerror}
\end{table}

\subsubsection{Key findings}

The results obtained for the active RFID system while transmitting in the pipe workshop in the best case scenario (with LOS) showed that the tags offer long range readings. This means that the system is suitable for monitoring in real-time wide areas of the workshop with just one reader (what also decreases the deployment costs). 

However, although the accuracy of the RSS-based location system is roughly 1\,m for distances of up 10\,m when applying multi-antenna techniques and Kalman filtering, it increases rapidly with further distances, what indicates that is necessary to make use of more sophisticated positioning algorithms.

The accuracy values obtained are in the same order of magnitude to those found in the literature (the most relevant examples can be seen in Table \ref{tab:comparison}). These examples can be examined as evidence of the behavior of the technologies in non-adverse environments, mainly offices and university buildings. 
Nevertheless, direct comparison with the behavior of the smart pipe system proposed is difficult because a coherent methodology does not exist between the different works: various technologies are applied, different algorithms or techniques, or distinct beacon spatial density.
One of the main factors that differentiate the environment presented with the ones exhibited by other publications is that the experimental conditions of our proposal, although sometimes similar regarding the size of the scenario, are quite adverse due to the massive presence of metals, thus, facing a truly challenging deployment.

\begin{table}[htb]
\centering
\caption{Main features of state-of-the-art indoor positioning systems.}
\resizebox{\textwidth}{!}{  
\begin{tabular}{|*{4}{c|}}
\hline
\large{\textbf{System}}&\large{\textbf{Scenario}}&\large{\textbf{Technology}}&\large{\textbf{Accuracy (average error)}} \\ 
\hline
LIPS \cite{LIPS} &Buildings&  Wi-Fi & $\thicksim$0.76\,m \\
\hline
SDM \cite{SDM} &Small building (598\,m$^{2}$)&  Wi-Fi & $\thicksim$3\,m \\
\hline
OIL \cite{OIL}& Area with four rooms & Wi-Fi &Order of meters\\
\hline
EZ \cite{EZ} & Small (486\,m$^{2}$) and big buildings (12.600\,m$^{2}$) &Wi-Fi & $\thicksim$ 2\,m (small) and 7\,m (big)\\
\hline
WiGEM \cite{WiGEM} & Small (600\,m$^{2}$) and medium (3.250\,m$^{2}$) buildings & Wi-Fi & <8\,m \\
\hline
WILL \cite{WILL} & Medium academic buildings (1.600\,m$^{2}$)& Wi-Fi& 86\% room level accuracy\\
\hline
UnLoc \cite{UnLoc} &Different setups (largest 4.000\,m$^{2}$) & Wi-Fi and inertial sensors&$\thicksim$1.69\,m\\
\hline
Zee \cite{Zee}& Medium sized building (2.275\,m$^{2}$)& Wi-Fi& $\thicksim$3\,m together with EZ or Horus.\\
\hline
LiFS \cite{LiFS}& Medium sized building (1.600\,m$^{2}$)&Wi-Fi& 89\% room level accuracy\\
\hline
Walkie-Markie \cite{Walkie-Markie}& Medium size office (3.600\,m$^{2}$) and shopping mall&Wi-Fi& $\thicksim$ 1.65\,m \\ 
\hline
RADAR \cite{RADAR}& Area 975\,$m^{2}$, more than 50 rooms&Wi-Fi& $\thicksim$ 3-5\,m \\ 
\hline
Horus \cite{Horus}&  4th floor of the Computer Science building&Wi-Fi&  $\thicksim$ 3-5\,m \\ 
\hline
AiRISTA Flow\cite{AiRISTA} &Asset tracking&Wi-Fi& $\thicksim$ 1\,m \\ 
\hline
IZat \cite{IZat} &Automotive &Wi-Fi, GPS, 4G& $\thicksim$ 5\,m \\ 
\hline
Ubisense \cite{Ubisense}& Smart factory&UWB&  $\thicksim$ 15\,cm \\ 
\hline
Dart \cite{Dart}&Manufacturing&UWB&$\thicksim$30\,cm \\ 
\hline
3D-LOCUS \cite{3D-LOCUS}&Laboratory &Ultrasound  & $\thicksim$ 8\,mm \\ 
\hline
Elpas \cite{Elpas}&Healthcare and commercial& IR, UHF and LF RFID&$\thicksim$ 1\,m \\ 
\hline
SpotON \cite{SpotON}&Laboratory & Active RFID&$\thicksim$ 3\,m \\  
\hline
Topaz \cite{TOPAZ}&Areas of $\thicksim$1000\,m$^{2}$ & Bluetooth+IR& 2-3\,m \\ 
\hline
Landmarc\cite{Landmarc}& Laboratory & Active RFID& <\,2\,m \\ 
\hline
Sparse distribution\cite{sparse}& Not specified& Passive RFID& <\,10\,cm \\ 
\hline
GPs \cite{GPPS}&Area of 1600\,m$^{2}$ (55 rooms) &Active RFID& $\thicksim$\,1.5\,m \\ 
\hline
Robotics-based \cite{Robot}&3rd floor of Duncan Hall at Rice University &Wi-Fi&  $\thicksim$\,1.5\,m \\
\hline
MultiLoc \cite{MultiLoc}&4th floor (more than 10 rooms) &Wi-Fi&  $\thicksim$\,2.7\,m \\
\hline
Zigbee IPS \cite{Zigbee} & Engineering building 7.26$\times$16.5 \,m &Zigbee & $\thicksim$\,3.01\,m \\
\hline
TIX \cite{TIX}& Office (1020\,m$^{2}$)  &Wi-Fi& $\thicksim$\,5.4\,m \\
\hline
\footnotesize{BLE fingerprinting}\cite{BLE}& Office (600\,m$^{2}$)&BLE& <\,2.6\,m, high beacon density\\
\hline
\footnotesize{GSM fingerprinting} \cite{GSMfingerprinting} & Large multi-floor buildings &GSM& $\thicksim$\,5\,m \\
\hline
NDI \cite{Optotrak} &Industry and healthcare&Infrared (3D)&  0.1\,mm \\
\hline
IRIS\_LPS \cite{IRIS_LPS}& Lecture hall (100\,m$^{2}$) &Infrared + stereo camara&  <\,16\,cm \\
\hline
Bat \cite{Bat} &Building (1000\,m$^{2}$), three floors&Ultrasound &  3\,cm \\
\hline
Cricket \cite{Cricket}&Different rooms &RF + Ultrasound &  10\,cm, orientation accuracy 3$\degree$ \\
\hline
Sonitor \cite{Sonitor}&Healthcare&Ultrasound,Wi-Fi and LF &  Room level accuracy \\
\hline
COMPASS \cite{COMPASS} & Office building (312\,m$^{2}$) &Wi-Fi, no real-time tracking &  $\thicksim$\,1.65\,m \\
\hline
OPT \cite{OPT}&Context aware app in personal network &IEEE 802.15.4  &1.5-3.8\,m \\ 
\hline
Easy Living \cite{EasyLiving} &In-home and in-offices &Multiple tech& Accuracy non guaranteed\\
\hline
 Beep \cite{Beep} & Room 20x9\,m: users can use their own devices&Sound source location& 0.4\,cm \\ 
\hline
\end{tabular}
\label{tab:comparison}
}
\end{table}

	
\subsection{Display module of the smart pipe system}

\begin{figure}[t]
    \centering
        \includegraphics[width=1\columnwidth] {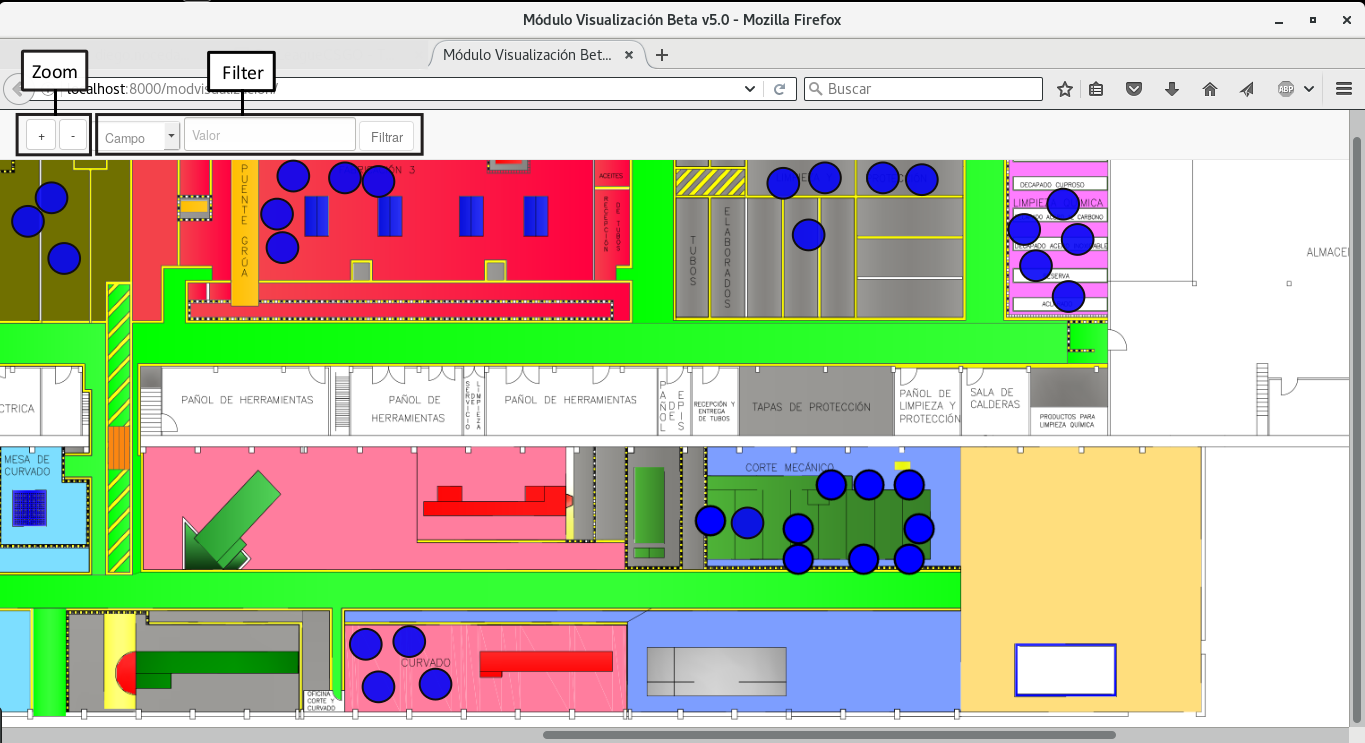}
\caption{Floor map of the workshop with located pipes (blue circles).} 
        \label{figure:VisualizationModule}
    \end{figure}

Once verified that constant monitoring can be only achieved with an active RFID system, additional tests were conducted for the ubiquitous real-time CPS system while transmitting in the pipe workshop. The results obtained can be seen by the operators thanks to the display module, which can be accessed through any device with a regular web browser (i.e. from PCs, Macs, smartphones, and tablets). 
The system shows the location of the pipes in the workshop in real time in the way shown in Figure \ref{figure:VisualizationModule}, where the blue circles represent the pipes or a set of pipes within a radius of 2\,m (this is useful for pallets that carry several pipes).

Considering that numerous pipes are displayed while moving through the workshop, a filter can be used to only show a specific pipe or a subset that meet certain criteria. Moreover, operators can zoom in and out throughout the floor map to watch specific areas.

\begin{figure}[H]
    \centering
        \includegraphics[width=1\columnwidth] {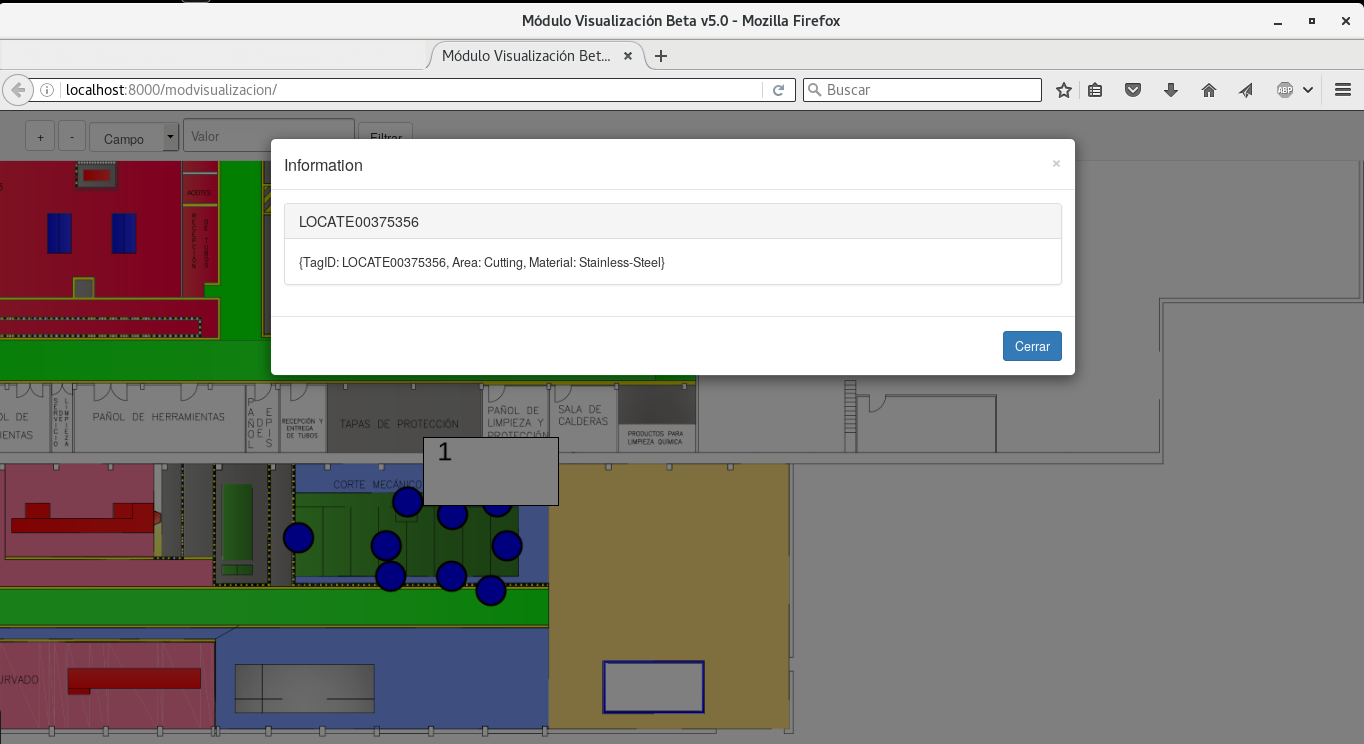}
		\caption{Example of the information shown to the operators about the basic characteristics of a pipe.} 
        \label{figure:FilteringPipesVisualizationModule}
    \end{figure}

Operators, in addition to viewing the pipes in a map, can access certain information about them. This functionality is illustrated in Figure \ref{figure:FilteringPipesVisualizationModule} and it implies obtaining data on specific parameters (i.e. pipe identifier, area or material).

Accuracy can also be studied from this display module: it can be observed the position oscillations of the pipes due to the RSS noise. Such a noise is drastically reduced with the techniques analyzed in the previous sections and gives a really stable representation of the state of the pipe workshop in locations with LOS.

\subsection{Automatic event detection using smart pipes}

\begin{figure}[!hbt]
    \centering
        \includegraphics[width=1\columnwidth] {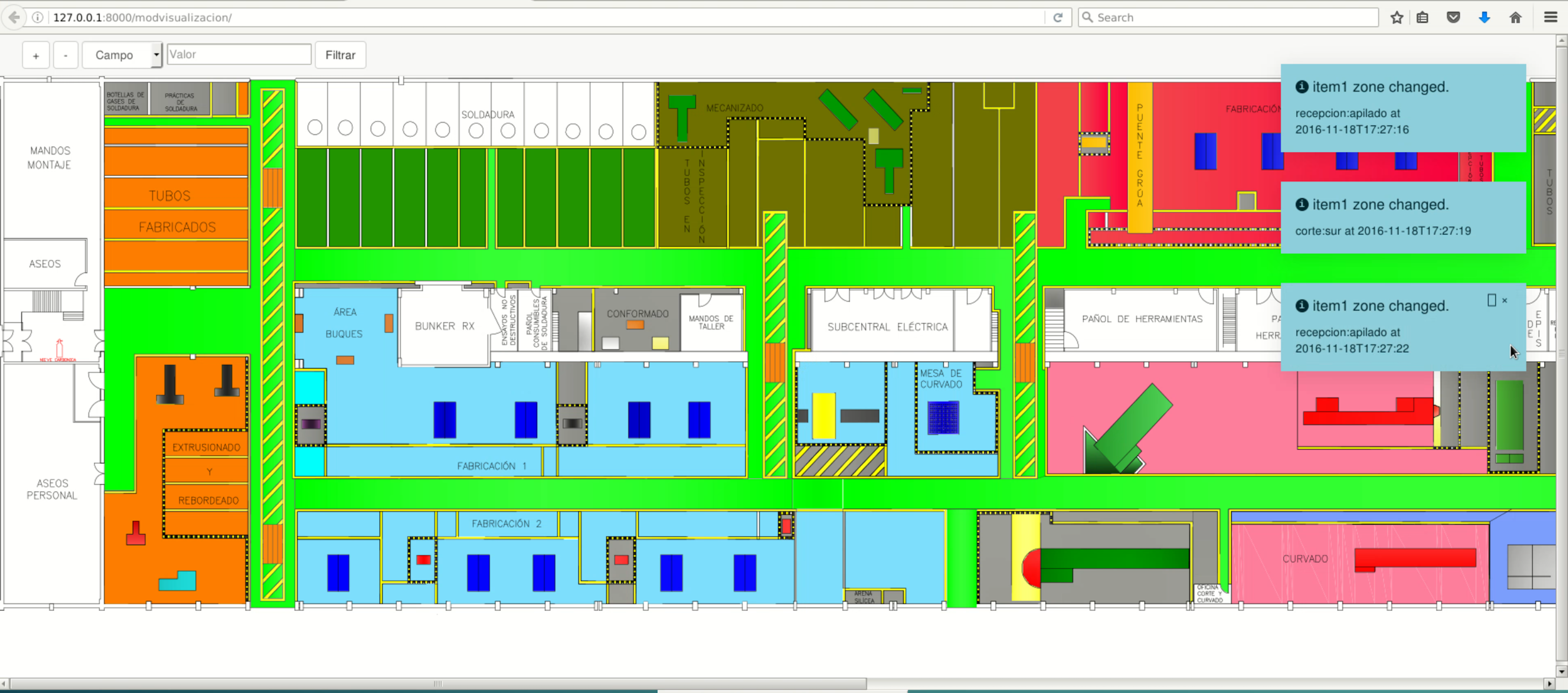}
		\caption{Notifications shown on the right upper part when a pipe crosses from one area to another.} 
        \label{figure:Notifications}
    \end{figure}

The concept of smart pipe is based on the fact that a pipe actually informs through its communications transceiver signal level about its position. Such an information, when accurate, allows for implementing a wide range of very useful services. The services are provided by the business intelligence module, which processes the position information and shows meaningful notifications through the display module. Figure \ref{figure:Notifications} illustrates an example of a smart service: when a pipe crosses from one area to another, a pop-up is shown on the upper right part of the screen indicating the event, the pipe identifier, and a timestamp.  This event warns the operators when a certain pipe is coming towards their work area so that they are prepared to receive it.

These notifications represent a simple but useful service. In future releases the software will incorporate other services like the measurement of the time spent by a pipe in certain areas (to create statistics about the performance of the different areas/operators), the quantification of the level of occupation of the stacking areas, or the possibility of triggering certain automatic behaviors (for instance, through the workshop machinery and robots) when a pipe reaches certain point.

\section{Conclusions} \label {SecCon}
This article described the selection and validation of the necessary technology to perform improved traceability and location of the pipes used in the construction of ships. The proposed smart pipe system is a novel example of the benefits of CPS, thus providing a reliable remote monitoring platform to leverage strategic applications and enhancements in the shipyard environment. The system is based on the concept of smart pipe, a kind of pipe able to transmit signals periodically that allows for providing useful services in a shipyard. This paper conducted a detailed analysis of the shipyard environment, the scenarios that involve pipes in their process, the operational and technical requirements, and a feasibility study to choose the best technology and communications architecture. Also, it has made a comprehensive technology validation.
Through multiple tests it was confirmed that:
\begin{itemize}
\item The passive RFID system, due to its limited range, is only suitable for situations where the aim is to control the movement between areas, but it would not be suitable for ubiquitous real-time control.

\item This passive system, when making use of an 'L-shaped' array of antennas, mitigates the low angle of reading, but it reduces the reading distance to almost a half.

\item In the case of active RFID readers, their use allows for reaching longer distances and get constant monitoring of tags in wide areas. However, the longer the distance, the less accurate the RSS-based distance estimations are.

\item Multi-antenna algorithms and Kalman filtering help to stabilize RSS and improve the accuracy of the positioning system.
\end{itemize}

Given all these factors, it is verified that the active RFID technology is the best positioned technology for implementing the CPS proposed. Further research needs to be carried out to explore even more accurate positioning algorithms in order to minimize the influence of the interference caused by the environment.

Although the present article was aimed at applying the latest research for building a smart pipe system, it also included the following original contributions, which have not been found together in the literature as of writing:

\begin{itemize}
\item It presented the novel concept of Shipyard 4.0, a shipyard built on the application of the principles of Industry 4.0.

\item It described in detail how a shipyard pipe workshop works and what are the requirements for building a smart pipe system.

\item It indicated how to build a positioning system from scratch in an environment as harsh in terms of communications as a shipyard. Additionally, it was not found in the literature any practical analysis on the application of RFID technology in any similar application and scenario.

\item It defined the concept of smart pipe and showed an example of its implementation and the architecture that surrounds it. 

\item It proposed the use of spatial diversity techniques to stabilize RSS values. It has not been found any other publication describing the application of such techniques in RFID systems in the way it is proposed in this paper.
\end{itemize}

To summarize, this article presented the foundations for enabling an affordable CPS for Shipyards 4.0. The system design proposed allows shipyards to collect more information on the pipes and to make better use of it. Furthermore, with this system working, the development of applications related to the monitoring of elements different than pipes (e.g. wearables for operators, tools, shared machines) is straightforward. Thus, the forthcoming applications will enable the Shipyard 4.0 to leverage smarter energy consumption, greater inbound/outbound logistics and information storage, workforce safety and control, and real-time yield optimization.


\acknowledgments{\textbf{Acknowledgments:}
This work is part of the Pipe Auto-ID research line of the Navantia-UDC Joint Research Unit and has been funded by it. 
}

\authorcontributions{\textbf{Author Contributions:} 
Diego Noceda-Davila and Tiago M. Fern\'andez-Caram\'es
conceived and designed the experiments;
Diego Noceda-Davila and Paula Fraga-Lamas performed the experiments; 
Paula Fraga-Lamas, Tiago M. Fern\'andez-Caram\'es, Manuel A. D\'iaz-Bouza, and Miguel Vilar-Montesinos analyzed the data; 
Paula Fraga-Lamas, Tiago M. Fern\'andez-Caram\'es, Diego Noceda-Davila, Manuel A. D\'iaz-Bouza, and Miguel Vilar-Montesinos wrote the paper.}

\conflictofinterests{\textbf{Conflicts of Interest:} The authors declare no conflict of interest. The founding sponsors had no role in the design of the study; in the collection, analyses, or interpretation of data; in the writing of the manuscript, and in the decision to publish the results.} 

\section*{\noindent Abbreviations}\vspace{6pt}\noindent 
The following abbreviations are used in this manuscript:\\

\noindent 
AoA: Angle of Arrival\\ 
BI: Business Intelligence\\
BLE: Bluetooth Low Energy\\
C4ISR: Command, Control, Communications, Computers, Intelligence, Surveillance and Reconnaissance\\
CPS: Cyber-Physical Systems\\
EGC: Equal Gain Combiner\\
HCI: Human-Computer Interfaces\\
IoT: Internet of Things\\
LAN: Local Area Network\\
LOS: Line Of Sight\\
RFID: Radio Frequency IDentification\\
REST: REpresentational State Transfer\\
RSS: Received Signal Strength\\
MES: Manufacturing Execution System\\
MIMO: Multiple-input Multiple-output\\
MRC: Maximum-Ratio Combiner\\
MSE: Minimum-Squared Error\\
SC: Selection Combiner\\
ScanC: Scanner Combiner\\
SSC: Switch-and-Stay Combiner\\
TDoA: Time Different of Arrival\\  
ToA: Time of Arrival\\ 
ToF: Time of Flight\\ 
WPAN: Wireless Personal Area Network\\ 
WSN: Wireless Sensor Networks\\
\\
\\
\\
\\

\bibliographystyle{mdpi}

\begin{thebibliography}{999} 

\bibitem{McKinsey2015}
McKinsey \& Company. \emph{Manufacturing's next act};
\newblock Cornelius, B.; Wee, D.
\newblock White paper; McKinsey \& Company: New York, United States, June 2015.

\bibitem{PwC}
PwC. \emph{2016 Global Industry 4.0 Survey. Industry 4.0: Building the digital enterprise};
\newblock Technical Report; PwC: London, United Kingdom, April 2016.

\bibitem{Wang2016}
Wang, S.; Wan, J.; Li, D.; Zhang, C.
\newblock Implementing Smart Factory of Industrie 4.0: An Outlook.
\newblock {\em International Journal of Distributed Sensor Networks..} {\bf
2016}, {\em 3159805},~1-10. 

\bibitem{cleaning}
Navarro, P. J.; Muro, J. S.; Alcover, P. M.; Fern\'{a}ndez-Isla, C.
\newblock Sensors systems for the automation of operations in the ship repair industry.
\newblock {\em Sensors.} {\bf
2013}, {\em 13},~12345-12374. 
 
\bibitem{Navantia} Navantia S.A. Available online: \url{https://www.navantia.es/index.php} (accessed on 31 October 2016).

\bibitem[Fa\'{\i}\~{n}a\em{et~al.}(May 2009)Fa\'{\i}\~{n}a, Souto, Deibe,L\'{o}pez-Pe\~{n}a, Duro,  Fern\'{a}ndez]{Faina2009} 
Fa\'{\i}\~{n}a, A.; Souto, D.; Deibe, A.; L\'{o}pez-Pe\~{n}a, F. ; Duro, R. J.; Fern\'{a}ndez, X. 
\newblock Development of a climbing robot for grit blasting operations in shipyards.
\newblock In Proceedings of the ICRA'09 2009 IEEE international conference on Robotics and Automation, New York, United States, 12--17 May 2009; pp. 200--205.


\bibitem[Kuss \em{et~al.}(June 2016)Kuss, Schneider, Dietz and Verl]{Kuss2016} 
Kuss, A.; Schneider, U.; Dietz, T.; Verl, A.
\newblock Detection of Assembly Variations for Automatic Program Adaptation in Robotic Welding Systems.
\newblock In Proceedings of the ISR 2016 47st International Symposium on Robotics, Munich, Germany, 21--22 June 2016; pp. 1--6.

\bibitem[Mun \em{et~al.}(July 2015) Mun, Nam, Lee, Doh, Park, Lee, Kim and Lee]{Mun2015} 
 Mun, S.; Nam, M.; Lee, J.; Doh, K.; Park, G.; Lee, H.; Kim, D.; Lee, J.
\newblock Sub-assembly welding robot system at shipyards.
\newblock In Proceedings of the 2015 IEEE International Conference on Advanced Intelligent Mechatronics (AIM), Busan, Korea, 7--11 July 2015; pp. 1502--1507.

\bibitem[Lee(2011)]{Lee2011}
Lee, D.; Ku, N.; Kim, T.-W.; Kim, J.; Lee, K.-Y.; Son, Y.-S.
\newblock Development and application of an intelligent welding robot system for shipbuilding.
\newblock {\em Robotics and Computer-Integrated Manufacturing.} {\bf
2011}, {\em 27},~377-388. 

\bibitem[Kim \em{et~al.}(Oct. 2001)Kim, Cho, and Kim]{NeuralNavigationRobots} 
Kim, M. Y.;  Cho, H. S.; Kim, J. 
\newblock Neural network-based recognition of navigation environment for intelligent shipyard welding robots.
\newblock In Proceedings of the 14th IEEE/RSJ International Conference on Intelligent Robots and Systems, Maui, United States, 29 October--3 November 2001; pp. 446--451.


\bibitem[Kim \em{et~al.}(Nov. 2000)Kim, Ko, Cho, and Kim]{VisualRecognitionEnvironment} 
Kim, M. Y.; Ko, K.; Cho, H. S.; Kim, J.  
\newblock Visual sensing and recognition of welding environment for intelligent shipyard welding robots.
\newblock In Proceedings of the 13th IEEE/RSJ International Conference on Intelligent Robots and Systems, Takamatsu, Japan, 31 October--5 November 2000; pp. 2159--2165.
 
\bibitem[Kawakubo \em{et~al.}(Jan. 2006)Kawakubo, Chansavang, Tanaka, and Iwasaki]{WirelessHumanPositioning} 
Kawakubo, S.; Chansavang, A.; Tanaka, S.; Iwasaki,  T. 
\newblock Wireless network system for indoor human positioning.
\newblock In Proceedings of the 1st International Symposium on Wireless Pervasive Computing, Phuket, Thailand, 16--18 January 2006.

\bibitem[Perez(2014)]{Perez2014}
P{\'e}rez-Garrido, C.; Gonz{\'a}lez-Casta\~no, F.J.; Chaves-D{\'i}eguez, D.; Rodr{\'i}guez-Hern{\'a}ndez, P.S. 
\newblock Wireless Remote Monitoring of Toxic Gases in Shipbuilding.
\newblock {\em Sensors.} {\bf
2014}, {\em 14},~2981-3000. 


\bibitem[Yang(2015)]{yang2015}
Yang, J.; Zhou, J.; Lv, Z.; Wei, W.; Song, H.
\newblock A Real-Time Monitoring System of Industry Carbon Monoxide Based on Wireless Sensor Networks.
\newblock {\em Sensors.} {\bf
2015}, {\em 15},~29535-29546. 

\bibitem[do Amaral Bichet \em{et~al.}(Mar. 2013)Amaral Bichet, Hasegawa, Sol{\'e}, and N{\'u}\~nez]{HyperEnvironmentTracking}
do Amaral Bichet, M. A.; Hasegawa,  E. K.; Sol{\'e}, R.; N{\'u}\~nez, A. 
\newblock Utilization of Hyper Environments for Tracking and Monitoring of Processes and Supplies in Construction and Assembly Industries.
\newblock In Proceedings of the Symposium on Computing and Automation for Offshore Shipbuilding (NAVCOMP), Rio Grande, Brazil, 14--15 March 2013; pp. 81--86.

\bibitem[Wong \em{et~al.}(Dec. 2013)Wong and Zheng]{metalNoiseFactorEffect}
Wong, S. F.; Zheng, Y. 
\newblock The effect of metal noise factor to RFID location system.
\newblock In Proceedings of the IEEE International Conference on Industrial Engineering and Engineering Management, Bangkok, Thailand, 10--13 December 2013; pp. 310--314. 

\bibitem[Deavours(Apr. 2010)Deavours]{improvingPerformanceRFID}
Deavours, D. D.
\newblock Improving the near-metal performance of UHF RFID tags.
\newblock In Proceedings of the IEEE International Conference on RFID, Orlando, United States, 14--16 April 2010; pp. 187--194.

\bibitem[Arumugan\em{et~al.}(Apr. 2008)Arumugan and Engels]{helicalPipesRFID}
Arumugan, D. D.; Engels, D. W. 
\newblock Characterization of RF Propagation in Helical and Toroidal Metal Pipes for Passive RFID Systems.
\newblock In Proceedings of the IEEE International Conference on RFID, Las Vegas, United States, 16--17 April 2008; pp. 269--276. 

\bibitem[Rao\em{et~al.}(Dec. 2010)Rao and Nikitin]{RFIDtagMetal}
Rao, K. V. S.; Lam, S. F.; Nikitin, P. V.
\newblock UHF RFID tag for metal containers.
\newblock In Proceedings of the Asia-Pacific Microwave Conference, Yokohama, Japan, 7--10 December 2010; pp. 179--182.
 
\bibitem[Bovelli\em{et~al.}(Sept. 2006)Bovelli, Neubauer and Heller]{RFIDTransponderAutoID}
Bovelli, S.; Neubauer, F.; Heller, C. 
\newblock Mount-on-Metal RFID Transponders for Automatic Identification of Containers.
\newblock In Proceedings of the 36th European Microwave Conference, Manchester, United Kingdom, 10--15 September 2006; pp. 726--728.
 
\bibitem[Jeong\em{et~al.}(Oct. 2011)Jeong and Son]{RFIDTagConcreteFloor}
Jeong, S. H.; Son, H. W. 
\newblock UHF RFID Tag Antenna for Embedded Use in a Concrete Floor.
\newblock {\em \mbox{IEEE Antennas and Wireless Propagation Letters}}, Oct. 2011.
{\bf 2011}, {\em 10}, pp. 1536--1225.

\bibitem[Heiss\em{et~al.}(June. 2013)Heiss and Hildebrant]{HighTempRFID}
Heiss, M.; Hildebrant, R.
\newblock High-Temperature UHF RFID Sensor Measurements in a Full-Metal Environment.
\newblock In Proceedings of the 2013 European Conference on Smart Objects, Systems and Technologies (SmartSysTech), Nuremberg, Germany, 12--12 June 2013.


\bibitem[QR(2014)]{QR2014}
Tarjan, L.; Senk, I.; Tegeltija, S.; Stankovski, S.; Ostojic, G.
\newblock A readability analysis for QR code application in a traceability system.
\newblock {\em Computers and Electronics in Agriculture.} {\bf
2014}, {\em 109},~1-11. 

\bibitem[RFID(2013)]{RFID2013}
Zhong, R. Y.; Dai,  Q.Y.; Qu, T.;Hu, G.J.; Huang, G. Q.  
\newblock RFID-enabled real-time manufacturing execution system for mass-customization production.
\newblock {\em Robotics and Computer-Integrated Manufacturing.} {\bf
2013}, {\em 29},~283-292. 

\bibitem{smartEpis} Barro-Torres, S. J.; Fern\'andez-Caram\'es, T. M.; P\'erez-Iglesias, H. J.; Escudero, C. J. "Real-Time Personal Protective Equipment Monitoring System", in {\em Computer Communications}, {\bf 2012}, {\em 8(3)}. 

\bibitem{smartsocket} Fern\'andez-Caram\'es, T.M. An Intelligent Power Outlet System for the Smart Home of the Internet of Things. {\em Int. J. Distrib. Sens. Netw. {\bf 2015}, 2015}, ~1-11.

\bibitem{freightcontainers} Barro-Torres, S. J.; Fern\'andez-Caram\'es, T. M.; Gonz\'alez-L\'opez, M.; Escudero-Casc\'on, C. J. "Maritime Freight Container Management System Using RFID", Proceedings of the Third International EURASIP Workshop on RFID Technology, La Manga del Mar Menor, Spain, Sep. 2010.

\bibitem[NFC(2015)]{NFC2015}
Ozdenizci, B.; Coskun, V.; Ok, K.
\newblock NFC Internal: An Indoor Navigation System.
\newblock {\em Sensors.} {\bf
2015}, {\em 15},~7571-7595. 

\bibitem[BLE(2016)]{BLE2016} 
Kitazawa, M.; Takahashi, S.; Takahashi, T. B.; Yoshikawa, A.; Terano, T.
\newblock Improving a Cellular Manufacturing System through Real Time-Simulation and-Measurement.
\newblock In Proceedings of the 2016 IEEE 40th Annual Computer Software and Applications Conference (COMPSAC), Atlanta, GA, 2016; pp. 117-122.


\bibitem[Maki \em{et~al.}(2016)Makki,Siddig,Saad,Cavallaro and Bleakleya]{TimeOfArrival2016}
Makki, A.; Siddig, A.; Saad, M.; Cavallaro, J. R.; Bleakley C. J. 
\newblock Indoor Localization Using 802.11 Time Differences of Arrival.
\newblock {\em IEEE Transactions on Instrumentation and Measurement.} {\bf 2016}, {\em 65}, 614--623.

\bibitem[Fingerprinting(2016)]{fingerprinting2016}
He, S.; Chan, S. H. G.
\newblock Wi-Fi Fingerprint-Based Indoor Positioning: Recent Advances and Comparisons.
\newblock {\em IEEE Communications Surveys \& Tutorials.} {\bf
2016}, {\em 18},~466-490. 

\bibitem[Infrared(2016)]{Infrared2016}
Mousa, M.; Zhang, X.; Claudel, C. 
\newblock Flash Flood Detection in Urban Cities Using Ultrasonic and Infrared Sensors.
\newblock {\em IEEE Sensors Journal.} {\bf
2016}, {\em 16},~7204-7216. 

\bibitem[Ultrasonic(2013)]{Ultrasonic2013} 
Ijaz, F.; Yang, H. K.; Ahmad, A. W.; Lee, C.
\newblock Indoor positioning: A review of indoor ultrasonic positioning systems.
\newblock In Proceedings of the 2013 15th International Conference on Advanced Communication Technology (ICACT), PyeongChang, Korea, 2013; pp. 1146-1150.

\bibitem[UWB(2016)]{UWB2016}
Alarifi, A.; Al-Salman, A.; Alsaleh, M.; Alnafessah, A.; Al-Hadhrami, S.; Al-Ammar, M.A.; Al-Khalifa, H.S. 
\newblock Ultra Wideband Indoor Positioning Technologies: Analysis and Recent Advances.
\newblock {\em Sensors.} {\bf
2016}, {\em 16},~707. 

\bibitem[ZigBee(2011)]{ZigBee2011}
Ruan, Q.; Xu, W.; Wang, G.
\newblock RFID and ZigBee based manufacturing monitoring system.
\newblock In Proceedings of the 2011 International Conference on Electric Information and Control Engineering (ICEICE), Wuhan, China, 2011; pp. 1672-1675.

\bibitem[Dash7(2015)]{Dash72015}
Ergeerts, G.; Nikodem, M.; Subotic, D.; Surmacz, T.; Wojciechowski, B.; De Meulenaere, P.; Weyn, M. 
\newblock DASH7 Alliance Protocol in Monitoring Applications.
\newblock In Proceedings of the 2015 10th International Conference on P2P, Parallel, Grid, Cloud and Internet Computing (3PGCIC), Krakow, Poland, 2015; pp. 623-628.


\bibitem[Ant+(2013)]{Ant+2013}
Belchior, R.; J{\'u}nior, D.; Monterio, A. 
\newblock ANT+ Medical Health Kit for Older Adults.
\newblock {\em Wireless Mobile Communication and Healthcare. Lecture Notes of the Institute for Computer Sciences, Social Informatics and Telecommunications Engineering.} {\bf
2013}, {\em 61},~20-29.

\bibitem[WirelessHART(2015)]{WirelessHART2015}
Horvath, P.; Yampolskiy, M.; Koutsoukos, X.
\newblock Efficient Evaluation of Wireless Real-Time Control Networks.
\newblock {\em Sensors.} {\bf
2015}, {\em 15},~4134-4153.

\bibitem[LoRa(2016)]{LoRa2016}
Augustin, A.; Yi, J.; Clausen, T.; Townsley, W.M.
\newblock A Study of LoRa: Long Range \& Low Power Networks for the Internet of Things.
\newblock {\em Sensors.} {\bf
2016}, {\em 16},~1466.

\bibitem[LoRaandSigFox(2015)]{LoRaandSigFox2015}
Margelis, G.; Piechocki, R.; Kaleshi, D.; Thomas, P.
\newblock Low Throughput Networks for the IoT: Lessons learned from industrial implementations.
\newblock In Proceedings of the 2015 IEEE 2nd World Forum on Internet of Things (WF-IoT), Milan, Italy, 14--16 December 2015; pp. 181-186.

\bibitem[RuBee(2014)]{RuBee2014} 
Boukhtouta, A.; Berger, J. 
\newblock Improving in-transit and in-theatre asset visibility of the Canadian Armed Forces supply chain network.
\newblock In Proceedings of the 2014 International Conference on Advanced Logistics and Transport (ICALT), Hammamet, Tunisia, 1--3 May 2014; pp. 149-154.


\bibitem{ICMCIS}
Fraga-Lamas, P.; Castedo-Ribas, L.; Morales-M{\'e}ndez, A.; Camas-Albar, J.M.
\newblock Evolving military broadband wireless communication systems: WiMAX, LTE and WLAN.
\newblock In Proceedings of the International Conference on Military
Communications and Information Systems (ICMCIS), Brussels, Belgium, 23--24 May
2016; pp.~1--8.

\bibitem{NATOReport}
Camas-Albar, J. M.; Morales-M{\'e}ndez, A.; Castedo-Ribas, L.; Fraga-Lamas, P.; Brown, C.; Tschauner, M.;  Hayri-Kucuktabak, M.
\newblock NATO Task Group ET-IST-068, IST (Information Systems Technology) panel of NATO STO (Science and Technology Organization). \emph{LTE vs. WiMAX for Military Applications};
\newblock Technical Report; NATO: Brussels, Belgium, 2015.

\bibitem{TacticalWimax}
Fraga-Lamas, P.; Camas, J. M.; Carro, A.;  Su{\'a}rez, P.; Castedo, L.;  Garc{\'i}a-Naya, J. A.; Morales, A. 
\newblock Mobile WiMAX for Next Generation Tactical Wireless Networks
\newblock In Information Systems Technology Panel Symposium on Emerged / Emerging 'Disruptive' Technologies (NATO IST-099 / RSY-024), Madrid, Spain, May 2011.

\bibitem{Nets4cars}
Fraga-Lamas, P.; Rodr{\'i}guez-Pi{\~n}eiro, J.; Garc{\'i}a-Naya, J. A.; Castedo, L. 
\newblock Unleashing the Potential of LTE for Next Generation Railway Communications
\newblock Communication Technologies for Vehicles (Proc. of the 8th International Workshop on Communication Technologies for Vehicles (Nets4Cars / Nets4Trains / Nets4Aircraft 2015)), Lecture Notes in Computer Science, vol. 9066, Sousse, Tunisia, May 2015; pp. 153--164.

\bibitem{AranduconSecurity}
Rodr{\'i}guez-Pi{\~n}eiro, J.; Fraga-Lamas, P.; Garc{\'i}a-Naya, J. A.; Castedo, L. 
\newblock Long Term Evolution security analysis for railway communications.
\newblock In IEEE Congreso de Ingenier{\'i}a en Electro-Electr{\'o}nica, Comunicaciones y Computaci{\'o}n (ARANDUCON 2012), Asunci{\'o}n, Paraguay, 28-30 November 2012.

\bibitem{AranduconSurvey}
Fraga-Lamas, P.; Rodr{\'i}guez-Pi{\~n}eiro, J.; Garc{\'i}a-Naya, J. A.; Castedo, L. 
\newblock A survey on LTE networks for railway services.
\newblock In IEEE Congreso de Ingenier{\'i}a en Electro-Electr{\'o}nica, Comunicaciones y Computaci{\'o}n (ARANDUCON 2012), Asunci{\'o}n, Paraguay, 28-30 November 2012.

\bibitem{1-Li2012} Li, H.; Chen, Y.; He, Z. The survey of RFID attacks and defenses, Proceedings of the 8th International Conference on Wireless Communications, Networking and Mobile Computing, Shanghai, China, Sep. 2012, pp. 1-4.

\bibitem{Fernandez2016} Fern\'andez-Caram\'es, T. M.; Fraga-Lamas, P.; Su\'arez-Albela, M.; Castedo, L. A methodology for evaluating security in commercial RFID systems. To be published in {\em Radio Frequency Identification}, 1st ed.; Crepaldi, P. C.; Pimenta, T. C.; INTECH: Rijeka, Croatia, 2016.

\bibitem{fraga2016}
Fraga-Lamas, P.; Fern{\'a}ndez-Caram{\'e}s, T. M.; Su{\'a}rez-Albela, M.; Castedo, L.; Gonz{\'a}lez-L{\'o}pez, M.
\newblock A Review on Internet of Things for Defense and Public Safety.
\newblock {\em Sensors.} {\bf
2016}, {\em 16},~1644.

\bibitem{Pantex}
B\&W Pantex. \emph{Advanced inventory and materials management at Pantex};
\newblock White Paper; B\&W Pantex: Amarillo, United States, 2011.

\bibitem{Rappaport2002} 
Rappaport, T. S. {\em Wireless Communications: Principles and Practice}, 2nd ed.; Prentice Hall: Upper Saddle River, New Jersey, United States, 2002.

\bibitem{Grewal2008} Grewal, M. H.; Andrews, A. P. {\em Kalman Filtering: theory and practice using Matlab}, 3rd ed.; John Wiley \& Sons: Hoboken, New Jersey, United States, 2008.

\bibitem[Bulten \em{et~al.}(2016)Bulten, Rossum and Haselager]{Bulten2016} 
Bulten, W.; Rossum, A. C. V.; Haselager, W. F. G.
\newblock Human SLAM, Indoor Localisation of Devices and Users.
\newblock In Proceedings of the IEEE First International Conference on Internet-of-Things Design and Implementation, Berlin, Germany, 4--8 Apr. 2016; pp. 221-222.

\bibitem[MIMO(2013)]{MIMO2013}
Clerckx, B.; Oestges, C. 
\newblock {\em MIMO Wireless Networks: Channels, Techniques and Standards for Multi-Antenna, Multi-User and Multi-Cell Systems}, 2nd ed.; Academic Press: Cambridge, Massachusetts, 2013. 

\bibitem[Fernandez\em{et~al.}(2007)] {Fernandez2007} 
Fern{\'a}ndez, T. M.; Rodas, J; Escudero, C. J.; Iglesia, D. I;
\newblock Bluetooth Sensor Network Positioning System with Dynamic Calibration.
\newblock In Proceedings of the 4th IEEE International Symposium on
Wireless Communication Systems, Trondheim, Norway, 16-19 Oct. 2007; pp. 45-49.

\bibitem{Speedway} Speedway Revolution R420 from Impinj. Available online: \url{http://www.impinj.com/products/readers/speedway-revolution} (accessed on 31 October 2016).

\bibitem{A6-UHFLongRange} A6-UHFLongRange. Available online: \url{http://www.nextpoints.com/es/productos-rfid/item/196-rugged-pda-a6-rfid-uhf.html} (accessed on 31 October 2016).

\bibitem{Omni-ID} Omni-ID products. Available online: \url{https://www.omni-id.com/industrial-rfid-tags} (accessed on 31 October 2016).

\bibitem{NPR} NPR ActiveTrack-2 New Edition. Available online: \url{http://www.nextpoints.com/es/productos-rfid/item/187-npr-active-track-2-new-edition.html} (accessed on 31 October 2016).

\bibitem{Active RuggedTag-175S} Active RuggedTag-175S. Available online: \url{http://www.nextpoints.com/es/productos-rfid/item/319-tag-rfid-activo-active-rugged-tag-175s.html} (accessed on 31 October 2016).
 
\bibitem{LIPS}  
Mascharka, D.; Manley, E. 
\newblock LIPS: Learning Based Indoor Positioning System using mobile phone-based sensors
\newblock In Proceedings of the 13th IEEE Annual Consumer Communications \& Networking Conference (CCNC), Las Vegas, NV, 2016, pp. 968-971.

\bibitem{SDM}  
Lim,H.;Kung, L.-C.; Hou,J.C.; Luo,H. 
\newblock Zero-configuration indoor localization over ieee 802.11 wireless infrastructure
\newblock {\em Wireless Network.}{\bf
2010}, {\em 16},~405-420.

\bibitem{OIL} 
Park, J.-G.; Charrow, B.; Curtis, D.; Battat, J.; Minkov, E.; Hicks, J.; Teller, S.;  Ledlie, J.
\newblock Growing an organic indoor location system,
\newblock In Proceedings of the 8th international conference on Mobile systems, applications, and services, 
San Francisco, California, USA, 15-18 June 2010; pp. 271-284.


\bibitem{EZ} 
Chintalapudi, K.; Padmanabha Iyer, A.; Padmanabhan, V. N. 
\newblock Indoor localization without the pain. 
\newblock In Proceedings of the sixteenth annual international conference on Mobile computing and networking, Chicago, Illinois, 20-24 Sep. 2010; pp. 173-184.
 
\bibitem{WiGEM} 
Goswami, A.;  Ortiz, L. E.; Das, S. R.
\newblock WiGEM: a learning-based approach for indoor localization.
\newblock In Proceedings of the  Seventh COnference on emerging Networking EXperiments and Technologies ACM CoNEXT, Tokyo, Japan, 6-9 December 2011; pp. 1-12.
 

\bibitem{WILL}  
Wu, C.;  Yang, Z.; Liu, Y.; Xi, W. 
\newblock WILL: Wireless Indoor Localization without Site Survey
\newblock {\em IEEE Transactions on Parallel and Distributed Systems.}{\bf 2013}, {\em 24},~839-848.

\bibitem{UnLoc}  
Wang, H.; Sen, S.; Elgohary, A.; Farid, M.; Youssef, M.;  Choudhury, R. R.  
\newblock No need to war-drive: Unsupervised indoor localization.
\newblock In Proceedings of the ACM 18th Annual International Conference on Mobile Computing and Networking (MobiCom 2012), Istambul, Turkey, 22-26 August 2012; pp. 197-210.
 

\bibitem{Zee} 
Rai, A.; Chintalapudi, K.K.; Padmanabhan, V.N.;  Sen, R.
\newblock Zee: Zero-effort crowdsourcing for indoor localization
\newblock In Proceedings of the ACM 18th Annual International Conference on Mobile Computing and Networking (MobiCom 2012), Istambul, Turkey,  22-26 August 2012; pp. 293-304.

\bibitem{LiFS} 
Yang, Z.; Wu, C.; Liu, Y. 
\newblock Locating in fingerprint space: wireless indoor localization with little human intervention.
\newblock In Proceedings of the ACM 18th Annual International Conference on Mobile Computing and Networking (MobiCom 2012), Istambul, Turkey,  22-26 August 2012; pp. 269-280.
 
  
\bibitem{Walkie-Markie}
Shen, G.; Chen, Z.; Zhang, P.;  Moscibroda, T.; Zhang, Y. 
\newblock Walkie-Markie: Indoor pathway mapping made easy.
\newblock In Proceedings of the 10th USENIX Conference on Networked Systems Design and Implementation, Berkeley, CA, USA, 2013; pp. 85-98.
 

\bibitem{RADAR} 
Bahl, P.; Padmanabhan, V. N. 
\newblock RADAR: An in-building RF-based user location and tracking system.
\newblock In Proceedings of the Nineteenth Annual Joint Conference of the IEEE Computer and Communications Societies (INFOCOM 2000), 26-30 March 2000, Vol. 2; pp. 775-784.
 
 
\bibitem{Horus} 
Youssef, M.; Agrawala, A.
\newblock Handling samples correlation in the horus system.
\newblock In Proceedings of the Twenty-third Annual Joint Conference of the IEEE Computer and Communications Societies (INFOCOM 2004), Hong Kong, China, 7-11 March 2004, Vol. 2; pp. 1023-1031.
 

\bibitem{AiRISTA} AiRISTA. Available online: \url{https://www.airistaflow.com/hardware/} (accessed on 31 October 2016).
 
\bibitem{IZat} IZat. Available online: \url{https://www.qualcomm.com/products/izat} (accessed on 31 October 2016).

\bibitem{Ubisense} Ubisense. Available online: \url{https://ubisense.net/en} (accessed on 31 October 2016).
 
\bibitem{Dart} Dart system. Available online: \url{https://www.zebra.com/us/en/solutions/location-solutions/enabling-technologies/dart-uwb.html} (accessed on 31 October 2016).
 
 
\bibitem{3D-LOCUS}
Prieto, J. C.;Jim\'enez, A. R.; Guevara, J.; Ealo, J. L.; Seco, F.; Roa, J. O.; Ramos, F. 
\newblock Performance evaluation of 3D-LOCUS advanced acoustic LPS.  
\newblock {\em IEEE transactions on instrumentation and measurement.}{\bf 2009}, {\em 58},~2385-2395.


\bibitem{Elpas} Elpas (Gdsystems). Available online: \url{http://www.gdsystems.com/staff-attack-alarms/elpas/} (accessed on 31 October 2016).

\bibitem{SpotON} 
 Hightower, J.; Want, R.; Borriello, G. 
\newblock SpotON: An indoor 3D location sensing technology based on RF signal strength. 
\newblock UW CSE 00-02-02 Technical report, University of Washington, Department of Computer Science and Engineering, Seattle, WA,~1, February 2000.
 
\bibitem{TOPAZ} Topaz local positioning (Tadlys). Available online: \url{http://www.tadlys.co.il/pages/Product_content.asp?iGlobalId=2} (accessed on 31 October 2016).

\bibitem{Landmarc} 
Ni, L. M.; Liu, Y.; Lau, Y. C.; Patil, A. P.
\newblock LANDMARC:Indoor location sensing using active RFID.
\newblock {\em Wireless Networks.}{\bf
2004}, {\em 10},~701-710.

\bibitem{sparse}
Yang, P.;  Wu, W.; Moniri, M.; Chibelushi, C. C. 
\newblock Efficient Object Localization Using Sparsely Distributed Passive RFID Tags.
\newblock {\em IEEE Transactions on Industrial Electronics.} {\bf
2013}, {\em 60},~5914-5924.

\bibitem{GPPS}
Seco, F.; Plagemann, C.; Jim\'enez, A. R.; Burgard, W. 
\newblock Improving RFID-based indoor positioning accuracy using Gaussian processes
\newblock In Proceedings of the 2010 International Conference on Indoor Positioning and Indoor Navigation, Zurich, Switzerland, 15-17 September 2010, pp. 1-8.

\bibitem{Robot}
Ladd, A. M.; Bekris, K. E.; Rudys, A.; Kavraki, L. E.; Wallach, D. S.
\newblock Robotics-Based Location Sensing Using Wireless Ethernet.
\newblock {\em Wireless Networks.} {\bf
2005}, {\em 11},~189-204.

\bibitem{MultiLoc}
Prasithsangaree, P.; Krishnamurthy, P.; Chrysanthis, P. 
\newblock On indoor position location with wireless LANs.
\newblock In Proceedings of the 13th IEEE International Symposium on Personal, Indoor and Mobile Radio Communications, 2002, pp. 720-724.

\bibitem{Zigbee} 
Luoh, L.
\newblock ZigBee-based intelligent indoor positioning system soft computing. 
\newblock {\em Soft Computing.} {\bf
2014}, {\em 18},~443-456.

\bibitem{TIX}
Gwon, Y.; Jain, R.  
\newblock Error characteristics and calibration-free techniques for wireless LAN-based location estimation. 
\newblock In Proceedings of the second international workshop on Mobility management \& wireless access protocols, Oct. 2004, pp. 2-9.
 
\bibitem{BLE}
Faragher, R.; Harle, R.
\newblock Location Fingerprinting With Bluetooth Low Energy Beacons.
\newblock {\em IEEE Journal on Selected Areas in Communications.} {\bf
2015}, {\em 33},~2418-2428.
 
\bibitem{GSMfingerprinting}  
Otsason, V.; Varshavsky, A.; LaMarca, A.; de Lara, E.
\newblock Accurate GSM indoor localization.
\newblock {\em UbiComp 2005, Lecture Notes Computer Science,
Springer-Varlag,.} {\bf
2005}, {\em 3660},~141-158.
 
\bibitem{Optotrak} Optotrak certus system. Available online: \url{ http://www.ndigital.com/msci/products/optotrak-certus/} (accessed on 31 October 2016).

\bibitem{IRIS_LPS} 
Aitenbichler, E.;  M\"uhlh\"auser, M. 
\newblock An IR Local Positioning System for Smart Items and Devices.
\newblock In Proceedings of the 23rd IEEE International Conference on Distributed Computing Systems Workshops (IWSAWC03), 19--22 May 2003.

\bibitem{Bat} Bat system. Available online: \url{http://www.cl.cam.ac.uk/research/dtg/attarchive/bat/
} (accessed on 31 October 2016).

\bibitem{Cricket}
Priyantha, N. B.  
\newblock The Cricket Indoor Location System.
\newblock PhD thesis, Massachusetts Institute of Technology (MIT): Cambridge, Massachusetts, United States, 2005.

\bibitem{Sonitor} Sonitor. Available online: \url{http://sonitor.com/} (accessed on 31 October 2016).

\bibitem{COMPASS}
King,  T.; Kopf, S.; Haenselmann, T.; Lubberger, C.; Effelsberg, W. 
\newblock COMPASS: A Probabilistic Indoor Positioning System Based on 802.11 and Digital Compasses.
\newblock In Proceedings of the First ACM Intl Workshop on Wireless Network Testbeds, Experimental evaluation and CHaracterization (WiN-TECH), Los Angeles, CA, USA, 29 September, 2006, pp. 34--40.

\bibitem{OPT} 
 An, X.; Wang, J.; Prasad, R. V.; Niemegeers, I. G. M. M.
 \newblock OPT: online person tracking system for context-awareness in wireless personal network.
 \newblock In Proceedings of the 2nd international workshop on Multi-hop ad hoc networks: from theory to reality (REALMAN '06), Florence, Italy, 26 May 2006; pp. 47-54.
 
\bibitem{EasyLiving}
Brumitt, B.; Meyers, B.; Krumm, J.; Kern, A.; Shafer, S. 
\newblock  Easyliving: Technologies for intelligent environments.
\newblock In Proceedings of the 2nd International Symposium on Handheld and Ubiquitous Computing,
Bristol, UK, 25-27 September 2000; pp. 12-29.

\bibitem{Beep}
Lopes, C. V.; Haghighat, A.;  Mandal, A.; Givargis, T.; Baldi, P. 
\newblock Localization of Off-the-Shelf Mobile Devices Using Audible Sound: Architectures, Protocols and Performance Assessment.
\newblock {\em ACM SIGMOBILE Mobile Computing and Communication Review.} {\bf
2006}, {\em 10},~2.




\end{thebibliography}
\renewcommand\bibname{References}

\end{document}